\documentclass{article}

\usepackage[a4paper]{geometry}

\usepackage[utf8]{inputenc}
\usepackage[american]{babel}
\usepackage{graphicx}
\usepackage{amsmath}
\usepackage{amssymb}
\usepackage{bm}
\usepackage{xcolor}
\usepackage{glossaries}
\usepackage{xr-hyper}
\usepackage{hyperref}
\usepackage{natbib}
\usepackage{placeins} %Float Barrier for Appendix
\usepackage{subcaption} %Include Picture and Plot
\usepackage{multirow} %Merge Table Cells
\usepackage{authblk}

\newcommand{\diag}{\operatorname{diag}}

\newcommand{\vect}{\operatorname{vec}}

\title{Multivariate Functional Additive Mixed Models}
\author[1]{Alexander Volkmann}
\author[1]{Almond Stöcker}
\author[2]{Fabian Scheipl}
\author[1]{Sonja Greven}
\affil[1]{Chair of Statistics, 
      School of Business and Economics,
      Humboldt-Universit\"at zu Berlin,
      Germany}
\affil[2]{Department of Statistics,
      Ludwig-Maximilians-Universit\"at M\"unchen, 
      Germany}
\date{}
\newacronym[longplural={functional principal component analyses}]{fpca}{FPCA}{functional principal component analysis}
\newacronym{flmm}{FLMM}{functional linear mixed model}
\newacronym{famm}{FAMM}{functional additive mixed model}
\newacronym{pls}{PLS}{penalized least squares}
\newacronym{reml}{REML}{restricted maximum likelihood}
\newacronym{fpc}{FPC}{functional principal component}
\newacronym{kl}{KL}{Karhunen-Loève}
\newacronym{mfpca}{MFPCA}{multivariate functional principal component analysis}
\newacronym{epg}{EPG}{electropalatographic}
\newacronym{aco}{ACO}{acoustic}
\newacronym{edf}{EDF}{effective degrees of freedom}
\newacronym{mse}{rrMSE}{root relative mean squared error}
\newacronym{umse}{urrMSE}{univariate root relative mean squared error}
\newacronym{mmse}{mrrMSE}{multivariate root relative mean squared error}
\newacronym{mfamm}{multiFAMM}{multivariate functional additive mixed model}
\newacronym{fmm}{FMM}{functional mixed model}
\newacronym{iid}{i.i.d.}{independent and identically distributed}
\newacronym{cb}{CB}{confidence band}
\newacronym{tv}{TV}{truncation via total variation}
\newacronym{uv}{UV}{truncation via univariate variation}
\glsdisablehyper

\begin{document}

\maketitle

\begin{abstract}
Multivariate functional data can be intrinsically multivariate like movement trajectories in 2D or complementary like precipitation, temperature, and wind speeds over time at a given weather station. We propose a multivariate functional additive mixed model (multiFAMM) and show its application to both data situations using examples from sports science (movement trajectories of snooker players) and phonetic science (acoustic signals and articulation of consonants). The approach includes linear and nonlinear covariate effects and models the dependency structure between the dimensions of the responses using multivariate functional principal component analysis. Multivariate functional random intercepts capture both the auto-correlation within a given function and cross-correlations between the multivariate functional dimensions. They also allow us to model between-function correlations as induced by e.g.\ repeated measurements or crossed study designs. Modeling the dependency structure between the dimensions can generate additional insight into the properties of the multivariate functional process, improves the estimation of random effects, and yields corrected confidence bands for covariate effects. Extensive simulation studies indicate that a multivariate modeling approach is more parsimonious than fitting independent univariate models to the data while maintaining or improving model fit.
\end{abstract}

\section{Introduction}

With the technological advances seen in recent years, functional data sets are increasingly multivariate. They can be multivariate with respect to the domain of a function, its codomain, or both. Here, we focus on multivariate functions with a one-dimensional domain $\bm{f} = (f^{(1)},..., f^{(D)}) \colon \mathcal{I} \subset \mathbb{R} \to \mathbb{R}^{D}$ with square-integrable components $f^{(d)} \in L^2(\mathcal{I}), d = 1,..., D$.  For this type of data, we can distinguish two subclasses: One has interpretable separate dimensions and can be seen as several complementary modes of a common phenomenon \citep[``multimodal'' data, cf.~][]{uludaug2014general}  as in the analysis of acoustic signals and articulation processes in speech production
in one of our data examples. The codomain then  simply is the Cartesian product $\mathcal{S} = \mathcal{S}^{(1)} \times ... \times \mathcal{S}^{(D)}$ of interpretable univariate codomains $\mathcal{S}^{(d)} \subset \mathbb{R}$. The other subclass is more ``intrinsically'' multivariate insofar as univariate analyses would not yield meaningful results. Consider for example two-dimensional movement trajectories as in one of our motivating applications, where the function measures Cartesian coordinates over time: for fixed trajectories, rotation or translation of the essentially arbitrary coordinate system would change the results of univariate analyses. For intrinsically multivariate functional data a multivariate approach is the natural and preferred mode of analysis, yielding interpretable results on the observation level. Even for multimodal functional data, a joint analysis may generate additional insight by incorporating the covariance structure between the dimensions. This motivates the development of statistical methods for multivariate functional data. We here propose multivariate functional additive mixed models to model potentially sparsely observed functions with flexible covariate effects and crossed or nested study designs. 

Multivariate functional data have been the interest in different statistical fields such as clustering \citep{jacques2014, park2017}, functional principal component analysis \citep{chiou2014, happ2018, backenroth2018, li2020}, and registration \citep{carroll2020, steyer2020}. There is also ample literature on multivariate functional data regression such as graphical models \citep{zhu2016}, reduced rank regression \citep{liu2020}, or varying coefficient models \citep{zhu2012, Li2017}. Yet, so far, there are only few approaches that can handle multilevel regression when the functional response is multivariate. In particular, \citet{goldsmith2016} propose a hierarchical Bayesian multivariate functional regression model that can include subject level and residual random effect functions to account for correlation between and within functions. They work with bivariate functional data observed on a regular and dense grid and assume \emph{a priori} independence between the different dimensions of the subject-specific random effects. Thus, they model the correlation between the dimensions only in the residual function. As our approach explicitely models the dependencies between dimensions for multiple functional random effects and also handles data observed on sparse and irregular grids on more than two dimensions, the model proposed by \citet{goldsmith2016} can be seen as a special case of our more general model class.

Alternatively, \citet{zhu2017} use a two stage transformation with basis functions for the multivariate functional mixed model. This allows the estimation of scalar regression models for the resulting basis coefficients that are argued to be approximately independent. The proposed model is part of the so called \gls{fmm} framework \citep{morris2017}. While \glspl{fmm} use basis transformations of functional responses (observed on equal grids) at the start of the analysis, we propose a multivariate model in the \gls{famm} framework, which uses basis representations of all (effect) functions in the model \citep{scheipl2015}. The differences between these two functional regression frameworks have been extensively discussed before \citep{greven2017, morris2017}.

The main advantages of our multivariate regression model, also compared to \citet{goldsmith2016} and \citet{zhu2017}, are that it is readily available for sparse and irregular functional data and that it allows to include multiple nested or crossed random processes, both of which are required in our data examples. Another important contribution is that our approach directly models the multivariate covariance structure of all random effects included in the model using multivariate \glspl{fpc} and thus implicitly models the covariances between the dimensions. This makes the model representation more parsimonious, avoids assumptions difficult to verify, and allows further interpretation of the random effect processes, such as their relative importance and their dominating modes. As part of the \gls{famm} framework, our model provides a vast toolkit of modeling options for covariate and random effects, of estimation and inference \citep{wood2017}. The proposed \gls{mfamm} extends the \gls{famm} framework combining ideas from multilevel modeling \citep{cederbaum2016} and multivariate functional data \citep{happ2018} to account for sparse and irregular functional data and different study designs. 

We illustrate the \gls{mfamm} on two motivating examples. The first (intrinsically multivariate) data stem from a study on the effect of a training programme for snooker players with a nested study design (shots within sessions within players) \citep{snooker2014}. The movement trajectories of a player's elbow, hand, and shoulder during a snooker shot are recorded on camera, yielding six-dimensional multivariate functional data (see Figure \ref{fig:snook_obs}). In the second data example, we analyse multimodal data from a speech production study with a crossed study design (speakers crossed with words) \citep{pouplier2016} on so-called ``assimilation'' of consonants. The two measured modes (acoustic and articulatory, see Figure \ref{fig:speech_obs}) are expected to be closely related but joint analyses have not yet incorporated the functional nature of the data.

These two examples motivate the development of a regression model for sparse and irregularly sampled multivariate functional data that can incorporate crossed or nested functional random effects as required by the study design in addition to flexible covariate effects. The proposed approach is implemented in \texttt{R} \citep{rsoftware} in package \texttt{multifamm}\citep{multifammpackage}. The paper is structured as follows: Section \ref{sec:MultiFAMM} specifies the \gls{mfamm} and section \ref{sec:Estimation} its estimation process. Section \ref{sec:application} presents the application of the \gls{mfamm} to the data examples and section \ref{sec:simulation} shows the estimation performance of our proposed approach in simulations. Section \ref{sec:discussion} closes with a discussion and outlook.

\section{Multivariate Functional Additive Mixed Model}
\label{sec:MultiFAMM}

\subsection{General Model}

Let $\bm{y}^{*}_i(t) = (y_i^{*(1)}(t),..., y_i^{*(D)}(t))^{\top}$ be the multivariate functional response of unit $i=1,...,N$ over $t \in \mathcal{I}$, consisting of dimensions $d=1,...,D$. Without loss of generality, we assume a common one-dimensional interval domain $\mathcal{I} = [0,1]$ for all dimensions, and square-integrable $y_i^{*(d)}\in L^2(\mathcal{I})$. Define $L^2_{D}(\mathcal{I}) := L^2(\mathcal{I}) \times ...\times L^2(\mathcal{I})$ so that $\bm{y}^{*}_i\in L^2_D(\mathcal{I})$. The underlying smooth function $\bm{y}_i^{*}$, however, is only evaluated at (potentially sparse or dimension-specific) points $\bm{y}_{it}^{*} = (y_{it}^{*(1)}, ..., y_{it}^{*(D)})^{\top}$ and the evaluation is subject to white noise, i.e.\ $\bm{y}_{it} = \bm{y}_{it}^{*} + \bm{\epsilon}_{it}$. The residual term $\bm{\epsilon}_{it}$ reflects additional uncorrelated white noise measurement error, following a $D$-dimensional multivariate normal distribution $\mathcal{N}_D$ with zero-mean and diagonal covariance matrix $\tilde{\bm{\Sigma}} = \diag(\sigma_{1}^2, ..., \sigma_{D}^2)$ with dimension specific variances $\sigma_{d}^2$.  We construct a multivariate functional mixed model as
\begin{align} \label{eq:multivariateFLMM}
	\begin{split}	
		\bm{y}_{it} &= \bm{y}_i^{*}(t) + \bm{\epsilon}_{it}= \bm{\mu}(\bm{x}_i, t) + \bm{U}(t)\bm{z}_i + \bm{\epsilon}_{it} \\
		&= \bm{\mu}(\bm{x}_i, t) + \sum_{j=1}^q\bm{U_j}(t)\bm{z}_{ij} + \bm{E}_i(t) + \bm{\epsilon}_{it}, \quad t \in \mathcal{I},		
	\end{split}
\end{align}
where
\begin{align*}
	\bm{U}_{j}(t) &= (\bm{U}_{j1}(t), ..., \bm{U}_{jV_{U_j}}(t)); j = 1,..., q,\\
	\bm{U}_{jv}(t) &\stackrel{\text{ind.c.}}{\sim} MGP\left(\bm 0, K_{U_{j}}\right); v = 1,..., V_{U_j}; \forall j = 1,...,q, \\ 
	\bm{E}_{i}(t) &\stackrel{\text{ind.c.}}{\sim} MGP\left(\bm 0, K_{E} \right); i = 1,..., N, \text{ and }\\
	\bm{\epsilon}_{it} &\stackrel{\text{i.i.d.}}{\sim} \mathcal{N}_D\left(\bm 0, \tilde{\bm{\Sigma}} = \diag(\sigma_{1}^2, ..., \sigma_{D}^2)\right) ; i = 1,..., N; \quad t \in \mathcal{I}.
\end{align*}
We assume an additive predictor $\bm{\mu}(\bm{x}_i, \cdot) = \sum_{l = 1}^p \bm{f}_l (\bm{x}_{i}, \cdot)$ of fixed effects, which consists of partial predictors  $\bm{f}_{l}(\bm{x}_{i}, \cdot) = (f_{l}^{(1)}(\bm{x}_{i}, \cdot), ..., f_{l}^{(D)}(\bm{x}_{i}, \cdot))^{\top}\in L^2_D(\mathcal{I}),\ l = 1,...,p$, that are multivariate functions depending on a subset of the vector of scalar covariates $\bm{x}_{i}$. This allows to include linear or smooth covariate effects as well as interaction effects between multiple covariates as in the univariate \gls{famm} \citep{scheipl2015}. Partial predictors may also depend on dimension specific subsets of covariates.

For random effects $\bm{U}$, we focus on model scenarios with $q$ independent multivariate functional random intercepts for crossed and/or nested designs. For group level $v = 1,\dots, V_{U_j}$ within grouping layer $j=1,\dots,q$, these take the value $\bm{U}_{jv} \in L^2_D(\mathcal{I})$. For each layer, the $\bm{U}_{j1}, ..., \bm{U}_{jV_{U_j}}$ present independent copies of a multivariate smooth zero-mean Gaussian random process. Analogously to scalar linear mixed models, the $\bm{U}_{jv}$ model correlations between different response functions $\bm{y}_i^{*}$ within the same group as well as variation across groups. By arranging them in a $(D \times V_{U_j})$ matrix $\bm{U}_j(t)$ per $t$, the $j$th random intercept can be expressed in the common mixed model notation in (\ref{eq:multivariateFLMM}) using appropriate group indicators $\bm{z}_{ij} = (z_{ij1}, \dots, z_{ijV_{U_j}})^\top$ for the respective design. 

Although technically a curve-specific functional random intercept, we distinguish the smooth residuals $\bm{E}_i\in L^2_D(\mathcal{I})$ in the notation, as they model correlation within rather than between response functions. We  write $\bm{E}_v \in  L^2_D(\mathcal{I}), v = 1,..., V_E$ with $V_E = N$. The $\bm{E}_i$ capture smooth deviations from the group-specific mean $\bm{\mu}(\bm{x}_i, \cdot) + \sum_{j=1}^{q}\bm{U}_j(\cdot)\bm{z}_{ij}$.

For a more compact representation, we can arrange all $\bm{U}_j(t)$ and $\bm{E}_i(t)$ together in a $(D \times (\sum_{j=1}^qV_{U_j} + N))$ matrix $\bm{U}(t)$ per $t$,  and the group indicators for all layers in a corresponding vector $\bm{z}_i = (\bm{z}_{i1}^\top, \dots, \bm{z}_{iq}^\top, \boldsymbol{e}_i^\top)^\top$ with $\boldsymbol{e}_i$ the $i$-th unit vector. The resulting model term $\bm{U}(t) \bm{z}_i$ then comprises all smooth random functions, accounting for all correlation between/within response functions $\bm{y}_i^{*}$ given the covariates $\bm{x}_i$ as required by the respective experimental design.

$\bm{E}_i$ and $\bm{U}_{jv}$  are independent copies (ind.\ c.) of random processes having multivariate $D\times D$ covariance kernels $K_E, K_{U_j}$, with univariate covariance surfaces $K_{E}^{(d,e)}(t, t') = \text{Cov}\left[E_i^{(d)}(t), E_i^{(e)}(t')\right]$ and $K_{U_{j}}^{(d,e)}(t, t') = \text{Cov}\left[U_{jv}^{(d)}(t), U_{jv}^{(e)}(t')\right]$ reflecting the covariance between the process dimensions $d$ and $e$ at $t$ and $t'$. We %will 
call these auto-covariances for $d = e$ and cross-covariance otherwise. The multivariate Gaussian processes are uniquely defined by their multivariate mean function, here the null function $\bf{0}$, and the multivariate covariance kernels $K_g$ and we write $MGP\left(\bm{0}, K_g\right), g \in \{U_1, \dots, U_q, E\}$. Note that vectorizing the matrix $\bm{U}(t)$ allows to formulate the joint distribution assumption $\vect(\bm{U}(t)) \sim MGP\left(\bm{0}, K_{U}\right)$ with $K_{U}(t, t')$ having a block-diagonal structure repeating each $K_{U_j}(t,t')$ for 
$V_{U_j}$ times and $K_E(t,t')$ for $N$ times.
 
We assume that the different sources of variation $\bm{U}_j(t),j = 1,...,q, \bm{E}_i(t)$, and $\bm{\epsilon}_{it}$ are mutually uncorrelated random processes to assure model identification. Assuming smoothness of the covariance kernel $K_E$ further guarantees that the residual process $\bm{E}_i(t)$ can be separated from the white noise $\bm{\epsilon}_{it}$, removing the error variance from the diagonal of the smooth covariance kernel \citep[e.g.,][]{yao2005functional}.

\subsection{FPC Representation of the Random Effects}

Model \eqref{eq:multivariateFLMM} specifies a univariate \gls{flmm} as given in \citet{cederbaum2016} for each dimension $d$. The main difference lies in the multivariate random processes that introduce dependencies between the dimensions. In order to avoid restrictive assumptions about the structure of these multivariate covariance operators, which would typically be very difficult to elicit \emph{a priori} or verify \emph{ex post}, we estimate them directly from the data. The main difficulty then becomes computationally efficient estimation, which is already costly in the univariate case. Especially for higher dimensional multivariate functional data, accounting for the cross-covariances can become a complex task, which we tackle with \gls{mfpca}. 

Given the covariance operators (see Section \ref{sec:Estimation}), we represent the multivariate random effects %included - SG 
in model \eqref{eq:multivariateFLMM} using truncated multivariate \gls{kl} expansions
\begin{equation}
\begin{aligned}
\label{eq:truncated_KL_expansion}
\bm{U}_{jv} (t) &\approx \sum_{m=1}^{M_{U_j}}\rho_{U_j v m}\bm{\psi}_{U_jm}(t),\;  j = 1,...,q;\, v =1, ..., V_{U_j},\\
\bm{E}_v(t) &\approx \sum_{m=1}^{M_E}\rho_{Ev m}\bm{\psi}_{Em}(t), \; v = 1, ..., N,
\end{aligned}
\end{equation}
where the orthonormal multivariate eigenfunctions $\bm{\psi}_{gm} = (\psi_{gm}^{(1)}, ..., \psi_{gm}^{(D)})^{\top}\in L^2_D(\mathcal{I})$, $m = 1,..., M_g$, $g\in\{U_1, ..., U_q, E\}$ of the corresponding covariance operators with truncation order $M_g$ are used as basis functions and the random scores $\rho_{gvm} \sim N(0, \nu_{gm})$  are \gls{iid} with ordered eigenvalues $\nu_{gm}$ of the corresponding covariance operator. Note that the assumption of Gaussianity for the random processes can be relaxed. For non-Gaussian random processes, the \gls{kl} expansion still gives uncorrelated (but non-normal) scores and estimation based on a \gls{pls} criterion (see Section \ref{subsec:FAMM}) remains reasonable.

Using \gls{kl} expansions gives a parsimonious representation of the multivariate random processes that is an optimal approximation with respect to the integrated squared error \citep[cf.][]{ramsay2005}, as well as interpretable basis functions capturing the most prominent modes of variation of the respective process. The distinct feature of this approach is that the multivariate \glspl{fpc} directly account for the dependency structure of each random process across the dimensions. If, by contrast, e.g.\ splines were used in the basis representation of the random effects, it would be necessary to explicitly model the cross-covariances of each random process in the model \citep[cf.][]{li2020}. Multivariate eigenfunctions, however, are designed to incorporate the dependency structure between dimensions and allow the assumption of independent (univariate) basis coefficients $\rho_{gvm}$ via the \gls{kl} theorem \citep[see e.g.~][]{happ2018}. This leads to a parsimonious multivariate basis for the random effects, where a typically small vector of scalar scores $\rho_{gvm}$ common to all dimensions represents nearly the entire information about these $D$-dimensional processes.

\section{Estimation}
\label{sec:Estimation}

We use a two-step approach to estimate the \gls{mfamm} and the respective multivariate covariance operators. In a first step (section \ref{subsec:EigenfunctionEstimation}), the D-dimensional eigenfunctions $\bm{\psi}_{gm}(t)$ with their corresponding eigenvalues $\nu_{gm}$ are estimated from their univariate counterparts following \citet{cederbaum2018} and \citet{happ2018}. These estimates are then plugged into \eqref{eq:truncated_KL_expansion} and we represent the \gls{mfamm} as part of the general \gls{famm} framework (section \ref{subsec:FAMM}) by suitable re-arrangement. We can view the estimated $\bm{\psi}_{gm}(t)$ simply as an empirically derived basis that parsimoniously represents the patterns in the observed data. While their estimation adds uncertainty, we are not interested in inferential statements for the variance modes and our simulations (see Section \ref{sec:simulation}) suggest that the estimated eigenfunctions are reasonable approximations that work well as a basis.

\subsection{Step 1: Estimation of the Eigenfunction Basis}
\label{subsec:EigenfunctionEstimation}

\subsubsection*{Step 1 (i): Univariate Mean Estimation}
\label{subsubsec:UniMean}

In a first step, we obtain preliminary estimates of the dimension specific means $\mu^{(d)}(\bm{x}_i, t) = \sum_{l=1}^{p}f_l^{(d)}(\bm{x}_{il}, t)$ using univariate \glspl{famm}. We model 
\begin{align}
y_{it}^{(d)}&= \mu^{(d)}(\bm{x}_i, t) + \epsilon^{(d)}_{it};\; d=1,\dots,D
\label{eq:uniMeanEstim}
\end{align}
independently for all $d$ with \gls{iid}~Gaussian random variables $\epsilon^{(d)}_{it}$. The estimation of $\mu^{(d)}(\bm{x}_i, t)$ proceeds analogously to the estimation of the \gls{mfamm} described in section \ref{subsec:FAMM}. It is based on the evaluation points of the $y_i^{*(d)}(t)$, whose locations on the interval $\mathcal{I}$ can vary across dimensions. Model \eqref{eq:uniMeanEstim} thus accommodates sparse and irregular multivariate functional data and implies a working independence assumption across scalar observations within and across functions. 

\subsubsection*{Step 1 (ii): Univariate Covariance Estimation}

This preliminary mean function is used to centre the data $\tilde{y}_{it}^{(d)} = y_{it}^{(d)}- \hat{\mu}^{(d)}(\bm{x}_i, t)$ in order to obtain noisy evaluations of the detrended functions $\tilde{y}_i^{*(d)}(t) = y_i^{*(d)}(t) - {\mu}^{(d)}(\bm{x}_i, t)$ for covariance estimation. \citet{cederbaum2016} already find that for this purpose, the working independence assumption within functions across evaluation points in \eqref{eq:uniMeanEstim} gives reasonable results. The expectation of the crossproducts of the centred functions then coincides with the auto-covariance, i.e.\ $\mathbb{E}\left(\tilde{y}_{it}^{(d)}\tilde{y}_{i't'}^{(d)}\right) \approx \text{Cov}\left[y_{it}^{(d)}, y_{i't'}^{(d)}\right]$. For the  independent random components specified in model \eqref{eq:multivariateFLMM}, this overall covariance decomposes additively into contributions from each random process as
\begin{align}
\label{eq:nested_random_effects_cov_estimation}
\mathbb{E}\left(\tilde{y}_{it}^{(d)}\tilde{y}_{i't'}^{(d)}\right) \approx \sum_{j=1}^q K_{U_j}^{(d,d)}(t, t')\delta_{v_jv_j'} + \big(K_{E}^{(d,d)}(t, t') + \sigma^2_{d}\delta_{tt'}\big)\delta_{ii'},
\end{align}
using indicators $\delta_{xx'}$ that equal one for $x=x'$ and zero otherwise. The indicator $\delta_{v_jv_j'}$ thus identifies if the curves in the crossproduct belong to the same group $v_j$ of the $j$th layer. Using $t$, $t'$, and the indicators $\delta_{v_jv_j'},\delta_{tt'}, \delta_{ii'}$ as covariates and the crossproducts of the centred data as responses, we can estimate the auto-covariances $K_{U_1}^{(d,d)}, ..., K_{U_q}^{(d,d)},$ and $K_{E}^{(d,d)}$ of the random processes using symmetric additive covariance smoothing \citep{cederbaum2018}. This extends the univariate approach proposed by \citet{cederbaum2016}. In particular, we also allow a nested random effects structure as required for the snooker training application in section \ref{subsec:snookerTraining} by specifying the indicator of the nested effect as the product of subject and session indicators. Note that estimating \eqref{eq:nested_random_effects_cov_estimation} also yields estimates of the dimension specific error variances $\sigma_d^{2}$ as a byproduct.

\subsubsection*{Step 1 (iii): Univariate Eigenfunction Estimation}

Based on the covariance kernel estimates, we apply separate univariate \glspl{fpca} for each random process by conducting an eigendecomposition of the respective linear integral operator. Practically, each estimated process- and dimension-specific auto-covariance is re-evaluated on a dense grid so that a univariate \gls{fpca} can be conducted. Alternatively, \citet{reiss2020tensor} provide an explicit spline representation of the estimated eigenfunctions. Eigenfunctions with non-positive eigenvalues are removed to ensure positive definiteness, and further regularization by truncation based on the proportion of variance explained is possible \citep[see e.g.][]{di2009multilevel, peng2009geometric, cederbaum2016}. However, we suggest to keep all univariate \glspl{fpc} with  positive eigenvalues for the computation of the \gls{mfpca} in order to preserve all important modes of variation and cross-correlation in the data.

\subsubsection*{Step 1 (iv): Multivariate Eigenfunction Estimation}

The estimated univariate eigenfunctions and scores are then used to conduct an \gls{mfpca} for each of the $g$ multivariate random processes separately. The \gls{mfpca} exploits correlations between univariate \gls{fpc} scores across dimensions to reduce the number of basis functions needed to sufficiently represent the random processes. We base the \gls{mfpca} on the following definition of a (weighted) scalar product
\begin{align}
\label{eq:weighted_scalar_product}
\langle\langle \bm{f}, \bm{g} \rangle\rangle := \sum_{d=1}^D w_d \int_{\mathcal{I}}f^{(d)}(t)g^{(d)}(t)dt, \quad \bm{f}, \bm{g} \in L^2_D(\mathcal{I}),
\end{align}
for the response space with positive weights $w_d, d = 1,...,D$ and the induced norm denoted by $\vert\vert\vert\cdot\vert\vert\vert$. 
The corresponding covariance operators $\Gamma_{g}: L^2_D(\mathcal{I}) \rightarrow L^2_D(\mathcal{I})$ of the multivariate random processes $\bm{U}_{jv}$ and $\bm{E}_v$ are then given by $(\Gamma_{g} \bm{f})(t) = \langle\langle \bm{f}, K_{g}(t, \cdot) \rangle\rangle$, $g\in\{U_1,..., U_q, E\}$.
The standard choice of weights in our applications is $w_1 = ... = w_D = 1$ (unweighted scalar product) but other choices are possible. Consider for example a scenario where dimensions are observed with different amounts of measurement error. If variation in dimensions with a large proportion of measurement error is to be downweighted, we propose to use $w_d = \frac{1}{\hat{\sigma}^2_d}$ with the dimension specific measurement error variance estimates $\hat{\sigma}^2_d$ obtained from \eqref{eq:nested_random_effects_cov_estimation}.

\citet{happ2018} show that estimates of the multivariate eigenvalues $\nu_{gm}$ of $\Gamma_g$ can be obtained from an eigenanalysis of a covariance matrix of the univariate random scores. The corresponding  multivariate eigenfunctions $\bm{\psi}_{gm}$ can be obtained as linear combinations of the univariate eigenfunctions with the weights given by the resulting eigenvectors. The estimates $\hat{\bm{\psi}}_{gm}$ are then substituted for the basis functions of the truncated multivariate \gls{kl} expansions of the random effects $\bm{U}_{jv}$ and $\bm{E}_v$ in \eqref{eq:truncated_KL_expansion}. Note that for each random process $g$, the maximum number of \glspl{fpc} is given by the total number of univariate eigenfunctions included in the estimation process of the \gls{mfpca} of $g$. To achieve further regularization and analogously to \citet{cederbaum2016}, we propose to choose truncation orders $M_g$ for each \gls{kl} expansion of the multivariate random processes using a prespecified proportion of explained variation.

\subsubsection*{Step 1 (v): Multivariate Truncation Order}

We offer two different approaches for the choice of truncation orders $M_g$ based on different variance decompositions (derivation in Appendix \ref{app_sec:vardecomp}):
\begin{gather}
\label{eq:total_variation_decomp}
\mathbb{E}\left(\vert\vert\vert \bm{y}_i - \bm{\mu}(\bm{x}_i) \vert\vert\vert^2 \right) = \sum_{d=1}^Dw_d\int_{\mathcal{I}}\mathrm{Var}\big(y_{i}^{(d)}(t)\big) dt  =\sum_{g}\sum_{m = 1}^{\infty}\nu_{gm} + \sum_{d=1}^Dw_d\sigma_d^2 \vert\mathcal{I}\vert, \\
\label{eq:univar_variation_decomp}
\text{and} \quad \int_{\mathcal{I}}\mathrm{Var}\big(y_{i}^{(d)}(t)\big) dt  = \sum_{g}\sum_{m = 1}^{\infty}\nu_{gm}\vert\vert\psi_{gm}^{(d)}\vert\vert^2 + \sigma_d^2 \vert\mathcal{I}\vert
\end{gather}
with $\vert \mathcal{I}\vert$ the length of the interval $\mathcal{I}$ (here equal to one) and $\vert\vert\cdot\vert\vert$ the $L^2$ norm. Multivariate variance decomposition \eqref{eq:total_variation_decomp} uses the (weighted) sum of total variation in the data across  dimensions. We select the \glspl{fpc} with highest associated eigenvalues $\nu_{gm}$ over all random processes $g$ until their sum reaches a prespecified proportion (e.g.\ 0.95) of the total variation, thus approximating the infinite sums in \eqref{eq:total_variation_decomp} with $M_g$ summands. For the approach based on the univariate variance  \eqref{eq:univar_variation_decomp}, we require $M_g$ to be the smallest truncation order for which at least a prespecified proportion of variance is explained on every dimension $d$. This second choice of $M_g$ might be preferable in situations where the variation is considerably different (in amount or structure) across dimensions, whereas the first approach gives a more parsimonious representation of the random effects. Note that both approaches can lead to a simplification of the  \gls{mfamm} if $M_g = 0$ is chosen for some $g$. The simulation results of section \ref{sec:simulation} suggest that increasing the number of \glspl{fpc} improves model accuracy which is why sensitivity analyses with regard to the truncation order are recommended.

\subsection{Step 2: Estimation of the multiFAMM}
\label{subsec:FAMM}

In the following, we discuss estimating the \gls{mfamm} given the estimated multivariate \glspl{fpc}. We base the proposed model on the general \gls{famm} framework of \citet{scheipl2015}, which models functional responses using basis representations. To make the extension of the \gls{famm} framework to multivariate functional data more apparent, the multivariate response vectors and the respective model matrices are stacked over dimensions, so that every block has the structure of a univariate \gls{famm} over all observations $i$. This gives concatenated basis functions with discontinuities between the dimensions. The fixed effects are modelled analogously to  the univariate case by interacting all covariate effects with a dimension indicator. The random effects are based on the parsimonious, concatenated multivariate \gls{fpc} basis.

\subsubsection*{Matrix Representation}

For notational simplicity we assume that the functions are evaluated on a fixed grid of time points $\boldsymbol{t} = (\boldsymbol{t}^{(1)}, ..., \boldsymbol{t}^{(D)})^{\top}$ with $\boldsymbol{t}^{(d)} = (\boldsymbol{t}_{1}^{(d)}, ..., \boldsymbol{t}_{N}^{(d)})$ and identical $\boldsymbol{t}_i^{(d)} \equiv (t_1,..., t_T)$ over all $N$ individuals and $D$ dimensions. However, our framework allows for sparse functional data using different grids per dimension and per observed function as in the two applications (Section \ref{sec:application}). Correspondingly, $\boldsymbol{y} = (\boldsymbol{y}^{(1)\top}, ..., \boldsymbol{y}^{(D)\top})^{\top}$ is the $DNT$-vector of stacked evaluation points with $\boldsymbol{y}^{(d)} = (\boldsymbol{y}_1^{(d)\top}, ..., \boldsymbol{y}_N^{(d)\top})^{\top}$ and $\boldsymbol{y}_i^{(d)} = (y_{i1}^{(d)},..., y_{iT}^{(d)})^{\top}$. Model \eqref{eq:multivariateFLMM} on this grid can be written as
\begin{align}
\label{eq:multiFAMM}
\bm{y} = \bm{\Phi\theta} + \bm{\Psi\rho} + \bm{\epsilon}
\end{align}
with $\bm{\Phi}, \bm{\Psi}$ the model matrices for the fixed and random effects, respectively, $\bm{\theta},\bm{\rho}$ the vectors of coefficients and random effect scores to be estimated, and $\bm{\epsilon} = (\bm{\epsilon}^{(1)\top}, ..., \bm{\epsilon}^{(D)\top})^{\top}$, $\bm{\epsilon}^{(d)} = (\epsilon_{11}^{(d)}, ..., \epsilon_{1T}^{(d)}, ..., \epsilon_{NT}^{(d)})^{\top}$ the vector of residuals. We have $\bm{\epsilon} \sim N(\bm{0}, \bm{\Sigma})$ with $\bm{\Sigma} =  \diag(\sigma_1^2,..., \sigma_D^2)\otimes \bm{I}_{NT}$, the Kronecker product denoted by $\otimes$, and the $(NT\times NT)$ identity matrix $\bm{I}_{NT}$.

We estimate $\bm{\theta}$ and $\bm{\rho}$ by minimizing the \gls{pls} criterion 
\begin{align}
\label{eq:pls}
(\bm{y} - \bm{\Phi\theta} - \bm{\Psi\rho})\bm{\Sigma}^{-1}(\bm{y} - \bm{\Phi\theta} - \bm{\Psi\rho})^{\top} + \sum_{l=1}^{p}\bm{\theta}_{l}^{\top}\bm{P}_{l}(\bm{\lambda}_{xl}, \bm{\lambda}_{tl})\bm{\theta}_{l} + \sum_{g}\lambda_{g}\bm{\rho}_{g}^{\top}\bm{P}_{g}\bm{\rho}_{g}
\end{align}
using appropriate penalty matrices $\bm{P}_{l}(\bm\lambda_{xl}, \bm\lambda_{tl})$ and $\bm{P}_{g}$ for the fixed effects and random effects, respectively, and smoothing parameters $\bm{\lambda}_{xl} = \left(\lambda_{xl}^{(1)}, ..., \lambda_{xl}^{(D)}\right),\bm{\lambda}_{tl} = \left(\lambda_{tl}^{(1)}, ..., \lambda_{tl}^{(D)}\right)$, and $\lambda_g$. The model and penalty matrices as well as the parameter vectors of \eqref{eq:multiFAMM} and \eqref{eq:pls} are discussed in detail below.

\subsubsection*{Modeling of Fixed Effects}
The block diagonal matrix $\bm{\Phi} = \diag\left(\bm{\Phi}^{(1)}, ..., \bm{\Phi}^{(D)}\right)$ models the fixed effects separately on each dimension as in a \gls{famm} \citep{scheipl2015}. The $(DNT \times b)$ matrix $\bm{\Phi}$ consists of the design matrices $\bm{\Phi}^{(d)} = (\bm{\Phi}_1^{(d)}\,|\,...\,|\,\bm{\Phi}_p^{(d)})$ that are constructed for the partial predictors $f_{l}^{(d)}(\bm{x}, \bm{t}^{(d)}), l = 1, ..., p$, which correspond to the $NT$-vectors of evaluations of the effect functions $f_l^{(d)}$. The vectors of scalar covariates $\bm{x}_i$ are repeated $T$ times to form the matrix of covariate information $\bm{x} = (\bm{x}_1,..., \bm{x}_1,..., \bm{x}_N)^{\top}$. We use the basis representations
\begin{align*}
f_l^{(d)}(\bm{x}, \bm{t}^{(d)}) \approx \bm{\Phi}_l^{(d)}\bm{\theta}_l^{(d)} = (\bm{\Phi}_{xl}^{(d)} \odot \bm{\Phi}_{tl}^{(d)})\bm{\theta}_l^{(d)} ,
\end{align*}
where $\bm{A}\odot\bm{B}$ denotes the row tensor product $(\bm{A}\otimes\bm{1}_b^{\top})\cdot(\bm{1}_a^{\top}\otimes\bm{B})$ of the $(h\times a)$ matrix $\bm{A}$ and the $(h\times b)$ matrix $\bm{B}$ with element-wise multiplication $\cdot$ and $\bm{1}_c$ the $c$-vector of ones. This modeling approach combines the $(NT \times b_{xl}^{(d)})$ basis matrix $\bm{\Phi}_{xl}^{(d)}$ with the $(NT \times b_{tl}^{(d)})$ basis matrix $\bm{\Phi}_{tl}^{(d)}$. These matrices contain the evaluations of suitable marginal bases in $\bm{x}$ and $\bm{t}^{(d)}$, respectively. For a linear effect, for example, the basis matrix $\bm{\Phi}_{xl}^{(d)}$ is specified as the familiar linear model design matrix $\bm{x}$ for the linear effect $f_l^{(d)}(\bm{x}, \bm{t}^{(d)}) = \bm{x}\bm{\beta}_l^{(d)}(\bm{t}^{(d)})$ with coefficient function $\bm{\beta}_l^{(d)}(\bm{t}^{(d)})$. For a nonlinear effect $f_l^{(d)}(\bm{x}, \bm{t}^{(d)})= g_l^{(d)}(\bm{x}, \bm{t}^{(d)})$, the basis matrix $\bm{\Phi}_{xl}^{(d)}$ contains an (e.g. B-spline) basis representation analogously to a scalar additive model. For the functional intercept, $\bm{\Phi}_{xl}^{(d)}$ is a vector of ones,  and we generally use a spline basis for $\bm{\Phi}_{tl}^{(d)}$. For a complete list of possible effect specifications with examples, we refer to \citet{scheipl2015}. The tensor product basis is weighted by the $b_{xl}^{(d)}b_{tl}^{(d)}$ unknown basis coefficients in $\bm{\theta}_l^{(d)}$. Stacking the vectors $\bm{\theta}_l^{(d)}$ gives $\bm{\theta}^{(d)} = (\bm{\theta}_1^{(d)\top}, ..., \bm{\theta}_p^{(d)\top})^{\top}$ and finally the $b$-vector $\bm{\theta} = (\bm{\theta}^{(1)\top},..., \bm{\theta}^{(D)\top})^{\top}$ with $b = \sum_{d}\sum_{l}b_{xl}^{(d)}b_{tl}^{(d)}$. 

Choosing the number of basis functions is a well known challenge in the estimation of nonlinear or functional effects. We introduce regularization by a corresponding quadratic penalty term in \eqref{eq:pls}. Let $\bm{\theta}_l$ contain the coefficients corresponding to the partial predictor $l$ and order it by dimensions. The penalty $\bm{P}_{l}(\bm{\lambda}_{xl}, \bm{\lambda}_{tl})$ is then constructed from the penalty on the marginal basis for the covariate effect, $\bm{P}_{xl}^{(d)}$, and the penalty on the marginal basis over the functional index, $\bm{P}_{tl}^{(d)}$. Specifically, $\bm{P}_{l}(\bm{\lambda}_{xl}, \bm{\lambda}_{tl})$ is a block-diagonal matrix with blocks for each $d$ corresponding to the Kronecker sums of the marginal penalty matrices $\lambda_{xl}^{(d)}\bm{P}_{xl}^{(d)}\otimes \bm{I}_{b_{tl}^{(d)}} + \lambda_{tl}^{(d)}\bm{I}_{b_{xl}^{(d)}} \otimes \bm{P}_{tl}^{(d)}$ \citep{wood2017}. A standard choice for these marginal penalty matrices given a B-splines basis representation are second or third order difference penalties, thus approximately penalizing squared second or third derivatives of the respective functions \citep{eilers1996}. For unpenalized effects such as a linear effect of a scalar covariate, the corresponding $\bm{P}_{xl}^{(d)}$ is simply a matrix of zeroes.

\subsubsection*{Modeling of Random Effects}

We represent the $DNT$-vectors $\bm{U}_j(\bm{t}) = (\bm{U}_j(\bm{t}^{(1)})^{\top}, ..., \bm{U}_j(\bm{t}^{(D)})^{\top})^{\top}$, $\bm{E}(\bm{t}) = (\bm{E}(\bm{t}^{(1)})^{\top}, ..., \bm{E}(\bm{t}^{(D)})^{\top})^{\top}$ with $\bm{U}_j(\bm{t}^{(d)})$, $\bm{E}(\bm{t}^{(d)})$ containing the evaluations of the univariate random effects for the corresponding groups and time points using the basis approximations 
\begin{align*}
\bm{U}_j(\bm{t}) \approx \bm{\Psi}_{U_j} \bm{\rho}_{U_j} = (\bm{\delta}_{U_j} \odot \tilde{\bm{\Psi}}_{U_j}) \bm{\rho}_{U_j}, \quad
\bm{E}(\bm{t}) \approx \bm{\Psi}_E \bm{\rho}_{E} = (\bm{\delta}_E \odot \tilde{\bm{\Psi}}_E) \bm{\rho}_{E}.
\end{align*}
The $v$th column in the $(DNT \times V_g), g \in \{U_1,..., U_q, E\}$ indicator matrix $\bm{\delta}_g$ indicates whether a given row is from the $v$th group of the corresponding grouping layer. Thus, the rows of the indicator matrix $\bm{\delta}_g$ contain  repetitions of the group indicators $\bm{z}_{ij}^{\top}$ and $\bm{e}_{i}^{\top}$ in model \eqref{eq:multivariateFLMM}. For the smooth residual, $\bm{\delta}_E$ simplifies to $\bm{1}_D\otimes(\bm{I}_N \otimes \bm{1}_T)$. The $(DNT \times M_g)$ matrix $\tilde{\bm{\Psi}}_g = (\tilde{\bm{\Psi}}_g^{(1)\top}| ... |\tilde{\bm{\Psi}}_g^{(D)\top})^{\top}$ comprises the evaluations of the $M_g$ multivariate eigenfunctions $\psi_{gm}^{(d)}(t)$ on dimensions $d = 1,...,D$ for the $NT$ time points contained in the $(NT\times M_g)$ matrix $\tilde{\bm{\Psi}}_g^{(d)}$. The $M_gV_g$ vector $\bm{\rho}_g = (\bm{\rho}_{g1}^{\top},...,\bm{\rho}_{gV_g}^{\top})^{\top}$ with $\bm{\rho}_{gv} = (\rho_{gv1},...,\rho_{gvM_g})^{\top}$ stacks all the unknown random scores for the functional random effect $g$. The $(DNT \times \sum_g M_gV_g)$ model matrix $\bm{\Psi} = (\bm{\Psi}_{U_1} | ... |\bm{\Psi}_{U_q} | \bm{\Psi}_E)$ then combines all random effect design matrices. Stacking the vectors of random scores in a $\sum_g M_gV_g$ vector $\bm{\rho}= (\bm{\rho}_{U_1}^{\top},..., \bm{\rho}_{U_q}^{\top}, \bm{\rho}_{E}^{\top})^{\top}$ lets us represent all functional random intercepts in the model via $\bm{\Psi}\bm{\rho}$.

For a given functional random effect, the penalty takes the form $\bm{\rho}_{g}^{\top}\bm{P}_{g}\bm{\rho}_{g} = \bm{\rho}_g^{\top} (\bm{I}_{V_g}\otimes \tilde{\bm{P}}_g)\bm{\rho}_g$, where $\bm{I}_{V_g}$ corresponds to the assumed independence between the $V_g$ different groups. The diagonal matrix $\tilde{\bm{P}}_g =\diag(\nu_{g1},..., \nu_{gM_g})^{-1}$ contains the (estimated) eigenvalues $\nu_{gm}$ of the associated multivariate \glspl{fpc}. This quadratic penalty is mathematically equivalent to a normal distribution assumption on the scores $\bm{\rho}_{gv}$ with mean zero and covariance matrix $\tilde{\bm{P}}_g^{-1}$, as implied by the \gls{kl} theorem for Gaussian random processes. Note that the smoothing parameter $\lambda_g$ allows for additional scaling of the covariance of the corresponding random process.

\subsubsection*{Estimation}

We estimate the unknown smoothing parameters in $\bm{\lambda}_{xl}, \bm{\lambda}_{tl}$, and $\lambda_g$ using fast \gls{reml}-estimation \citep{wood2017}. The standard identifiability constraints of \glspl{famm} are used \citep{scheipl2015}. In particular, in addition to the constraints for the fixed effects, the multivariate random intercepts are subject to a sum-to-zero constraint over all evaluation points as given by e.g.\ \citet{refundpackage}.

We propose a weighted regression approach to handle the heteroscedasticity assumption contained in $\bm{\Sigma}$. We weigh each observation proportionally to the inverse of the estimated univariate measurement error variances $\hat{\sigma}_d^2$ from the estimation of the univariate covariances \eqref{eq:nested_random_effects_cov_estimation}. Alternatively, updated measurement error variances can be obtained from fitting separate univariate \glspl{famm} on the dimensions using the univariate components of the multivariate \glspl{fpc} basis. In practice, we found that the less computationally intensive former option gives reasonable results.

As our proposed model is part of the \gls{famm} framework, inference for the \gls{mfamm} is readily available based on inference for scalar additive mixed models \citep{wood2017}. Note, however, that all inferential statements do not incorporate uncertainty due to the estimated multivariate eigenfunction bases, nor in the chosen smoothing parameters. The estimation process readily provides i.a.\ standard errors for the construction of point-wise univariate \glspl{cb}.

\subsection{Implementation}

We provide an implementation of the estimation of the proposed \gls{mfamm} in the \texttt{multifamm} \texttt{R}-package \citep{multifammpackage}. It is possible to include up to two functional random intercepts in $\bm{U}(t)$, which can have a nested or crossed structure, in addition to the curve-specific random intercept $\bm{E}_i(t)$. While including e.g.\ functional covariates is conceptually straightforward \citep[see][]{scheipl2015}, our implementation is restricted to scalar covariates and interactions thereof. We provide different alternatives for specifying the multivariate scalar product, the multivariate cut-off criterion, and the covariance matrix of the white noise error term. Note that the estimated univariate error variances have been proposed as weights for two separate and independent modeling decisions: as weights in the scalar product of the \gls{mfpca} and as regression weights under heteroscedasticity across dimensions.

\section{Applications}
\label{sec:application}

We illustrate the proposed \gls{mfamm} for two different data applications corresponding to intrinsically multivariate and multimodal fuctional data. The presentation focuses on the first application with a detailed description of the multimodal data application in Appendix \ref{APPsec:ca_data}. We provide the data and the code to produce all analyses in the github repository \textit{\href{https://github.com/alexvolkmann/multifammPaper}{multifammPaper}}.

\subsection{Snooker Training Data}
\label{subsec:snookerTraining}

\subsubsection*{Data Set and Preprocessing}

\begin{figure}
\begin{subfigure}{.37\textwidth}
\includegraphics[scale=0.55]{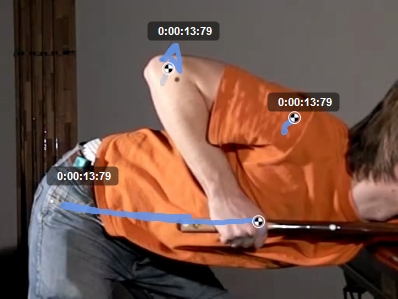}
\end{subfigure}
\begin{subfigure}{.5\textwidth}
\includegraphics[scale=0.50]{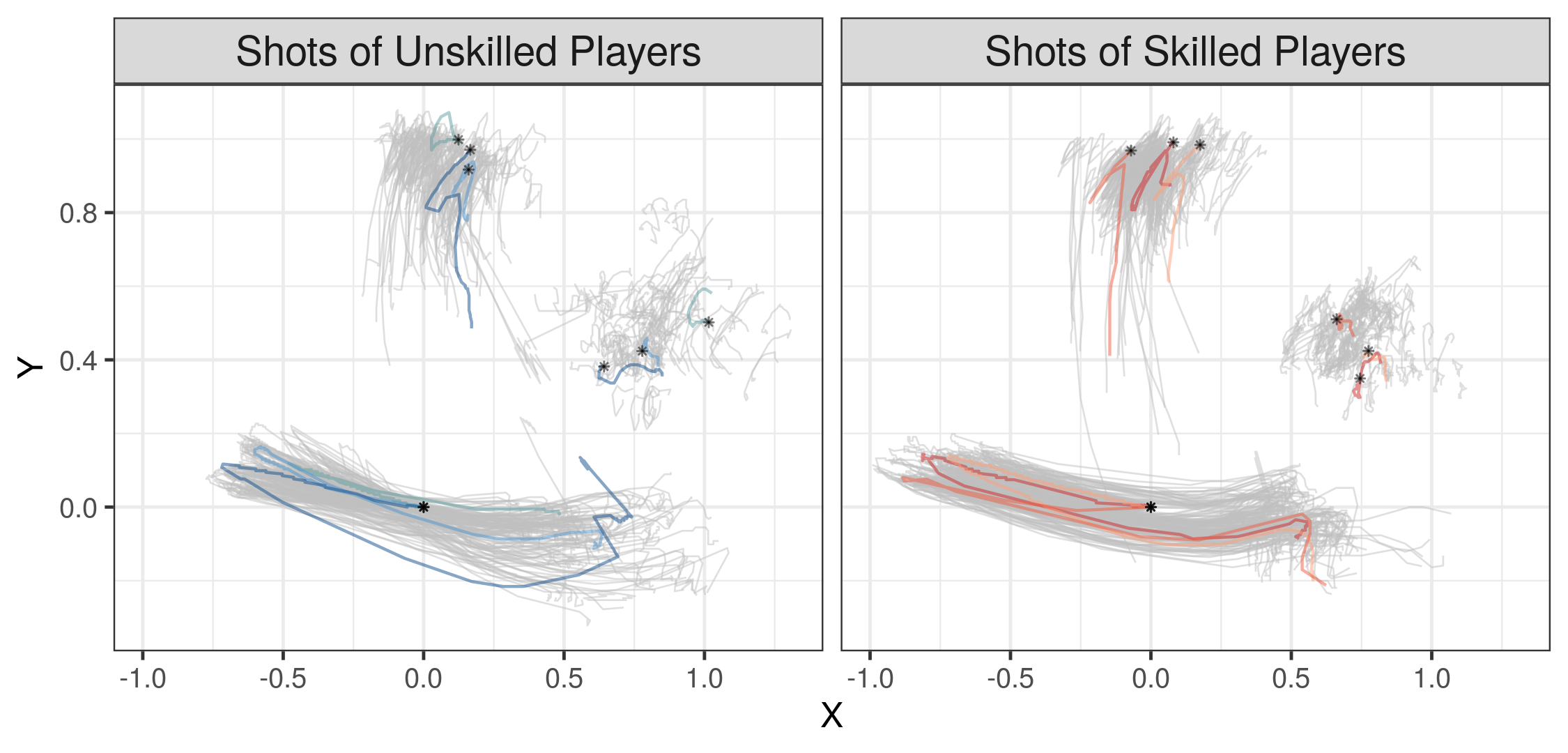}
\end{subfigure}
\caption{\textit{Left:} Screenshot of software for tracking (blue lines) the points of interest (circles). \textit{Right:} Two-dimensional trajectories of the snooker training data set (grey curves, right). For both groups of skilled and unskilled participants, three randomly selected observations are highlighted. Every colour corresponds to one multivariate observation, i.e.\ one observation consists of three trajectories: elbow (top), shoulder (right), hand (bottom). The start of the exemplary trajectories are marked with a black asterisk with the hand trajectory centred at the origin.}
\label{fig:snook_obs}
\end{figure}

In a study by \citet{snooker2014}, 25 recreational snooker players split into two groups, one of which had instructions to follow a self-administered training schedule over the next six weeks consisting of exercises aimed at improving snooker specific muscular coordination. The second was a control group. Before and after the training period, both groups were recorded on high speed digital camera under similar conditions to investigate the effects of the training on their snooker shot of maximal force. In each of the two recording sessions, six successful shots per participant were videotaped. The recordings were then used to manually locate points of interest (a participant's shoulder, elbow, and hand) and track them on a two-dimensional grid over the course of the video. This yields a six dimensional functional observation per snooker shot $\bm{y}^{*} = (y^{*(\text{elbow.x})}, ..., y^{*(\text{shoulder.y})}): \mathcal{I} = [0, 1] \to \mathbb{R}^{6}$, i.e.\ a two-dimensional movement trajectory for each point of interest (see Figure \ref{fig:snook_obs}).	

In their starting position (hand centred at the origin), the snooker players are positioned centrally in front of the snooker table aiming at the cue ball. From their starting position, the players draw back the cue, then accelerate it forwards and hit the cue ball shortly after their hands enter the positive range of the horizontal $x$ axis. After the impulse onto the cue ball, the hand movement continues until it is stopped at a player's chest. \citet{snooker2014} identify two underlying techniques that a player can apply: dynamic and fixed elbow. With a dynamic elbow, the cue can be moved in an almost straight line (piston stroke) whereas additionally fixing the elbow results in a pendular motion (pendulum stroke). In both cases, the shoulder serves as a fixed point and should be positioned close to the snooker table.

We adjust the data for differences in body height and relative speed \citep{steyer2020} and apply a coarsening method to reduce the number of redundant data points, thereby lowering computational demands of the analysis. Appendix \ref{app_sec:snooker_analysis} provides a detailed description of the data preprocessing. As some recordings and evaluations of bivariate trajectories are missing, the final data set contains 295 functional observations with a total of 56,910 evaluation points. These multivariate functional data are irregular and sparse, with a median of 30 evaluation points per functional observation (minimum 8, maximum 80) for each of the six dimensions.

\subsubsection*{Model Specification}

We estimate the following model
\begin{align}
\label{eq:snooker_model}
\bm{y}_{ijht}= \bm{\mu}(\bm{x}_{ij}, t) + \bm{B}_i(t) + \bm{C}_{ij}(t) + \bm{E}_{ijh}(t) + \bm{\epsilon}_{ijht},
\end{align}
with $i = 1,..., 25$ the index for the snooker player, $j = 1, 2$ the index for the session, $h = 1,..., H_{ij}$ the index for the typically six snooker shot repetitions in a session, and $t \in [0,1]$ relative time. Correspondingly, $\bm{B}_i(t)$ is a subject-specific random intercept, $\bm{C}_{ij}(t)$ is a nested subject-and-session-specific random intercept, and $\bm{E}_{ijh}(t)$ is the shot-specific random intercept (smooth residual). The nested random effect $\bm{C}_{ij}(t)$ is supposed to capture the variation within players between sessions (e.g.\ differences due to players having a good or bad day). Different positioning of participants with respect to the recording equipment or the snooker table as well as shot to shot variation are captured by the smooth residual $\bm{E}_{ijh}(t)$. The white noise measurement error $\bm{\epsilon}_{ijht}$ is assumed to follow a zero-mean multivariate normal distribution with covariance matrix $\sigma^2\bm{I}_6$, as all six dimensions are measured with the same set-up. The additive predictor is defined as
\begin{align*}
\bm{\mu}(\bm{x}_{ij},t) 
&= \bm{f}_0(t) + \texttt{skill}_{i}\cdot \bm{f}_1(t) +\texttt{group}_{i} \cdot \bm{f}_2(t) + \texttt{session}_{j}\cdot \bm{f}_3(t) \\ & \quad + \texttt{group}_{i}\cdot \texttt{session}_{j}\cdot\bm{f}_4(t).
\end{align*}
The dummy covariates $\texttt{skill}_i$ and $\texttt{group}_{i}$ indicate whether player $i$ is an advanced snooker player and belongs to the treatment group (i.e.\ receives the training programme), respectively. Note that the snooker players self-select into training and control group to improve compliance with the training programme, which is why we include a group effect in the model. The dummy covariate $\texttt{session}_{j}$ indicates whether the shot $j$ is recorded after the training period. The effect function $\bm{f}_4(t)$ can thus be interpreted as the treatment effect of the training programme.

Cubic P-splines with first order difference penalty, penalizing deviations from constant functions over time, with 8 basis functions are used for all effect functions in the preliminary mean estimation as well as in the final \gls{mfamm}. For the estimation of the auto-covariances of the random processes, we use cubic P-splines with first order difference penalty on five marginal basis functions. We use an unweighted scalar product \eqref{eq:weighted_scalar_product} for the \gls{mfpca} to give equal weight to all spatial dimensions, as we can assume that the measurement error mechanism is similar across dimensions. Additionally, we find that hand, elbow, and shoulder contribute roughly the same amount of variation to the data, cf.\ Table \ref{APPENDIXtab:snooker_varcontr} in Appendix \ref{app_subsec:snooker_analysis}, where we also discuss potential weighting schemes for the \gls{mfpca}. The multivariate truncation order is chosen such that 95\% of the (unweighted) sum of variation \eqref{eq:total_variation_decomp} is explained.

\subsubsection*{Results}

The \gls{mfpca} gives sets of five (for $\bm{C}$ and $\bm{E}$) and six (for $\bm{B}$) multivariate \glspl{fpc} that explain 95\% of the total variation. The estimated eigenvalues allow to quantify their relative importance. Approximately 41\% of the total variation (conditional on covariates) can be attributed to the nested subject-and-session-specific random intercept $\bm{C}_{ij}(t)$, 33\% to the subject-specific random intercept $\bm{B}_i(t)$, 14\% to the shot-specific $\bm{E}_{ijh}(t)$, and 7\% to white noise. This suggests that day to day variation within a snooker player is larger than the variation between snooker players. Note that these proportions are based on estimation step 1 (see Section \ref{subsec:EigenfunctionEstimation}).

\begin{figure}
\centering
\includegraphics[width=0.84\textwidth]{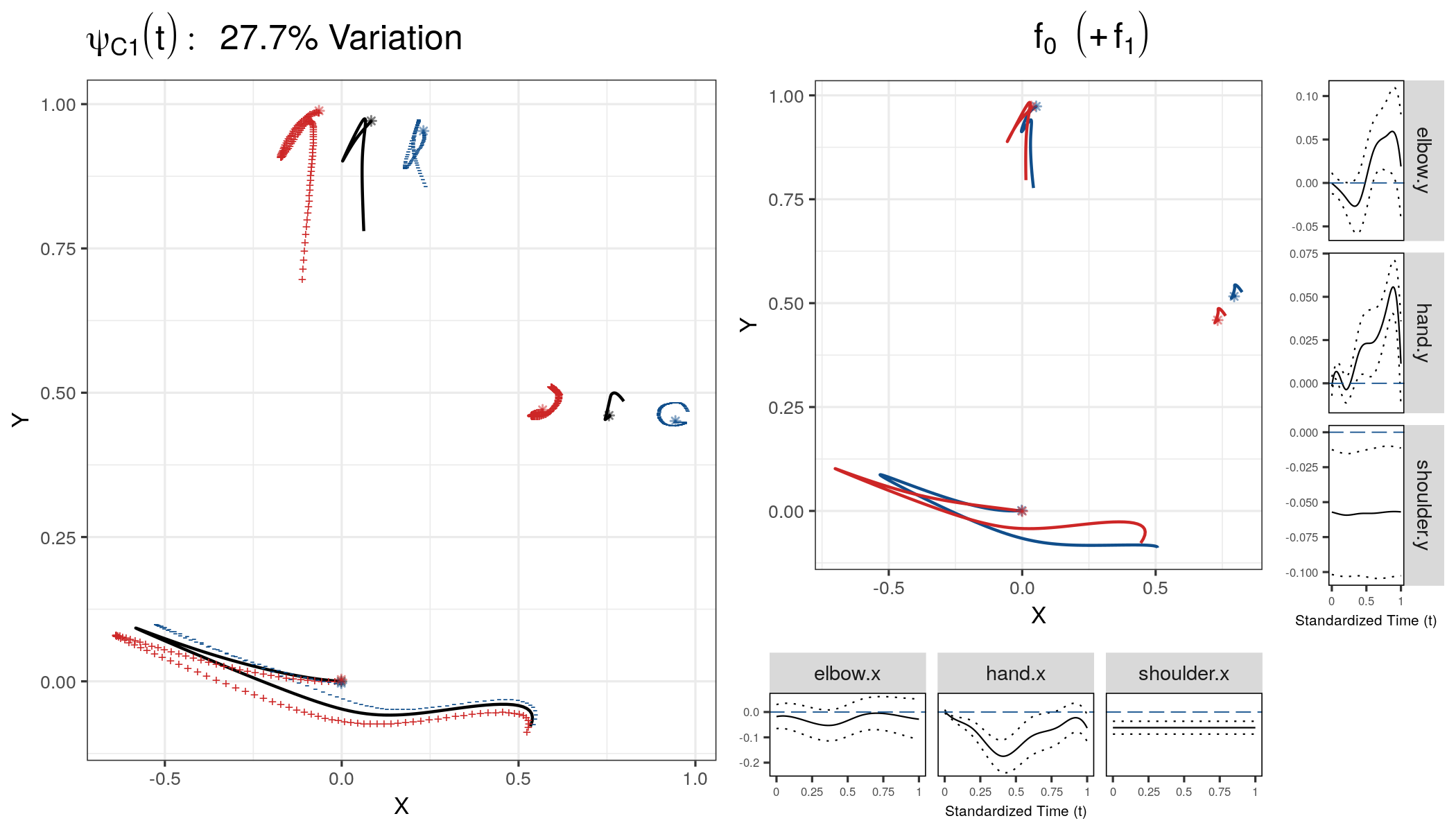}
\caption{\textit{Left:} Dominant mode ($\bm{\psi}_{C1}$) of the subject-and-session-specific  random effect, explaining $27.7\%$ of total variation and shown as mean trajectory (black solid) plus (red $+$) or minus (blue $-$) $2\sqrt{\nu_{C 1}}$ times the first \gls{fpc}. An asterisk marks the start of a trajectory. \textit{Right:} Estimated covariate effect functions for skill. The central plot shows the effect of the coefficient function (red) on the two-dimensional trajectories for the reference group (blue). The marginal plots show the estimated univariate effect functions (black) with pointwise 95\% \glspl{cb} (dotted) and the baseline (blue dashed).}
\label{fig:eval_snooker}
\end{figure}
 	
The left plot of Figure \ref{fig:eval_snooker} displays the first \gls{fpc} for $\bm{C}$, which explains about $28\%$ of total variation. A suitable multiple of the \glspl{fpc} is added ($+$) to and subtracted ($-$) from the overall mean function (black solid line, all covariate values set to $0.5$). We find that the dominant mode of the random subject-and-session specific effect influences the relative positioning of a player's elbow, shoulder, and hand, thus suggesting a strong dependence between the dimensions. \citet{snooker2014} argue from a theoretical viewpoint that the ideal starting position should place elbow and hand in a line perpendicular to the plane of the snooker table (corresponding to the x axis). The most prominent mode of variation captures deviations from this ideal starting position found in the overall mean. The next most important \gls{fpc} $\bm{\psi}_{B1}$ of the subject-specific random effect, which explains about $15\%$ of total variation, represents a subject's tendency towards the piston or pendulum stroke (see Appendix Figure \ref{APPENDIXfig:snooker_fpc_B}). This additional insight into the underlying structure of the variance components might be helpful for e.g.\ developing personalized training programmes.

The central plot on the right of Figure \ref{fig:eval_snooker} compares the estimated mean movement trajectory for  advanced snooker players (red) to that in the reference group (blue). It suggests that more experienced players tend towards the dynamic elbow technique, generating a hand trajectory resembling a straight line (piston stroke). Uncertainties in the trajectory could be represented by pointwise ellipses, but inference is more straightforward to obtain from the univariate effect functions. The marginal plots display the estimated univariate effects with pointwise 95\% confidence intervals. Even though we find only little statistical evidence for increased movement of the elbow (horizontal-left and vertical-top marginal panels), the hand and shoulder movements (horizontal centre and right,  vertical centre and bottom) strongly suggest that the skill level indeed influences the mean movement trajectory of a snooker player. Further results indicate that the mean hand trajectories might slightly differ between treatment and control group at baseline as well as between sessions ($\bm{f}_2(t)$ and $\bm{f}_3(t)$, see Appendix Figure \ref{APPENDIXfig:snooker_cov_023}). The estimated treatment effect $\bm{f}_4(t)$ (Appendix Figure \ref{APPENDIXfig:snooker_cov_1_4}), however, suggests that the training programme did not change the participants' mean movement trajectories substantially. Appendix \ref{app_subsec:snooker_analysis} contains a detailed discussion of all model terms as well as some model diagnostics and sensitivity analyses.

\subsection{Consonant Assimilation Data}

\subsubsection*{Data Set and Model Specification}

\citet{pouplier2016} study the assimilation of the German /s/ and /sh/ sounds such as the final consonant sounds in ``Kürbis'' (English example: ``haggis'')  and ``Gemisch'' (English example: ``dish''), respectively. The research question is how these sounds assimilate in fluent speech when combined across words such as in ``Kürbis-Schale'' or ``Gemisch-Salbe'', denoted as /s\#sh/ and /sh\#s/ with \# the word boundary. The 9 native German  speakers in the study repeated a set of 16 selected word combinations five times. Two different types of functional data, i.e.\ \gls{aco} and \gls{epg} data, were recorded for each repetition to capture the acoustic (produced sound) and articulatory (tongue movements) aspects of assimilation over (relative) time $t$ within the consonant combination. 

\begin{figure}
\centering
\includegraphics[width=0.8\textwidth]{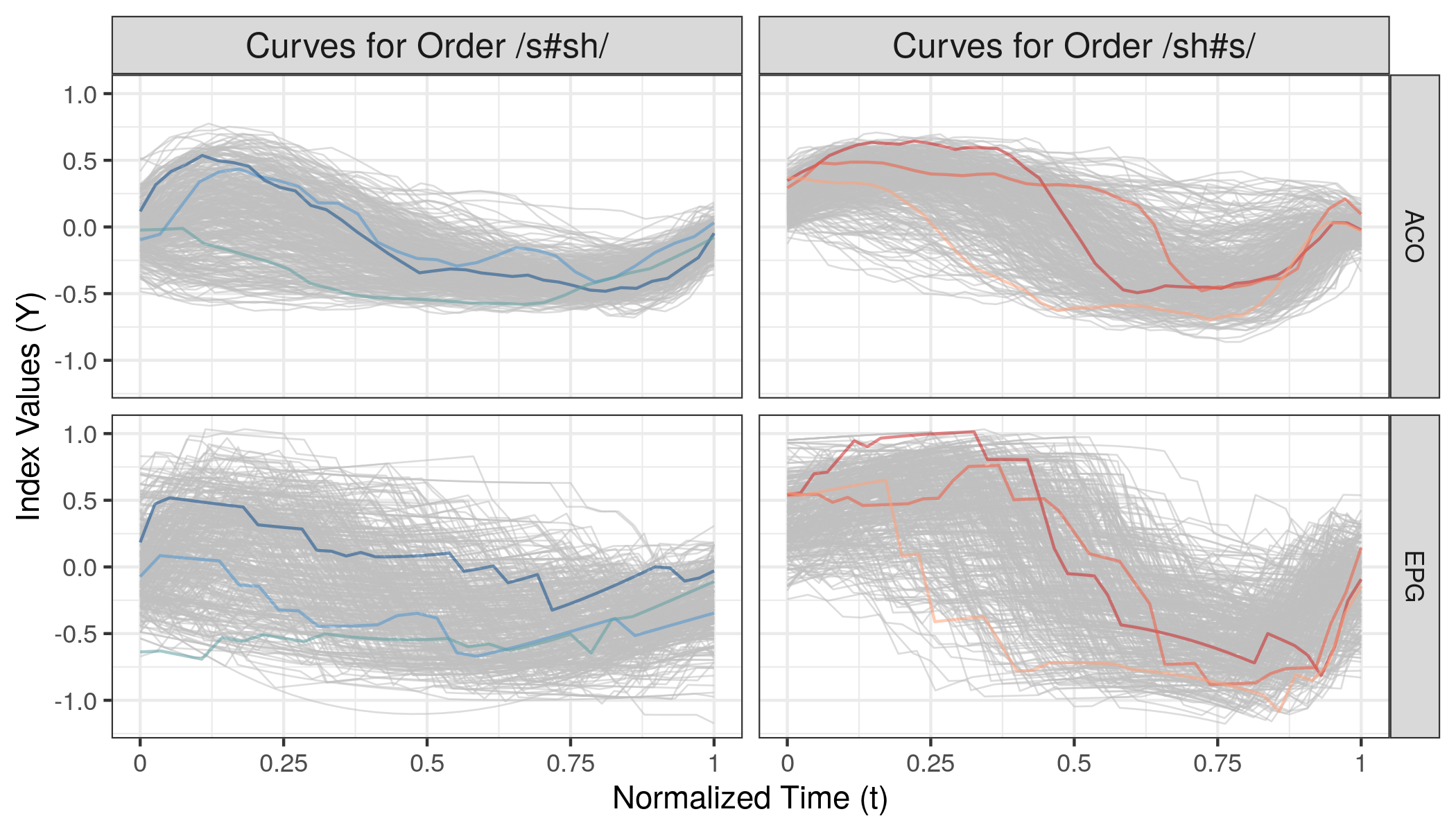}
\caption{Index curves of the consonant assimilation data set for both \gls{aco} and \gls{epg} data as a function of standardized time $t$ (grey curves). For every consonant order, three randomly selected observations have been highlighted. Every colour corresponds to one multivariate observation, i.e.\ one observation consists of two index curves.}
\label{fig:speech_obs}
\end{figure}

Each functional index varies roughly between $+1$ and $-1$ and measures how similar the articulatory or acoustic pattern is to its reference patterns for the first ($+1$) and second ($-1$) consonant at every observed time point \citep{cederbaum2016}. Without assimilation, the data are thus expected to shift from positive to negative values in a sinus-like form (see Figure \ref{fig:speech_obs}). The data set contains 707 bivariate functional observations with differently spaced grids of evaluation points per curve and dimension, with the number of evaluation points ranging from 22 to 59 with a median of 35. Note that the consonant assimilation data are unaligned as registration of the time domain would mask transition speeds between the consonants, which are an interesting part of assimilation.

For comparability, we follow the model specification of \citet{cederbaum2016}, who analyse only the \gls{aco} dimension and ignore the second mode \gls{epg}. Our specified multivariate model is similar to \eqref{eq:snooker_model} with $i = 1,..., 9$ the speaker index, $j = 1, ..., 16$ the word combination index, $h = 1,..., H_{ij}$ the repetition index and $t \in [0,1]$ relative time. Note that the nested effect $\bm{C}_{ij}(t)$ is replaced by the crossed random effect $\bm{C}_{j}(t)$ specific to the word combinations. The additive predictor $\bm{\mu}(\bm{x}_{j}, t)$ now contains eight partial effects: the functional intercept plus main and interaction effects of scalar covariates describing characteristics of the word combination such as the order of the consonants /s/ and /sh/. The white noise measurement error $\bm{\epsilon}_{ijht}$ is assumed to follow a zero-mean bivariate normal distribution with diagonal covariance matrix $\diag(\sigma^2_{\text{ACO}}, \sigma^2_{\text{EPG}})$. The basis and penalty specifications follow the univariate analysis in \citet{cederbaum2016}. Given different sampling mechanisms, we also compare the \gls{mfamm} based on weighted and unweighted scalar products for the \gls{mfpca}.

\subsubsection*{Results}

The multivariate analysis supports the findings of \citet{cederbaum2016} that assimilation is asymmetric (different mean patterns for /s\#sh/ and /sh\#s/). Overall, the estimated fixed effects are similar across dimensions as well as comparable to the univariate analysis. Hence, the multivariate analysis indicates that previous results for the acoustics are consistently found also for the articulation. Compared to univariate analyses, our approach reduces the number of \gls{fpc} basis functions and thus the number of parameters in the analysis. The \gls{mfamm} can improve the model fit and can provide smaller \glspl{cb} for the \gls{aco} dimension compared to the univariate model in \citet{cederbaum2016} due to the strong cross-correlation between the dimensions. We find similar modes of variation for the multivariate and the univariate analysis as well as across dimensions. In particular, the word combination-specific random effect $\bm{C}_{j}(t)$ is dropped from the model as much of the between-word variation is already explained by the included fixed effects. The definition of the scalar product has little effect on the estimated fixed effects but changes the interpretation of the \glspl{fpc}. Appendix \ref{APPsec:ca_data} contains a more in depth description of this application.

\section{Simulations}
\label{sec:simulation}

\subsection{Simulation Set-Up}

We conduct an extensive simulation study to investigate the performance of the \gls{mfamm} depending on different model specifications and data settings (over 20 scenarios total), and to compare it to univariate regression models as proposed by \citet{cederbaum2016}, estimated on each dimension independently. Given the broad scope of analysed model scenarios, we refer the interested reader to Appendix \ref{app_sec:simulation} for a detailed report and restrict the presentation here to the main results.

We mimic our two presented data examples (Section \ref{sec:application}) and simulate new data based on the respective \gls{mfamm}-fit. Each scenario consists of model fits to 500 generated data sets, where we randomly draw the number and location of the evaluation points, the random scores, and the measurement errors according to different data settings. The accuracy of the estimated model components is measured by the \gls{mse} based on the unweighted multivariate norm but otherwise as defined by \citet{cederbaum2016}, see Appendix \ref{app_subsec:sim_description}. The \gls{mse} takes on (unbounded) positive values with smaller values indicating a better fit.

\subsection{Simulation Results}

\begin{figure}
\includegraphics[width=\textwidth]{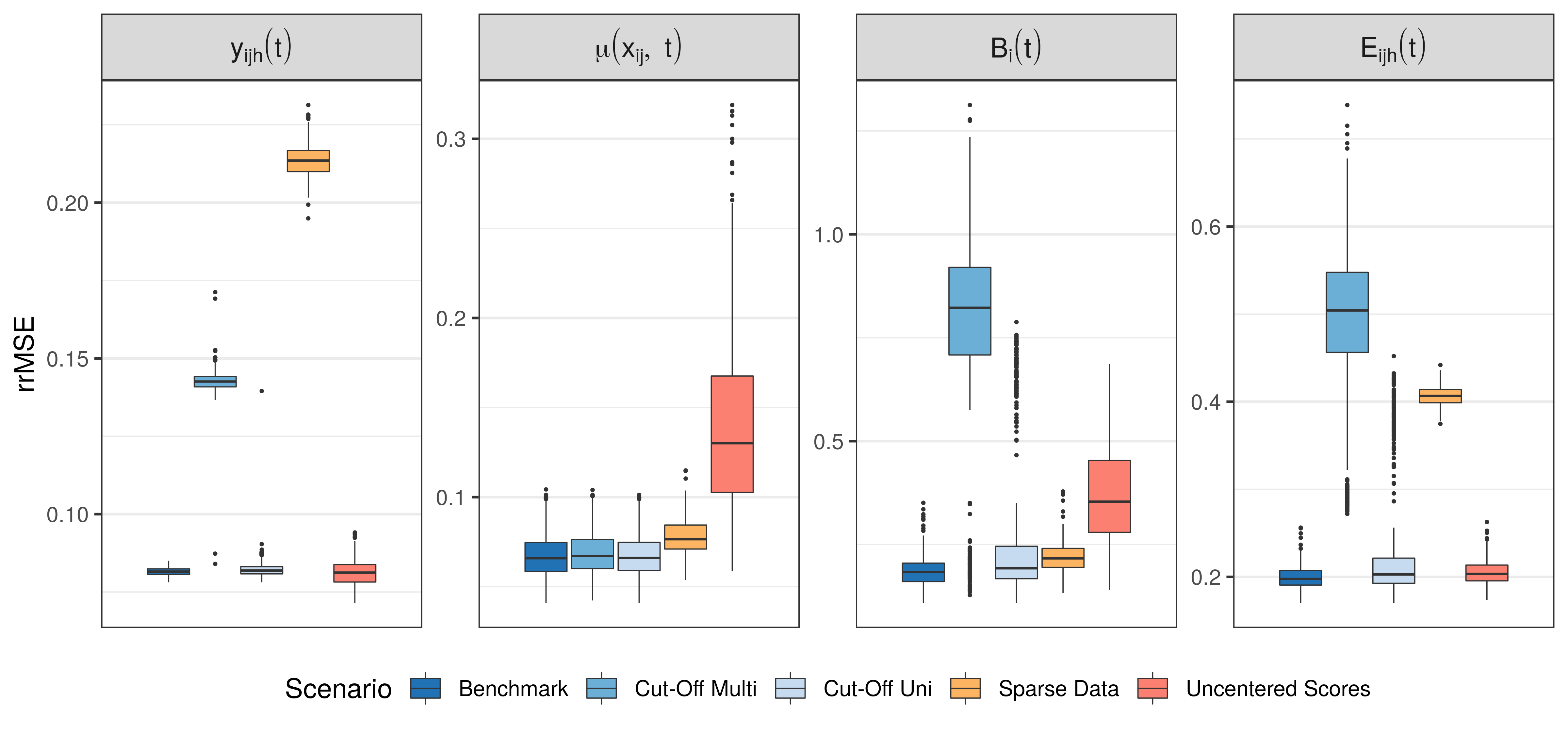}
\caption{\gls{mse} values of the fitted curves $\bm{y}_{ijh}(t)$, the mean $\bm{\mu}(\bm{x}_{ij}, t)$, and the random effects $\bm{B}_{i}(t)$ and $\bm{E}_{ijh}(t)$ for different modeling scenarios. Scenarios coloured with blue correspond to different model specifications in the same data setting.}
\label{fig:sim_eval_multi}
\end{figure}

Figure \ref{fig:sim_eval_multi} compares the \gls{mse} values over selected modeling scenarios based on the consonant assimilation data. We generate a benchmark scenario (dark blue), which imitates the original data without misspecification of any model component. In particular, the number of \glspl{fpc} is fixed to avoid truncation effects. Comparing this scenario to the other blue scenarios illustrates the importance of the number of \glspl{fpc} in the accuracy of the estimation. Choosing the truncation order via the proportion of univariate variance explained (Cut-Off Uni) as in \eqref{eq:univar_variation_decomp} gives models with roughly the same number of \glspl{fpc} (mean $\bm{B}:2.8, \bm{E}:5$) as is used for the data generation ($\bm{B}:3, \bm{E}:5$). The cut-off criterion based on the multivariate  variance (Cut-Off Mul) given by \eqref{eq:total_variation_decomp} results in more parsimonious models (mean $\bm{B}:2.15, \bm{E}:4$) and thus considerably higher \gls{mse} values. The increased variation in the \gls{mse} values can also be attributed to variability in the truncation orders (cf. Appendix Figure \ref{fig:app_sim_nfpc_y}), leading to a mixture distribution. Comparing the benchmark scenario to more sparsely observed functional data (ceteris paribus) suggests a lower estimation accuracy for the Sparse Data scenario, especially for the curve-specific random effect $\bm{E}_{ijh}(t)$ and resultingly the fitted curves $\bm{y}_{ijh}(t)$, but pooling the information across functions helps the estimation of $\bm{\mu}(\bm{x}_{ij}, t)$ and $\bm{B}_{i}(t)$. In particular, the estimation of the mean $\bm{\mu}(\bm{x}_{ij}, t)$ is quite robust against the increased uncertainty of these three scenarios. Only when the random scores are not centred and decorrelated as in the benchmark scenario do we find an increase in \gls{mse} values for the mean (Uncentred Scores). This corresponds to a departure from the modeling assumptions likely to occur in practice when only few levels of a random effect are available (here for the subject-specific $\bm{B}_{i}(t)$). The model then has difficulties to correctly separate the intercept in $\bm{\mu}(\bm{x}_{ij}, t)$ and the random effects $\bm{B}_{i}(t)$. The empirical (non-zero) mean of the $\bm{B}_{i}(t)$ is then absorbed by the intercept in $\bm{\mu}(\bm{x}_{ij}, t)$, resulting in higher \gls{mse} values for both of these model terms. However, this shift does not affect the overall fit to the data $\bm{y}_{ijh}(t)$ nor the estimation of the other fixed effects (cf. Appendix Figure \ref{fig:app_sim_eff_umse}). Note that the \gls{mse} values of the Sparse Data and Uncentred Scores scenarios are based on slightly different normalizing constants (i.e.\ different true data) and cannot be directly compared except for the mean.

Our simulation study thus suggests that basing the truncation orders on the proportion of explained variation on each dimension \eqref{eq:univar_variation_decomp} gives parsimonious and well fitting models. If interest lies mainly in the estimation of fixed effects, the alternative cut-off criterion based on the total variation in the data \eqref{eq:total_variation_decomp} allows even more parsimonious models at the cost of a less accurate estimation of the random effects and overall model fit. Furthermore, the results presented in Appendix \ref{app_sec:simulation} show that the mean estimation is relatively stable over different model scenarios including  misspecification of the measurement error  variance structure or of the multivariate scalar product, as well as in scenarios with strong heteroscedasticity across dimensions. In our benchmark scenario, the \glspl{cb} cover the true effect  $89-94\%$ of the time but coverage can further decrease with additional uncertainty e.g.\ about the number of \glspl{fpc}. Overall, the covariance structure such as the leading \glspl{fpc} can be recovered well, also for a nested random effect such as in the snooker training application. The comparison to the univariate modeling approach suggests that the \gls{mfamm} can improve the mean estimation but is especially beneficial for the prediction of the random effects while reducing the number of parameters to estimate. In some cases like strong heteroscedasticity, including weights in the multivariate scalar product might further improve the modeling.

\section{Discussion}
\label{sec:discussion}

The proposed multivariate functional regression model is an additive mixed model, which allows to model flexible covariate effects for sparse or irregular multivariate functional data. It uses \gls{fpc} based functional random effects to model complex correlations within and between functions and dimensions. An important contribution of our approach is estimating the parsimonious multivariate \gls{fpc} basis from the data. This allows us to account not only for auto-covariances, but also for non-trivial cross-covariances over dimensions, which are difficult to adequately model using alternative approaches such as  parametric covariance functions like the Matèrn family or penalized splines, which imply a parsimonious covariance only within but not necessarily between functions. As a \gls{famm}-type regression model, a wide range of covariate effect types is available, also providing pointwise \glspl{cb}. Our applications show that the \glspl{mfamm} can give valuable insight into the multivariate correlation structure of the functions in addition to the mean structure.

An apparent benefit of multivariate modeling is that it allows to answer research questions simultaneously relating to different dimensions. In addition, using multivariate \glspl{fpc} reduces the number of parameters compared to fitting comparable univariate models while improving the random effects estimation by incorporating the cross-covariance in the multivariate analysis. The added computational costs are small: For our multimodal application, the multivariate approach prolongs the computation time by only five percent (104 vs.\ 109 minutes on a 64-bit Linux platform). 

We find that the average point-wise coverage of the point-wise \glspl{cb} can in some cases lie considerably below the nominal value. There are two main reasons for this: One, the \glspl{cb} presented here do not incorporate the uncertainty of the eigenfunction estimation nor of the smoothing parameter selection. Two, coverage issues can arise in (scalar) mixed models, if effect functions are estimated as constant when in truth they are not \citep[e.g.][]{wood2017, Grevencomment}. To resolve these issues, further research on the level of scalar mixed models might be needed. A large body of research covering \gls{cb} estimation for functional data \citep[e.g.][]{goldsmith2013corrected, choi2018, liebl2019fast} suggests that the construction of \glspl{cb} is an interesting and complex problem, also outside of the \gls{famm} framework.

It would be interesting to extend the \gls{mfamm} to more general scenarios of multivariate functional data such as observations consisting of functions with different dimensional domains, e.g.\ functions over time and images as in \citet{happ2018}.  This would require adapting the estimation of the univariate auto-covariances for spatial arguments $t, t'$. Exploiting properties of dense functional data, such as the block structure of design matrices for functions observed on a grid, could help to reduce computational cost in this case. Future research could further generalize the covariance structure of the \gls{mfamm} by allowing for additional covariate effects. In our snooker training application, for example, a treatment effect of the snooker training might show itself in the form of reduced intra-player variance \citep[cf.][]{backenroth2018}. Ideas from distributional regression could be incorporated to jointly model the mean trajectories and covariance structure conditional on covariates.

\section*{Acknowledgements}
We thank Timon Enghofer, Phil Hoole, and Marianne Pouplier for providing access to their data and for fruitful discussions. We also thank Lisa Steyer for contributing the data registration of the snooker training data and two anonymous reviewers for their helpful suggestions. Sonja Greven, Almond Stöcker, and Alexander Volkmann were funded by grant GR 3793/3-1 from the German research foundation (DFG).
Fabian Scheipl was funded by the German Federal Ministry of Education and Research (BMBF) under Grant No. 01IS18036A.

\bibliography{lit}

\newpage
\appendix

\section{Derivation of Variance Decompositions}
\label{app_sec:vardecomp}

This is the derivation of the variance decompositions \eqref{eq:total_variation_decomp} and \eqref{eq:univar_variation_decomp} from the model equations \eqref{eq:multivariateFLMM} and \eqref{eq:truncated_KL_expansion}. Following the argumentation of \cite{cederbaum2016}, the variation of the response can be decomposed per dimension using iterated expectations as 
\begin{align*}
\int_{\mathcal{I}}\mathrm{Var}\big(y_{i}^{(d)}(t)\big) dt  = & \sum_{j=1}^q\sum_{v=1}^{V_{U_j}} z_{ijv}^2 \int_{\mathcal{I}}\mathrm{Var}\big(U_{jv}^{(d)}(t)\big) dt +  \int_{\mathcal{I}}\mathrm{Var}\big(E_{i}^{(d)}(t)\big) dt +\int_{\mathcal{I}}\mathrm{Var}\big(\epsilon_{it}^{(d)}\big) dt \\
     = & \sum_{j=1}^{q}\sum_{m = 1}^\infty\nu_{U_jm}\underbrace{\int_{\mathcal{I}}\psi_{U_jm}^{(d)}(t)\psi_{U_jm}^{(d)}(t)dt}_{\textstyle=\vert\vert\psi_{U_jm}^{(d)}\vert\vert^2} \\
     & + \sum_{m = 1}^\infty\nu_{Em}\underbrace{\int_{\mathcal{I}}\psi_{Em}^{(d)}(t)\psi_{Em}^{(d)}(t)dt}_{\textstyle=\vert\vert\psi_{Em}^{(d)}\vert\vert^2} + \int_{\mathcal{I}}\sigma_{d}^2dt \\
     = & \sum_{j=1}^{q}\sum_{m = 1}^\infty\nu_{U_jm}\vert\vert\psi_{U_jm}^{(d)}\vert\vert^2 + \sum_{m = 1}^\infty\nu_{Em}\vert\vert\psi_{Em}^{(d)}\vert\vert^2 + \sigma_d^2 \vert\mathcal{I}\vert
\end{align*}
assuming in the second equality that each observation is in exactly one group in the $j$th grouping layer, i.e.\ exactly one indicator $z_{ijv} = z_{ijv}^2, v = 1, \dots, V_{U_j}$ is one for each $i$ and $j$, and also using in the second equality that the scores are uncorrelated across eigenfunctions, in particular $\text{Cov}(\rho_{U_jvm}, \rho_{U_jvm'}) = 0$ if $m \neq m'$.

Similarly, the total variance as measured by the sum of the univariate variances can be decomposed as
\begin{align*}
\mathbb{E}\left(\vert\vert\vert \bm{y}_i - \bm{\mu}(\bm{x}_i) \vert\vert\vert^2 \right) = & \sum_{d=1}^{D}w_d\int_{\mathcal{I}}\mathrm{Var}\big(y_{i}^{(d)}(t)\big) dt \\ = & \sum_{j=1}^q \sum_{d=1}^{D}\sum_{v=1}^{V_{U_j}} z_{ijv}^2 w_d\int_{\mathcal{I}}\mathrm{Var}\big(U_{jv}^{(d)}(t)\big) dt +  \sum_{d=1}^{D}w_d\int_{\mathcal{I}}\mathrm{Var}\big(E_{i}^{(d)}(t)\big) dt \\
& + \sum_{d=1}^{D}w_d\int_{\mathcal{I}}\mathrm{Var}\big(\epsilon_{it}^{(d)}\big) dt \\
     = & \sum_{j=1}^{q}\sum_{m = 1}^\infty\nu_{U_jm}\underbrace{\sum_{d=1}^{D}w_d\int_{\mathcal{I}}\psi_{U_jm}^{(d)}(t)\psi_{U_jm}^{(d)}(t)dt}_{\textstyle=\vert\vert\vert\bm{\psi}_{U_jm}\vert\vert\vert^2 = 1} \\
     & + \sum_{m = 1}^\infty\nu_{Em}\underbrace{\sum_{d=1}^{D}w_d\int_{\mathcal{I}}\psi_{Em}^{(d)}(t)\psi_{Em}^{(d)}(t)dt}_{\textstyle=\vert\vert\vert\bm{\psi}_{Em}\vert\vert\vert^2 = 1} + \sum_{d=1}^{D}w_d\int_{\mathcal{I}}\sigma_{d}^2dt \\
     = & \sum_{j=1}^{q}\sum_{m = 1}^\infty\nu_{U_jm} + \sum_{m = 1}^\infty\nu_{Em} + \sum_{d=1}^{D}w_d\sigma_d^2 \vert\mathcal{I}\vert.
\end{align*}

\newpage
\FloatBarrier
\section{Additional Results for the Snooker Data}
\label{app_sec:snooker_analysis}

\subsection{Data Description}

In this section, we provide additional information of the study by \cite{snooker2014}. The participants of the snooker training study volunteered if they wanted to take part in the training group. The training schedule for this treatment group recommended to perform a training two to three times a week which consisted of three exercises that are supposed to improve snooker specific muscular coordination. Note that these were autonomous trainings so that there is no reliable information if the participants followed the recommendations. 

\begin{figure}
\centering
\includegraphics[width=0.8\textwidth]{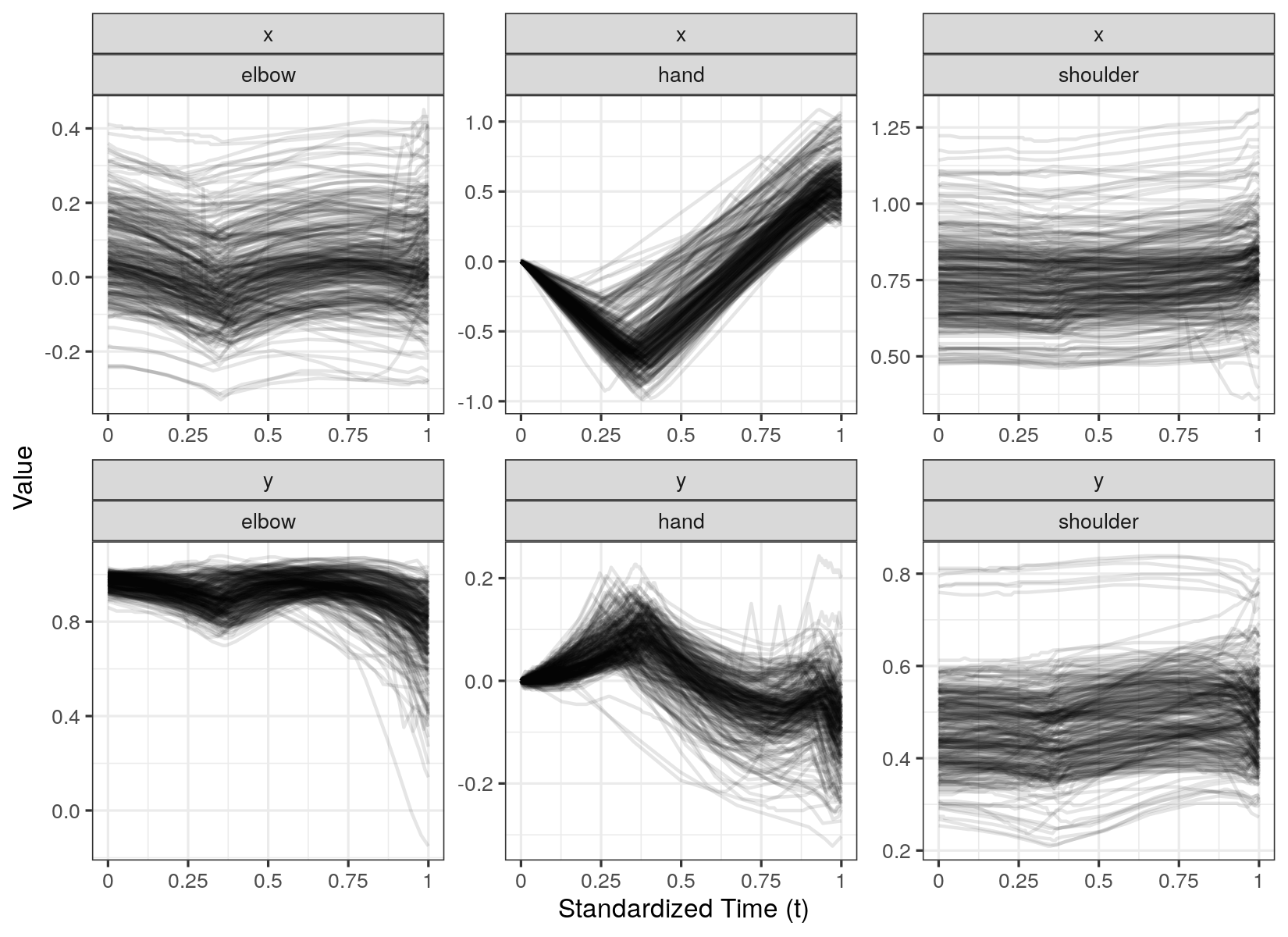}
\caption{Univariate functions of the snooker trajectories as functions of the standardized time $t$. The value of the function corresponds to the position of the elbow, hand, and shoulder on the $x$ and $y$ axes at the associated time point.}
\label{APPENDIXfig:snooker_univ_obs}
\end{figure}

In order to analyse the snooker shot of maximal force, the participants were asked to place themselves in the typical playing position. Yet no exact measure of the distance between the tip of their cue and the cue ball is available and thus the time of the impulse onto the cue ball cannot be exactly determined from the data. The participants were videotaped and the open source software Kinovea (\url{https://www.kinovea.org}) was used to manually locate points of interest (a participant's shoulder, elbow, and hand) and track them on a two-dimensional grid over the course of the video. Figure \ref{APPENDIXfig:snooker_univ_obs} shows the univariate functions that make up the two-dimensional trajectories of the participants' elbow, hand, and shoulder. The univariate functions for the x-axis of the hand show that for most observed curves, the time of impact might lie around $0.75$ of the relative time of the snooker shot. That is when the hand moves into the positive range on the $x$ dimension. For their shot, the snooker players can either try to fix their elbow or move the elbow dynamically. In order to move the hand in a straight line (piston stroke), the elbow has to move dynamically down and up when the hand is drawn back and accelerated towards the cue ball. The latter part of the movement trajectory (after the cue hits the cue ball) does not impact the snooker shot itself. Note that a long final downwards movement of an elbow trajectory indicates that the hand is not stopped at the chest but is rather pulled through towards the shoulder. This can give insight into a player's stance at the snooker table, e.g.\ if the upper body is close to the table or if it is more erect.

Note that the univariate functions of Figure \ref{APPENDIXfig:snooker_univ_obs} show different characteristics over the different dimensions, and a first-order difference penalty penalizing deviations from a constant function seems to be a sensible default choice. 

\subsection{Data Preprocessing}

In order to account for differences in body height between the participants, we first rescale the observed coordinate locations by the median distance between hand and elbow. As we are interested in the movement trajectory irrespective of phase variation, i.e.\ independent of the exact timing of different parts of the stroke, we apply an elastic registration approach to the data which aligns the two dimensional trajectory of the hand to its respective mean trajectory across all players and shots \citep{steyer2020}. We then reparameterize the time for the elbow and shoulder trajectories according to the results of the hand alignment. Other registration approaches are possible and we provide both the original and the reparameterized time of the data. Due to the high frame rate of the high speed camera and a comparatively rough grid of the tracking software, the resulting multivariate functional data are dense with redundant information. The following subsection describes the coarsening method in detail. The coarsening reduces the data set from roughly 400,000 to only 56,910 scalar observations. Note that 5 functional observations are lost due to technical problems during the recording. Even bivariate trajectories within one multivariate functional observation can have different numbers of scalar observations due to inconsistencies in the tracking programme.

\subsubsection*{Coarsening Method for the Snooker Data}
\label{app_sec:snooker_analysis:coarse}

Functional response variables as given by the snooker trajectories typically exhibit very high autocorrelation. Thus, up to a certain point, removing single curve measurements will inflict only very limited information loss while reducing  computational cost in the model fit at least linearly. 
We therefore coarsen the snooker trajectories as a preprocessing step, as they were originally sampled ``overly'' densely in that sense. In particular, there are also evaluation points that do not carry additional information as they measure the same location as the evaluation points right before and after (e.g. because the hand does not move noticeably over these three or more time points). The coarsening should a) retain original data points via subsampling to avoid pre-smoothing losses and b) be optimal in the sense that as much information as possible is preserved given the subsample size.

\begin{figure}
	\centering
	\includegraphics{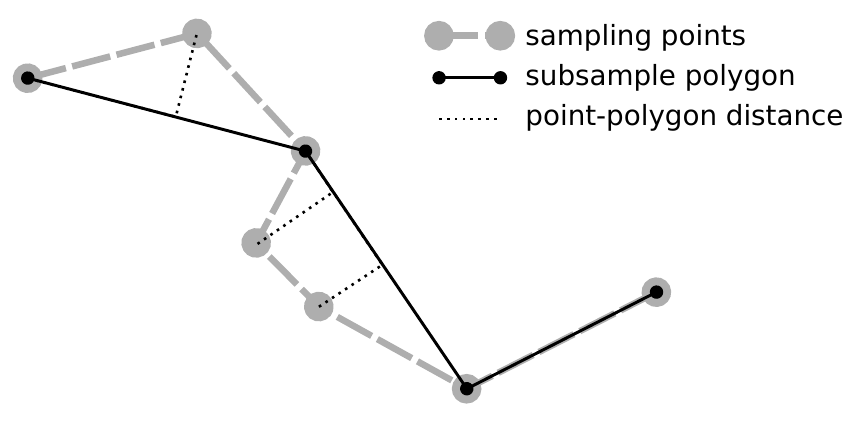}
	\caption{An example subsample polygon of length four through some curve measurements. The polygon canonically connects points in the subsample according to the order of the indices $t_i$ of the curve measurements $y(t_i)$ in the subsample.}
	\label{APPENDIXfig:coarsen1}
\end{figure}

Finding an optimal subsample, with respect to a general criterion and for all possible subsamples, is however typically very computationally demanding in itself. We therefore propose a fast greedy coarsening algorithm recursively discarding single points from the sample. The ultimate aim of the algorithm is to find a subsample polygon (Fig. \ref{APPENDIXfig:coarsen1}) representing the whole data as well as possible. 

\newcommand{\idx}{\mathcal{T}}
\newcommand{\score}{\Delta}

\begin{figure}
	\centering
	\includegraphics{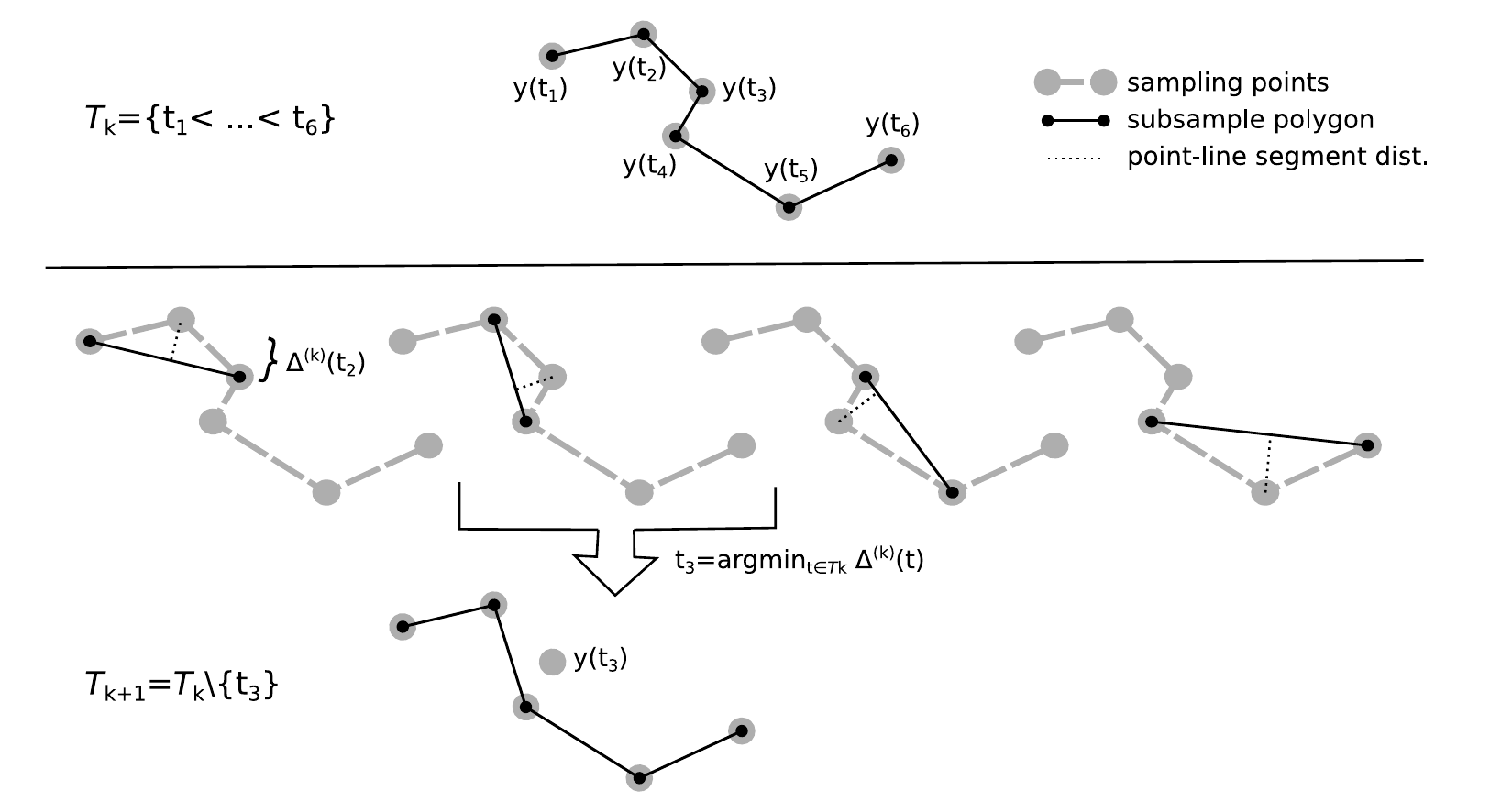}
	\caption{One iteration of the proposed fast greedy coarsening algorithm: Given the current subsample polygon (top), the loss $\score^{(k)}(t)$ inflicted when approximating $y(t)$ by the line segment between its neighbors is computed for each index $t_2,\dots,t_5$ (middle). The data point with the least approximation loss is discarded (bottom). Note that $\score^{(k+1)}(t)$ has to be re-computed only for the two adjacent points. All others can be passed forward.}
	\label{APPENDIXfig:coarsen2}
\end{figure}

One step of the coarsening algorithm described in the following is illustrated in Figure \ref{APPENDIXfig:coarsen2}. Let $\idx_0 = \{t_1, \dots, t_m\} \subset \mathcal{I}$ be the set of evaluation points $t_1 < \dots < t_m$ of a curve $y:\mathcal{I} \rightarrow \mathbb{R}^2$.  
We select a sequence of subsamples $\idx_0 \supset \idx_1 \supset \idx_2 \supset \dots$ by recursively discarding the point $t$ from $\idx_k$ where $y(t)$ is closest to the line segment $[y(t_-), y(t_+)]$ between its adjacent points at $t_- = \max \{s\in \idx_k : s < t\}$ and $t_+ = \min \{s\in \idx_k : s > t\}$. 
Note that this is only well-defined for $t \in \mathring \idx_k = \idx_k \setminus \{\min \idx_k, \max \idx_k\}$, in the ``interior'' of $\idx_k$, and the first and last point $t_1$ and $t_m$ are kept throughout the algorithm. 

In more detail, we proceed in the $k$th step of the algorithm as follows 
\begin{enumerate}
	\item $\forall t\in \mathring \idx_k$ compute the step-to-step quadratic error 
	\begin{align*}
	\score^{(k)}(t) = \min_{p\in[y(t_-), y(t_+)]} (y(t), p)^2.
	\end{align*}
	For computing the quadratic distance $\score^{(k)}(t)$, we have to distinguish the case where the projection of $y(t)$ on the line through $y(t_-)$ and $y(t_+)$ lies between the two points and the case where it lies on either side of them. To do so, define $y'(t) = y(t) - y(t_-)$ and $y'_+ = y'(t_+)$, as well as the unit vector $u = \frac{y'_+}{\|y'_+\|}$ pointing from $y(t_-)$ to $y(t_+)$. Then we can compute 
	\begin{align*}
	\score^{(k)}(t) = \begin{cases}
\|y'(t)\|^2 & \text{ if } \langle y'(t), u\rangle \leq 0\\
\|y'(t) - \langle y'(t), u\rangle u\|^2 & \text{ if } \|y'_+\| \geq \langle y'(t), u\rangle > 0 \\
\|y'(t) - y'_+\|^2 & \text{ if }  \langle y'(t), u\rangle > \|y'_+\| 
\end{cases}.
	\end{align*}
	\item Choose $t^{(k)} \in \operatorname{argmin}_{t\in\mathring \idx_k} \score^{(k)}(t)$ and set $\idx_{k+1} = \idx_k \setminus \{t^{(k)}\}$.
\end{enumerate} 

This procedure may be repeated until $\idx_{k+1}$ is of the desired sample size. However, we consider an error based stopping criterion more convenient in most cases.
We suggest to specify a threshold for the cumulative step-to-step error $S_k = \sum_{\kappa=1}^k \score^{(\kappa)}(t^{(\kappa)})$ or a relative version $R_k = S_k / \bar\score$ of it, relative to $\bar \score = \frac{1}{m-2} \sum_{\mathring \idx} \score^{(\infty)}(t)$ the mean quadratic distance to the line segment $[y(t_1), y(t_m)]$ between the two only points left at the ultimate subsample $\idx_\infty$.

In our experience, a threshold $R^* = 10^{-4}$ for $R_k$ yields a close to lossless subsample from a visual point of view. For the model, $R^* = 0.003$ was chosen to limit computational demands.
While the \gls{mfamm} would allow for dimension specific evaluations of the curves, the warping algorithm applied as part of the preprocessing does not. Thus, we decided to apply the coarsening algorithm to the hand trajectories, which we consider the most informative component with the best signal to noise ratio. Unselected observed time points for each hand trajectory are then also dropped for the corresponding trajectories of the shoulder and elbow.
As a result, the coarsened data retain the characteristic of the snooker trajectory data such that for a given time point, we observe the location of (almost) all points of interest (i.e.\ the grid of evaluation points is identical over the dimensions for a given observation).

\subsection{Additional Results of the Analysis}
\label{app_subsec:snooker_analysis}

\subsubsection*{Analysis of the Variance Components}

The \gls{mfamm} estimates a total of 61 univariate \glspl{fpc} (20 each for $B$ and $C$, and 21 for $E$), where each process is represented by three to five univariate \glspl{fpc} on each dimension. Table \ref{APPENDIXtab:snooker_varcontr} shows that the three points of interest (hand, elbow, and shoulder) all show a similar amount of variation. The source of variation, however, differs in interpretation. While the variation in the hand trajectories stems from differing movements, the variation in the shoulder mainly reflects different positioning. Applying higher weights to dimensions associated with the hand, for example, can shift the focus of the \gls{mfpca} to favor movement based variation of the hand trajectories in the analysis. From Table \ref{APPENDIXtab:snooker_varcontr} it is also apparent that movements or positions along the horizontal $x$-axis contribute more to the variation in the data than movements or positions along the vertical $y$-axis. This also suggests that variation in the $x$ direction is the driving factor of the \gls{mfpca}. By assigning higher weights in the scalar product to dimensions associated with the $y$-axis, the analyst can ``distort'' the natural grid of observations to balance out the different variation in the axes. Note that we use an unweighted scalar product in the analysis. Also keep in mind that the variation in Table \ref{APPENDIXtab:snooker_varcontr}  is calculated based on estimation step 1.

\begin{figure}
\centering
\includegraphics[width=0.8\textwidth]{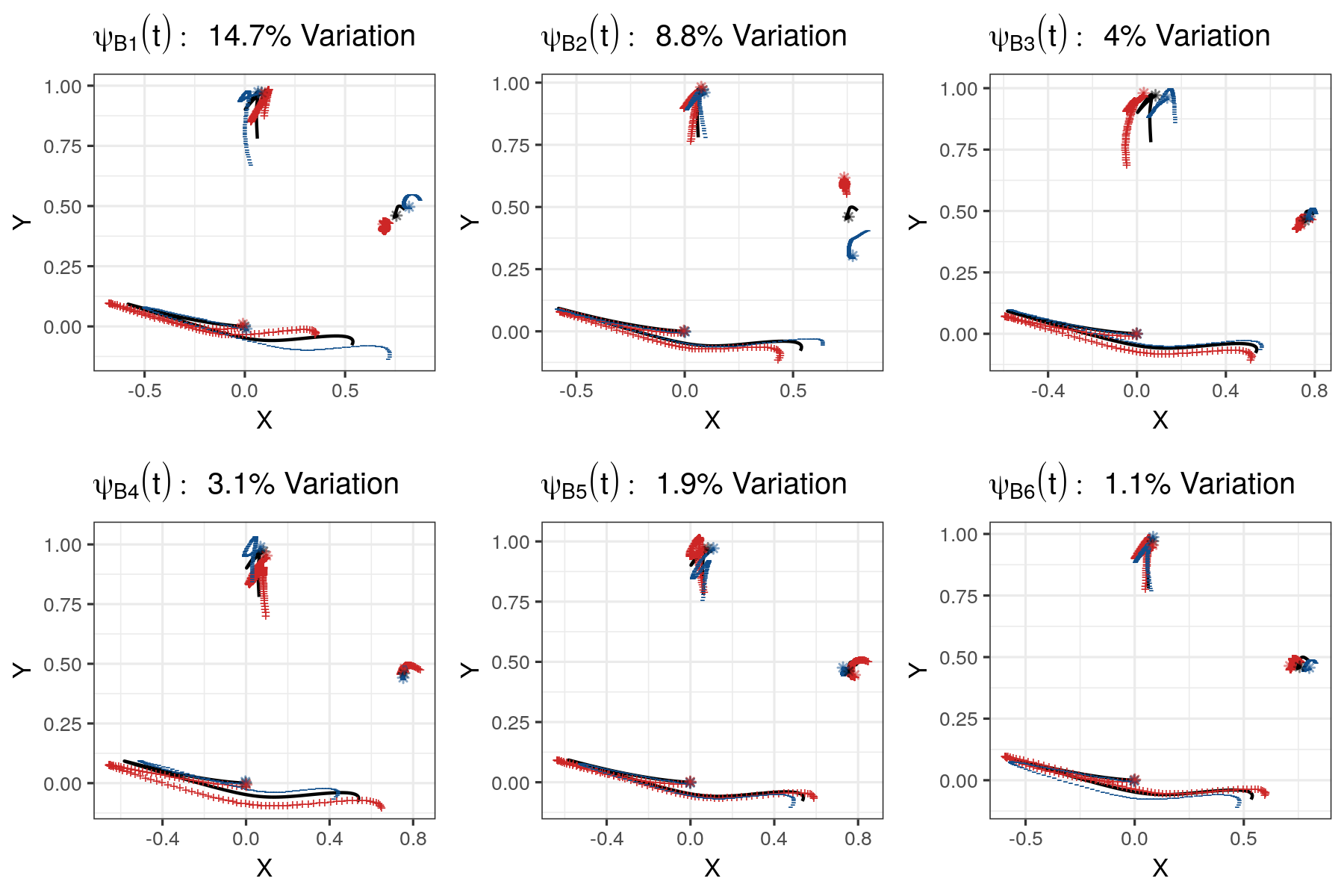}
\caption{\glspl{fpc} for the subject-specific functional random effect $\bm{B}_i(t)$ with the respective proportion of variance explained. The black solid line represents the mean trajectory to which a suitable multiple ($2\sqrt{\nu_{B\cdot}}$) of the \gls{fpc} is added (red $+$) and subtracted (blue $-$). The start of the trajectories are marked with an asterisk.}
\label{APPENDIXfig:snooker_fpc_B}
\end{figure}

\begin{figure}
\centering
\includegraphics[width=0.8\textwidth]{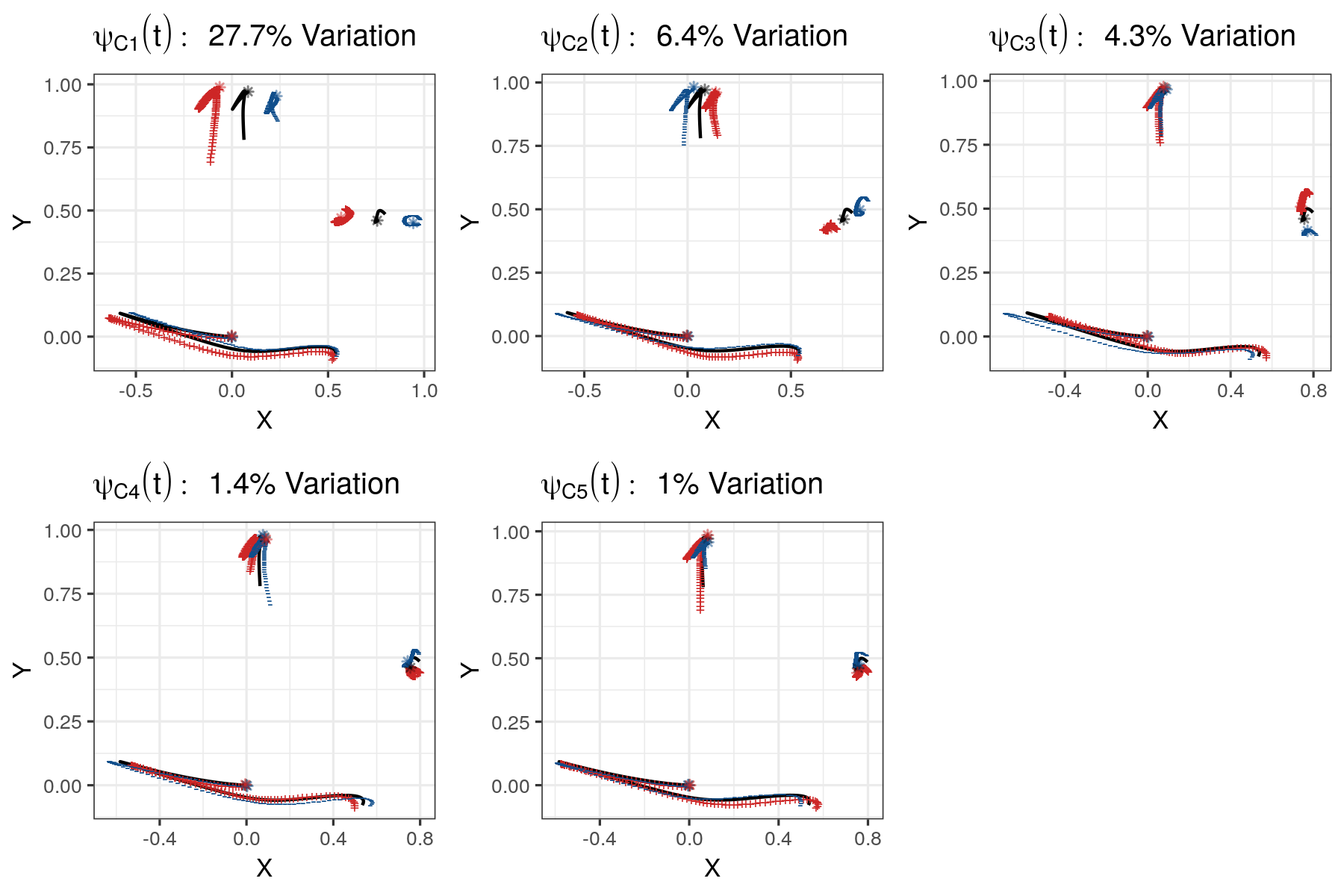}
\caption{\glspl{fpc} for the subject-and-session-specific functional random effect $\bm{C}_{ij}(t)$ with the respective proportion of variance explained. The black solid line represents the mean trajectory to which a suitable multiple ($2\sqrt{\nu_{C\cdot}}$) of the \gls{fpc} is added (red $+$) and subtracted (blue $-$). The start of the trajectories are marked with an asterisk.}
\label{APPENDIXfig:snooker_fpc_C}
\end{figure}

Based on the cut-off criterion \eqref{eq:total_variation_decomp}, 16 multivariate \glspl{fpc} are chosen to explain $95\%$ of total variation. For a similar amount of variance explained, 23 univariate \glspl{fpc} would be needed, which would give a more complex model that contains redundancies and ignores the multivariate nature of the data. Figures \ref{APPENDIXfig:snooker_fpc_B} and \ref{APPENDIXfig:snooker_fpc_C} show i.a.\ the two most prominant modes of variation in the snooker training data. As described in the main part, the \gls{fpc} $\bm{\psi}_{C1}$ seems to represent variation in the starting position (explaining about $27\%$ of total variation. The second leading \gls{fpc} $\bm{\psi}_{B1}$  (explaining about $15\%$ of total variation) shows subject-specific variation related to the two snooker techniques identified by \cite{snooker2014}: The red trajectory ($+$) shows that the elbow moves first down and then up to draw the hand back and accelerate it towards the cue ball. This then vertically contracts the hand trajectory compared to fixing the elbow during the acceleration phase (blue $-$), which allows the hand to swing more freely (pendulum stroke). We also find that players with a personal tendency towards the pendulum stroke seem to not stop their hands at their chest. Note that for $\bm{\psi}_{B1}$ in Figure \ref{APPENDIXfig:snooker_fpc_B}, the red trajectory is overlapping and thus masks the down-up-and-down movement of the elbow.

\begin{table}
	\centering 
  	\caption{Total variation of the centred responses per univariate dimension and overall. The variation is calculated as the sum of non-negative univariate eigenvalues and the dimension-specific measurement error variance as given from Step $1$ in the estimation of the \gls{mfamm} as described in Section \ref{subsec:EigenfunctionEstimation}}
  	\label{APPENDIXtab:snooker_varcontr} 
\begin{tabular}{l|cccccc|c} 
& elbow.x & elbow.y & hand.x & hand.y & shoulder.x & shoulder.y & Total \\\hline           
Variation & 0.012 & 0.004 & 0.015 & 0.001 & 0.014 & 0.008 & 0.055 \\
Proportion & 22\% & 7\% & 28\% & 2\% & 26\% & 14\% & 100\%
\end{tabular}
\end{table}

\begin{figure}
\centering
\includegraphics[width=0.8\textwidth]{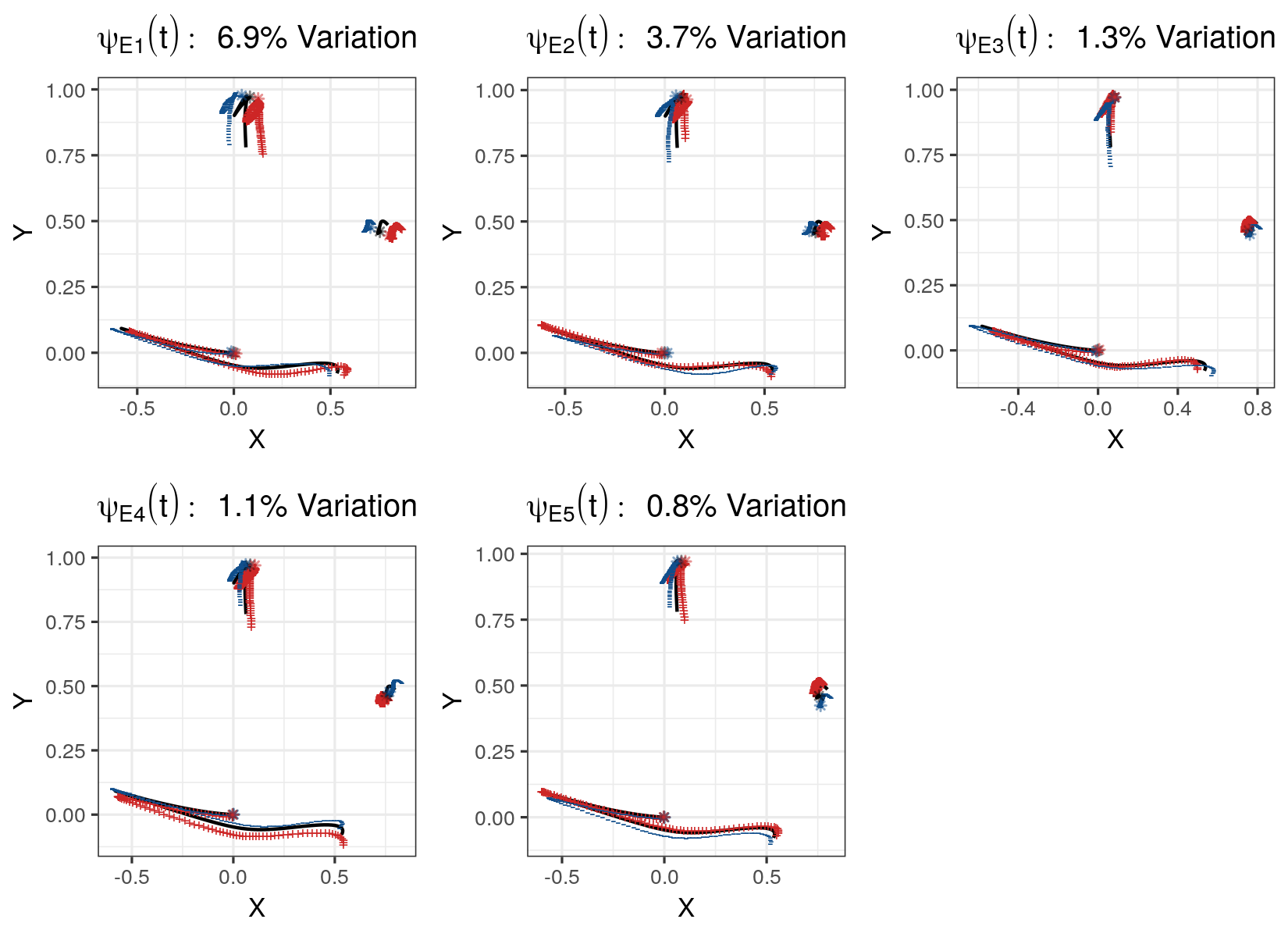}
\caption{\glspl{fpc} for the curve-specific functional random effect $\bm{E}_{ijh}(t)$ with the respective proportion of variance explained. The black solid line represents the mean trajectory to which a suitable multiple ($2\sqrt{\nu_{E\cdot}}$) of the \gls{fpc} is added (red $+$) and subtracted (blue $-$). The start of the trajectories are marked with an asterisk.}
\label{APPENDIXfig:snooker_fpc_E}
\end{figure}

Figure \ref{APPENDIXfig:snooker_fpc_B} shows the multivariate \glspl{fpc} of the subject-specific functional random intercept included in the \gls{mfamm}. $\bm{\psi}_{B2}(t)$ (explaining about 9\% of the total variation) seems to capture variation in the positioning of the shoulder, size differences of the upper arm, and the final part of the movement trajectories after the cue ball is hit. Figure \ref{APPENDIXfig:snooker_fpc_C} shows the multivariate \glspl{fpc} of the subject-and-session-specific random intercept. The second \gls{fpc} (explaining about $6\%$ of the total variation) shifts the relative positioning of the elbow and shoulder. Figure \ref{APPENDIXfig:snooker_fpc_E} shows the multivariate \glspl{fpc} of the curve-specific random intercept. The leading \gls{fpc} (explaining about $7\%$ of total variation) captures variation in the starting position of elbow and shoulder, and in the final hand movement.

\subsubsection*{Analysis of the Estimated Effect Functions}

\begin{figure}
\centering
\includegraphics[width=0.8\textwidth]{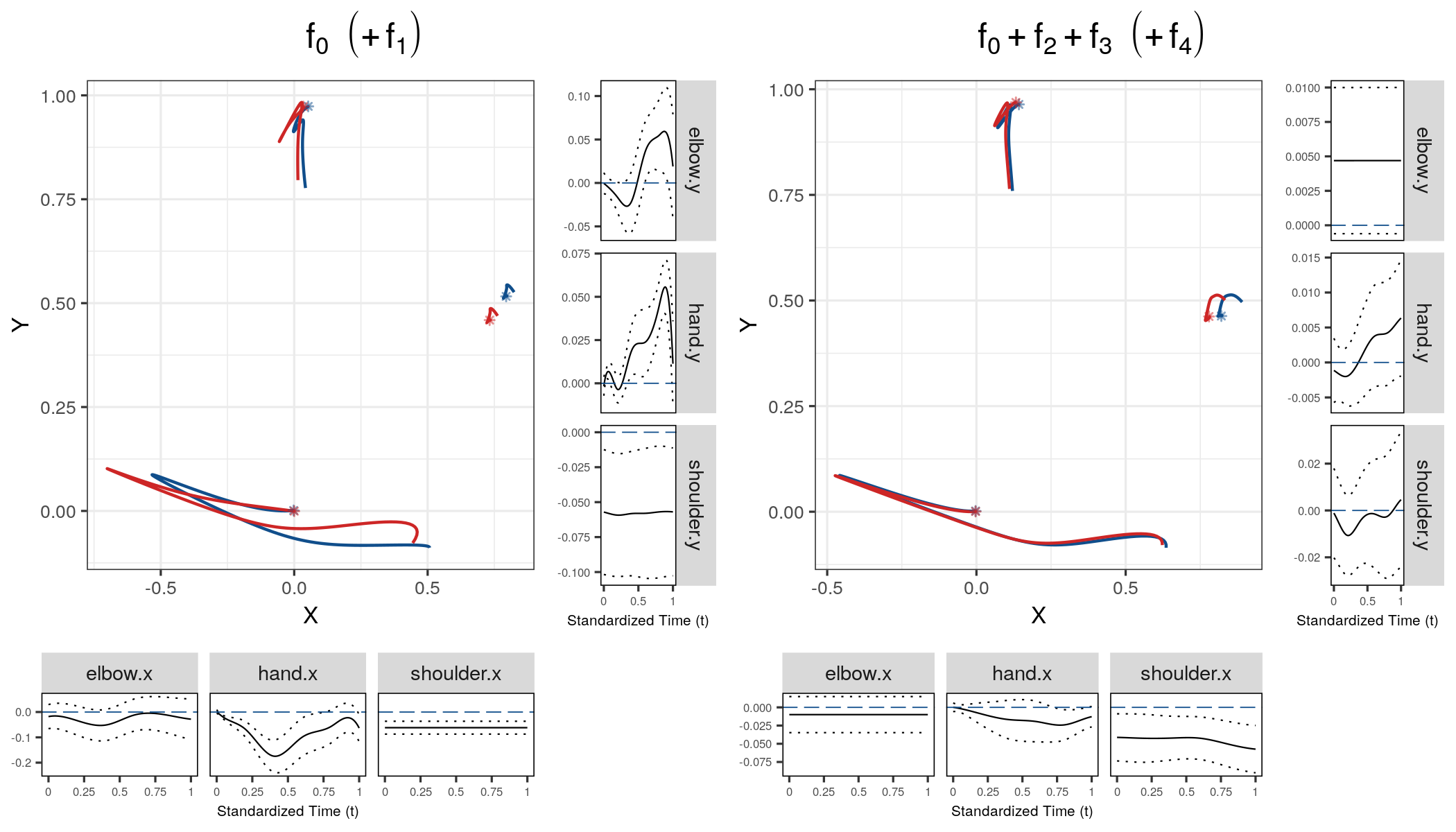}
\caption{Estimated covariate effect functions for skill (left) and treatment effect (right). The central plot shows the effect of the coefficient function in parenthesis in the title (red) on the two dimensional trajectories for the reference (without paranthesis, blue). The start of the trajectories are marked with an asterisk. The marginal plots show the estimated univariate effect functions (black) with pointwise 95\% confidence intervals (dotted) and the reference (blue dashed).}
\label{APPENDIXfig:snooker_cov_1_4}
\end{figure}

The left panel of Figure \ref{APPENDIXfig:snooker_cov_1_4} shows the estimated covariate effect functions for the covariate \texttt{skill}. In addition to the effect described in the main part, we point out that for skilled players, the shoulder is positioned closer to the table as well as to the body all other things being equal. The movement trajectories in the right panel of Figure \ref{APPENDIXfig:snooker_cov_1_4} are composed of the estimated effect functions of intercept, treatment group, and session (blue trajectories), so that the interaction effect can be separated and interpreted (red trajectories). We do not find strong evidence for differences in the displayed mean trajectories, nor in the univariate effects (marginal plots). This suggests that the training programme did not considerably change the participants' mean movement trajectories.

\begin{figure}
\includegraphics[width=\textwidth]{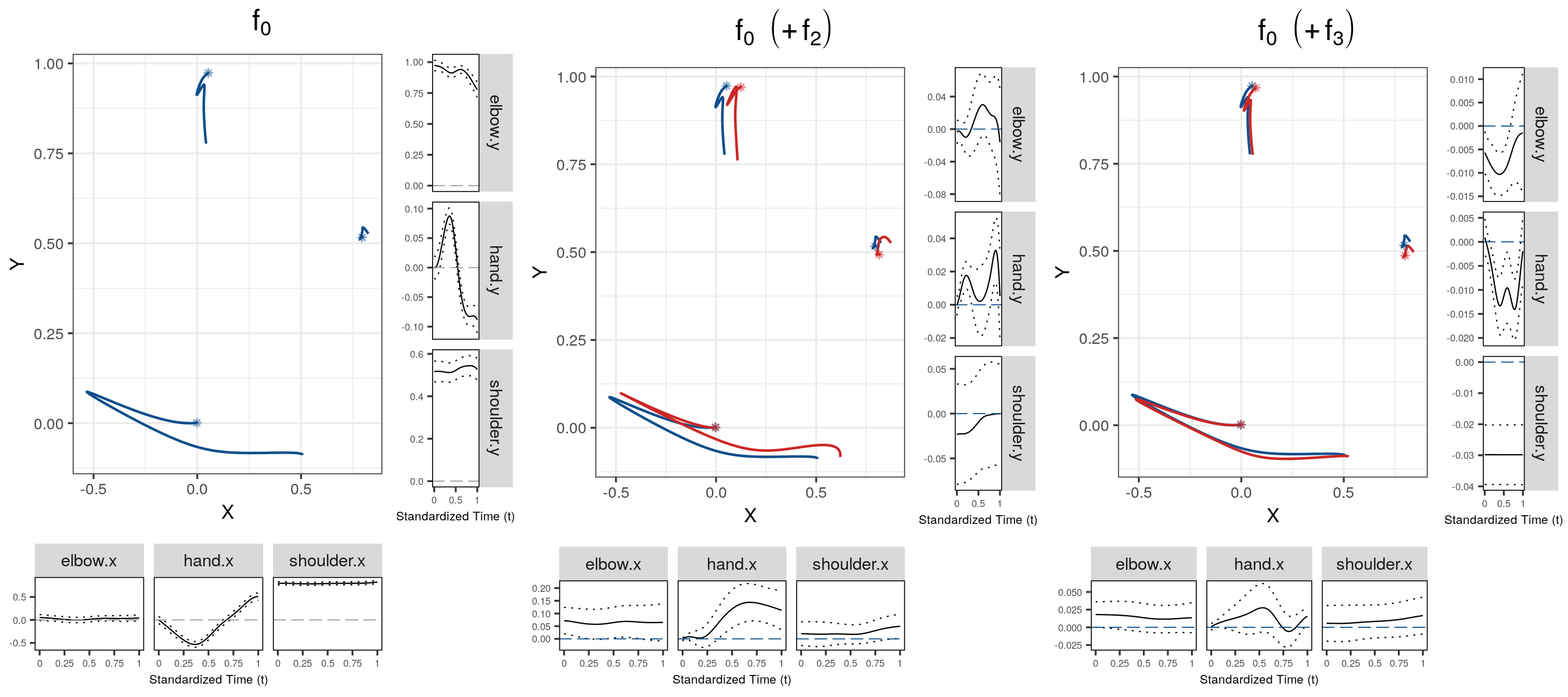}
\caption{Estimated functional fixed effects of intercept (left), covariate treatment group (center), and covariate session (right). The main plots show the effect of the coefficient function in parenthesis (red) on the two dimensional trajectories for the intercept (blue). The start of the trajectories are marked with an asterisk. The marginal plots show the estimated univariate effect functions (black) with pointwise 95\% confidence intervals (dotted) and the reference (grey/blue dashed).}
\label{APPENDIXfig:snooker_cov_023}
\end{figure}

Figure \ref{APPENDIXfig:snooker_cov_023} shows the estimated intercept (left) and effect functions of the covariates for treatment group (center) and session (right). The intercept (scalar plus functional) gives the mean movement trajectories (dark blue) in the reference group, i.e.\ an unskilled snooker player in the control group with a shot in the first session. In the middle panel, the red trajectories show the estimated effect of the treatment group added to the intercept. The marginal plots around the movement trajectories show the estimated univariate effects and their pointwise 95\% confidence intervals. We find only minor differences between movement trajectories in treatment and control group. The hand trajectories of the treatment group seem to be slightly higher and further along the $x$-axis than the control group, all other things being equal. The right panel compares the mean trajectory for the reference group of players in the first (blue) to the second (red) session. The univariate estimated covariate effects on the $y$-axis seem to indicate a slight shift in the vertical direction of the trajectories. Keep in mind that the trajectories of the hand have been centred to the origin before the analysis.

\subsubsection*{Model Diagnostics and Sensitivity Analysis}

\begin{figure}
\centering
    \includegraphics[width=0.9\textwidth]{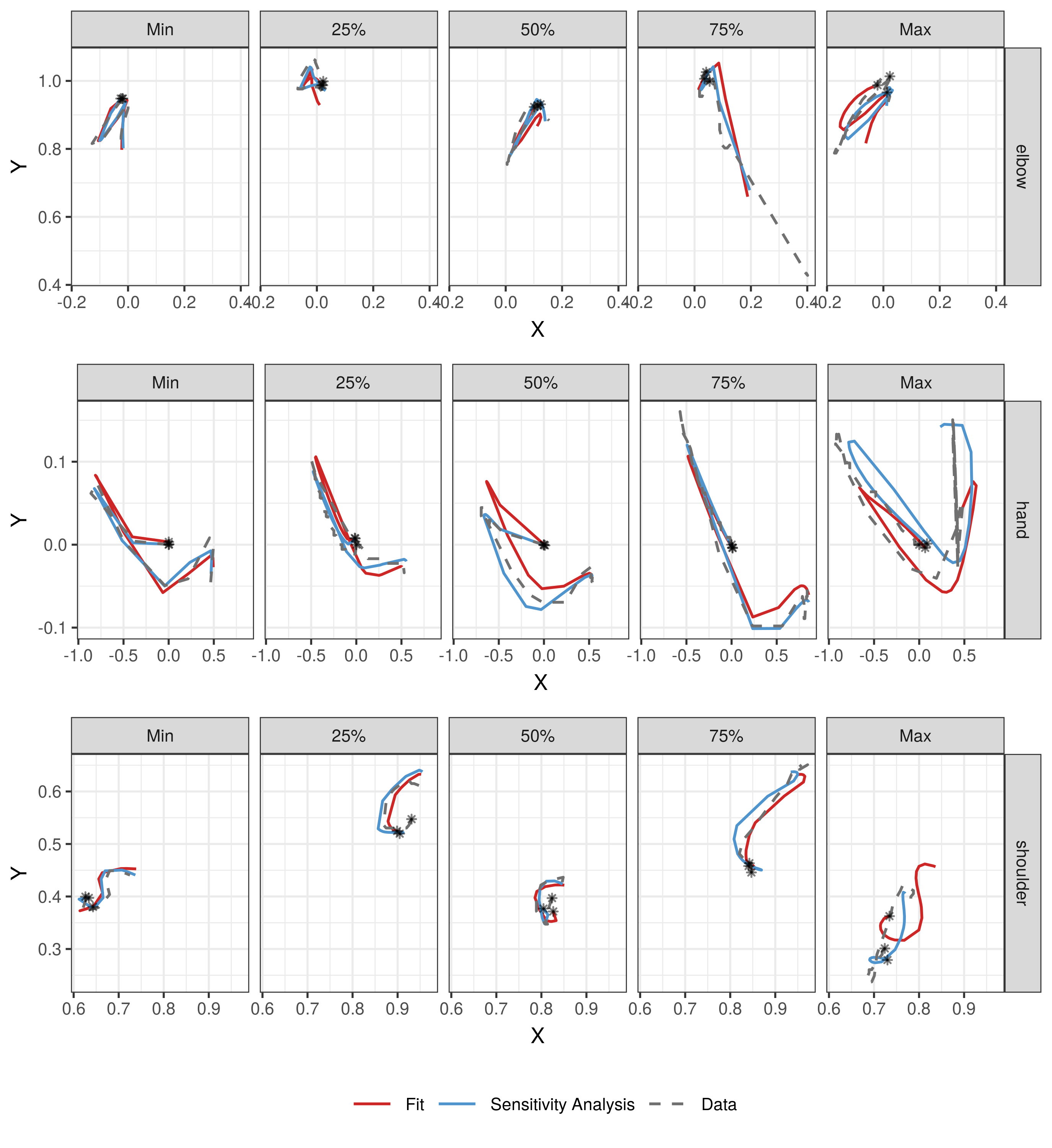}
    \caption{Fitted (red) and observed (blue) snooker training trajectories for selected observations (grey dashed). The qantiles given in the subtitles correspond to the contribution to the \gls{mmse}.}
    \label{APPENDIXfig:snooker_diagn_fit}
\end{figure}

We use the \gls{umse} defined in Section \ref{sec:simulation} as a criterion for model fit between the different dimensions. Table \ref{APPENDIXtab:snooker_umse_fit} shows that the model fits comparatively well on all dimensions except for the $x$-axis of the elbow and the $y$-axis of the hand. In order to get an impression of the quality of the model fit, Figure \ref{APPENDIXfig:snooker_diagn_fit} shows selected fitted trajectories in solid red together with the observed trajectories in dashed grey. We choose to present the quantiles from the contribution of each observation to the \gls{mmse}, i.e.\ quantiles from $\vert\vert\vert\bm{\zeta}_i- \hat{\bm{\zeta}}_i\vert\vert\vert^2$ of the multivariate functional trajectory $\bm{\zeta}_i, i = 1,..., N$ and the corresponding estimate $\hat{\bm{\zeta}}_i$ (see Section \ref{sec:simulation}).

\begin{figure}
\centering
\includegraphics[width=0.8\textwidth]{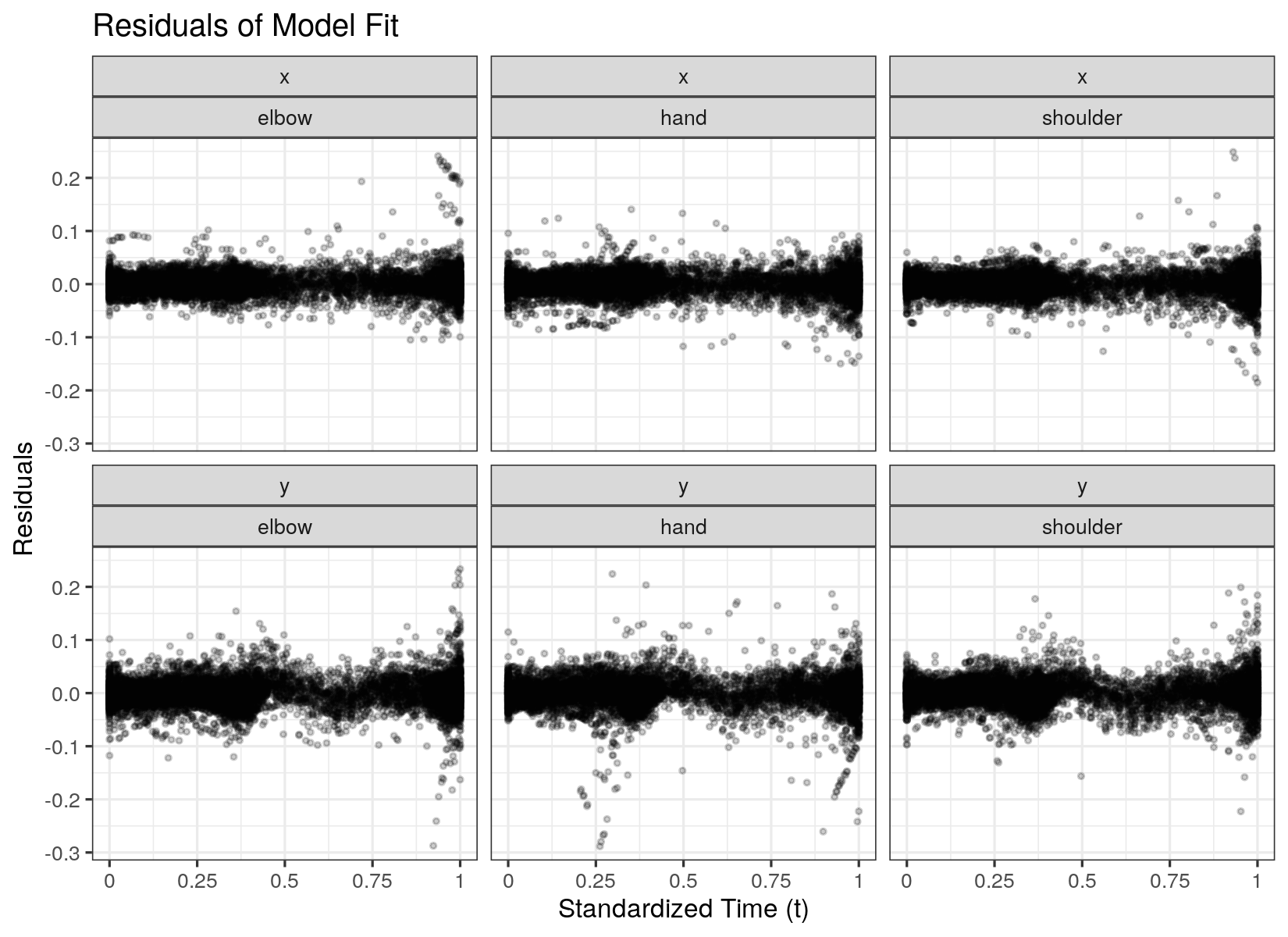}
\includegraphics[width=0.8\textwidth]{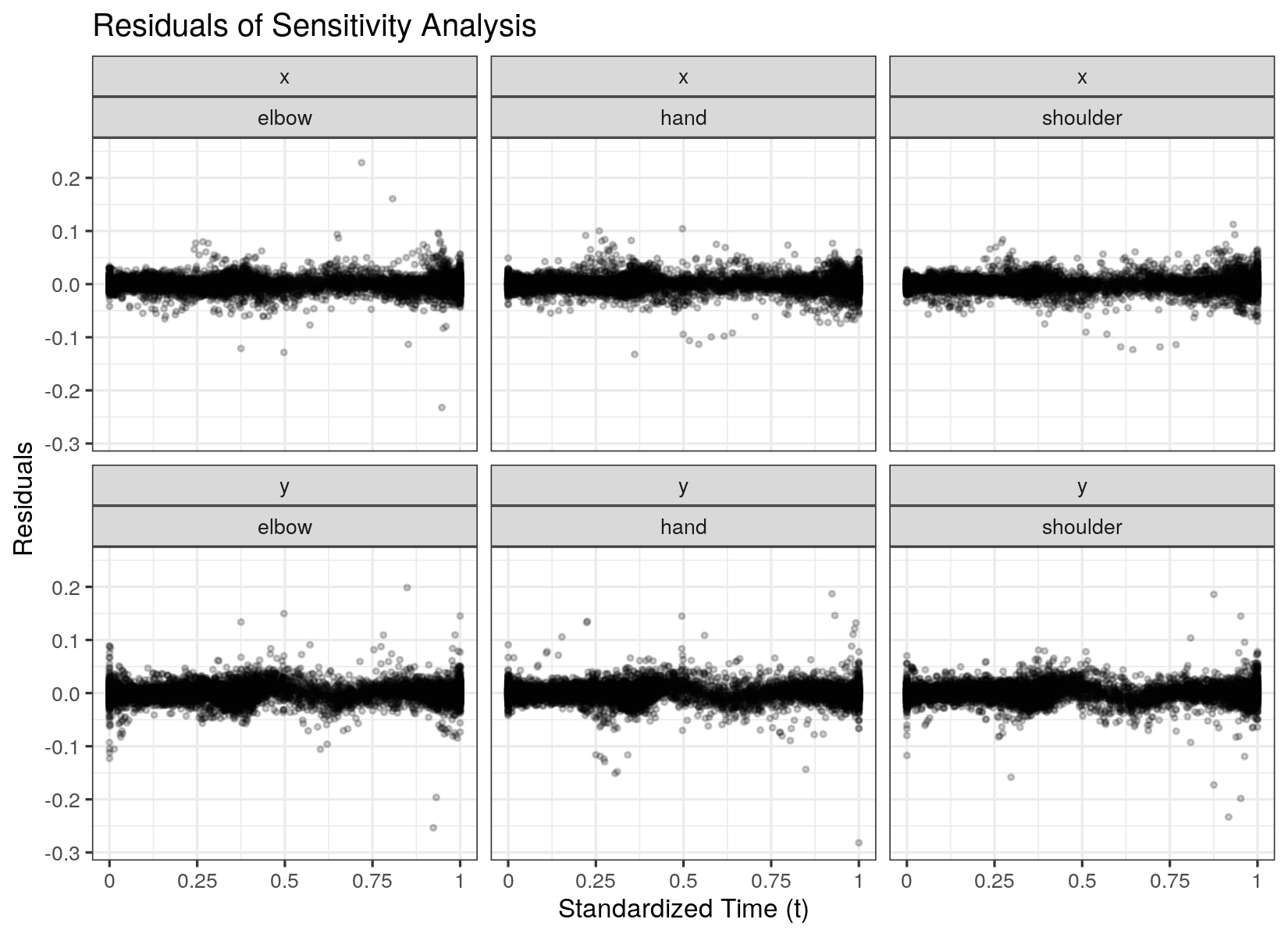}
\caption{Model residuals per dimension over time. The top panels correspond to the model presented in the main analysis, the lower panels to the model from the sensitivity analysis. Five observations have been removed from the lower panels so that the scale of the plots is identical.}
\label{APPENDIXfig:snooker_resids}
\end{figure}

\begin{figure}
\centering
\includegraphics[width=0.8\textwidth]{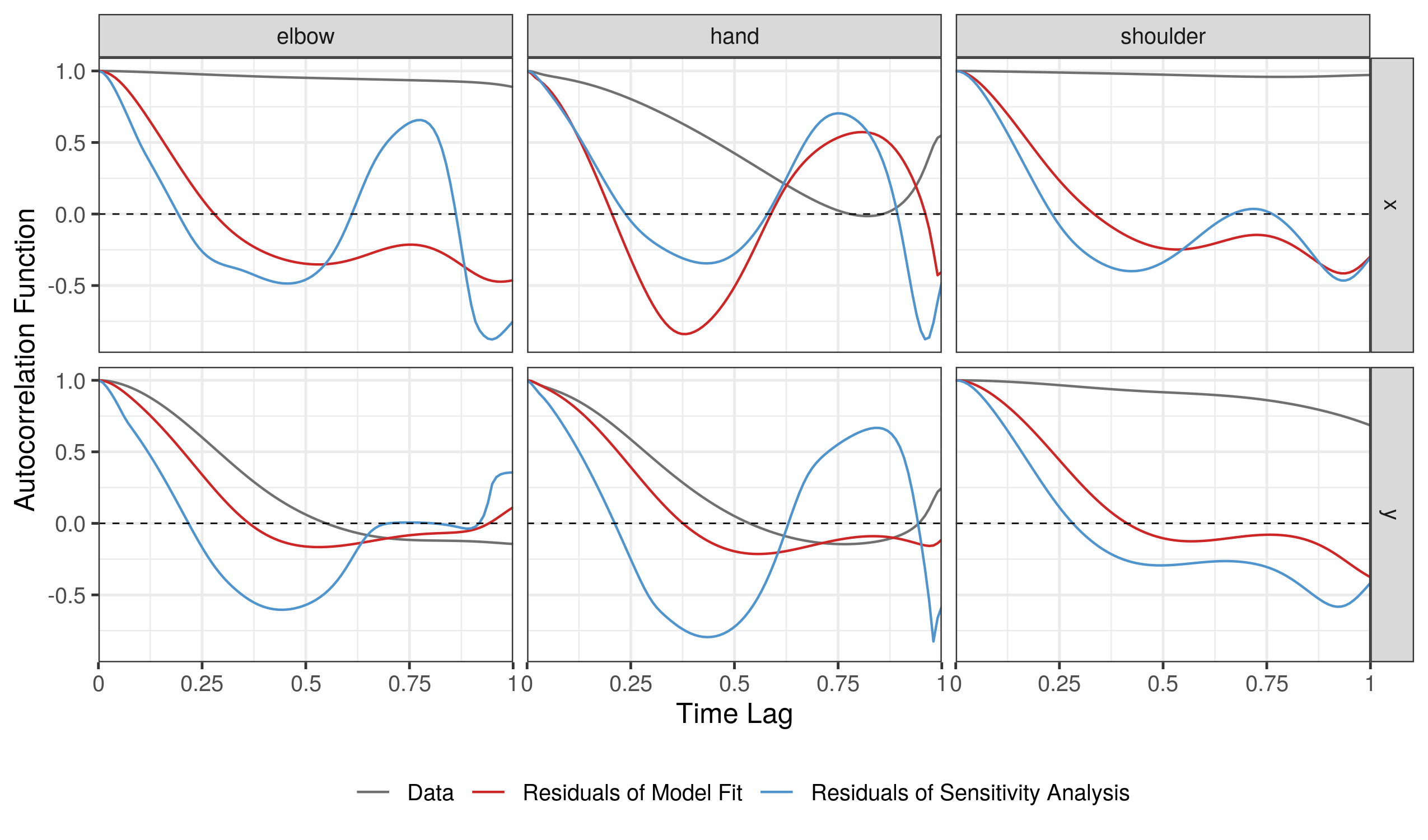}
\caption{Approximated mean autocorrelation function over the time lag for the observed data (grey), the fit of the model presented in the main analysis (red), and the model from the sensitivity analysis (blue). The dashed line marks zero autocorrelation.}
\label{APPENDIXfig:snooker_acf}
\end{figure}

The presented fitted trajectories suggest that the estimates might not always overlap with the observed trajectories, which suggests residual structure in the model residuals. The top panels of Figure \ref{APPENDIXfig:snooker_resids} plot the scalar residuals over time for each of the dimensions. Especially on the $y$-axis, we can discern patterns in the residuals even though this type of graphic is prone to overplotting. In order to investigate residual autocorrelation, we apply the following ad-hoc method, which allows us to approximate the well-known concept of an autocorrelation function in time series analysis. We first use fast symmetric additive covariance smoothing \citep{cederbaum2018} to estimate a smoothed correlation matrix for the residuals. Then we calculate the means of the off-diagonals, which corresponds to an approximated mean autocorrelation for a given time lag. Figure \ref{APPENDIXfig:snooker_acf} shows the results based on the residuals for each dimension separately. Overall, we find that the model residuals (red) show clear signs of autocorrelation, in particular up to a time lag of $0.25$. However, compared to the autocorrelation of the original data (solid grey line), we see a considerable reduction, which can be attributed to the functional random effects in the model.

Table \ref{APPENDIXtab:snooker_pred_var} also underlines the importance of the random effects in modeling the snooker trajectories. We use the predictor variance as an indicator for quantifying the effect of the fixed and random effects on the model fit. Separately on each dimension, we calculate the variance of the partial predictors on the data and give the respective proportions in Table \ref{APPENDIXtab:snooker_pred_var}. We find that with exception of the reference mean ($\bm{f}_0(t)$), the partial predictors of the fixed effects contribute generally little to the overall predictor variance (highest proportions for $\bm{f}_1(t)$ with around $5\%$ on most dimensions). Even the reference mean contributes little to the predictor variance on the shoulder and the $x$-axis of the elbow. Consequently, we find that most variance in the partial predictors can be assigned to the random effects. This suggests that the snooker training data might be dominated by random processes with little explanatory power of the available covariates.

\begin{table}
	\centering 
  	\caption{\gls{umse} values for the model fit on the snooker training data.}
  	\label{APPENDIXtab:snooker_umse_fit} 
\begin{tabular}{l|cccccc} 
& elbow.x & elbow.y & hand.x & hand.y & shoulder.x & shoulder.y \\\hline           
Main Model &  0.172 & 0.028 & 0.088 & 0.370 & 0.025 & 0.039 \\
Sensitivity Analysis & 0.121 & 0.019 & 0.084 & 0.233 & 0.019 & 0.025 \\
\end{tabular}
\end{table}

\begin{table}
	\centering 
  	\caption{Proportion of predictor variance for each dimension.}
  	\label{APPENDIXtab:snooker_pred_var} 
\begin{tabular}{l|cccccccc} 
& $f_0^{(d)}(t)$ & $f_1^{(d)}(t)$ & $f_2^{(d)}(t)$ & $f_3^{(d)}(t)$ & $f_4^{(d)}(t)$ & $B_i^{(d)}(t)$ & $C_{ij}^{(d)}(t)$ & $E_{ijh}^{(d)}(t)$ \\ \hline
elbow.x & 0.018 & 0.020 & 0.053 & 0.003 & 0.001 & 0.210 & 0.120 & 0.574 \\ 
elbow.y & 0.381 & 0.048 & 0.006 & 0.002 & 0.000 & 0.144 & 0.261 & 0.157 \\ 
hand.x & 0.862 & 0.022 & 0.014 & 0.000 & 0.000 & 0.022 & 0.050 & 0.030 \\ 
hand.y & 0.755 & 0.037 & 0.012 & 0.004 & 0.000 & 0.103 & 0.076 & 0.012 \\ 
shoulder.x & 0.007 & 0.047 & 0.013 & 0.001 & 0.019 & 0.184 & 0.144 & 0.585 \\ 
shoulder.y & 0.009 & 0.068 & 0.007 & 0.018 & 0.001 & 0.105 & 0.512 & 0.280 \\ 
\end{tabular}
\end{table}

We conduct a sensitivity analysis, where we increase the prespecified proportion of variance to $0.99$. Additionally, we assume heteroscedasticity in the model given the residual plot in Figure  \ref{APPENDIXfig:snooker_resids}. The model then estimates different measurement error variances with a predominantly large variance on dimension $hand.x$. The model almost doubles the number of \glspl{fpc} and includes 30 multivariate \glspl{fpc} (B: $9$, C: $10$, E:$11$). The estimation of fixed effects hardly change (results not shown) but we find a better fit to the data (compare Table \ref{APPENDIXtab:snooker_umse_fit} and Figure \ref{APPENDIXfig:snooker_diagn_fit}). The scalar residuals in the lower panels of Figure \ref{APPENDIXfig:snooker_resids} seem to exhibit less structure for the model of the sensitivity analysis but we can still discern patterns in the residuals. Considering the residual autocorrelation, Figure \ref{APPENDIXfig:snooker_acf} overall shows a reduction of autocorrelation for small time lags compared to the main model. For larger time lags we find a somewhat surprising increase of the approximated autocorrelation function. Including a large number of comparatively trivial \glspl{fpc} might thus overall improve model fit while introducing new dependencies in the residuals. Given that our main interest lies in the leading \glspl{fpc} and the fixed effects, we conclude that the sensitivity analysis yields similar results with added computational complexity.

\newpage
\FloatBarrier
\section{Detailed Analysis of the Consonant Assimilation Data}
\label{APPsec:ca_data}
\subsection{Data and Model Description}
\label{APPsubsec:phonetic_description}

\subsubsection*{Data Description}
In this section we present more information on the analysis of the consonant assimilation data. The \gls{aco} data were recorded via microphone at 32,768 Hz. In addition, all speakers wore a custom-fit palate sensor with 62 electrodes that measured the area where the tongue touched the palate during the articulation in high temporal resolution (\gls{epg} data).  These primary data were summarized and transformed into a functional index over the time of articulation for the two dimensions \gls{aco} and \gls{epg} separately. Each functional index measures how similar the articulatory or acoustic pattern is to its reference patterns for the first and second consonant at every observed time point. These similarity indices vary roughly between $+1$ and $-1$, where the value $+1$ indicates patterns close to the reference for the final target consonant (i.e.\ the consonant ending the first word) and $-1$ for the initial target consonant (i.e.\ the consonant beginning the second word). Due to the specifics of the index calculation the index values can lie slightly outside the interval $[-1,1]$. The curves on the \gls{aco} dimension are generally smoother than on the \gls{epg} dimension, which reflects the index calculation: The \gls{aco} signal is much richer in information because 256 continuously valued points are considered in the calculation of the similarity index for every time point while only 62 binary points enter the index calculation for the \gls{epg} signal. Details on the data generation and functional index computation can be found in \cite{pouplier2016} and \cite{cederbaum2016}.

\subsubsection*{Model Specification}
For this application, we follow the model specification of \cite{cederbaum2016}, who analyse only the \gls{aco} dimension of the data and ignore the second mode of the consonant assimilation data. Our specified multivariate model is 
\begin{align}
\label{eq:phonetic_model}
\bm{y}_{ijh}(t) = \bm{\mu}(\bm{x}_{ij}, t) + \bm{B}_i(t) + \bm{C}_{j}(t) + \bm{E}_{ijh}(t) + \bm{\epsilon}_{ijht},
\end{align}
with $i = 1,..., 9$ the speaker index, $j = 1, ..., 16$ the word combination index, and $h = 1,..., H_{ij}$ the repetition index. Again, $\bm{B}_i(t)$ and $\bm{E}_{ijh}(t)$ are the person-specific and curve-specific random intercepts. $\bm{C}_{j}(t)$ is the word combination-specific random intercept, which gives a crossed random effects structure. Note that here, the curve-specific random component $\bm{E}_{ijh}(t)$ also captures speaker and word interactions. Taking the different smoothness of the dimensions into account, the white noise measurement error $\bm{\epsilon}_{ijht}$ is assumed to follow a zero-mean bivariate normal distribution with diagonal covariance matrix $\diag(\sigma^2_{\text{ACO}}, \sigma^2_{\text{EPG}})$.

The additive predictor of the \gls{mfamm} is specified as
\begin{align*}
\bm{\mu}(\bm{x}_{ij},t) &= \bm{f}_0(t) + \texttt{order}_{j}\cdot\bm{f}_1(t) + \texttt{stress1}_{j}\cdot\bm{f}_2(t) + \texttt{stress2}_{j}\cdot\bm{f}_3(t) \\
 & \quad + \texttt{vowel}_{j}\cdot\bm{f}_4(t) + \texttt{order}_{j}\cdot\texttt{stress1}_{j}\cdot\bm{f}_5(t) \\
 & \quad + \texttt{order}_{j}\cdot\texttt{stress2}_{j}\cdot\bm{f}_6(t) + \texttt{order}_{j}\cdot\texttt{vowel}_{j}\cdot\bm{f}_7(t),
\end{align*}
with dummy covariates $\texttt{order}_{j}, \texttt{stress1}_{j}, \texttt{stress2}_{j}$, and $\texttt{vowel}_{j} $ indicating whether in the word combination /sh/ is followed by /s/ (instead of the reference /s\#sh/), the final target syllable is not stressed, the initial target syllable is not stressed, and the vowel context (i. e. the adjacent vowel for each consonant sound of interest) is /a\#i/ (instead of the reference /i\#a/ as in ``Gem\textbf{i}sch S\textbf{a}lbe''), respectively. The functional intercept $\bm{f}_0(t)$ and the effect function $\bm{f}_1(t)$ and their deviation from a sinus-like form or zero, respectively, are especially interesting for studying assimilation.

For comparability, we follow the univariate analysis in \cite{cederbaum2016} by specifying cubic P-splines with third order difference penalty with eight and five (marginal) basis functions for the covariate effect functions and auto-covariance estimation, respectively. The \gls{mfpca} is based on an unweighted scalar product. We choose the multivariate \gls{mfpca} truncation orders so that at least 95\% of the total variation in the data is explained as presented in \eqref{eq:total_variation_decomp}. To handle the heteroscedasticity, we use the weighted regression approach. Given the different sampling mechanisms on the dimensions, Appendix \ref{app_subsec:phon_m_wei} also contains an alternative analysis based on a weighted scalar product for the \gls{mfpca}.

\subsection{Results of the Model Presented in the Main Part}
\label{app_subsec:phon_main}

\subsubsection*{Analysis of the Variance Components}

The univariate decomposition of the auto-covariances yields five univariate \glspl{fpc} for the random components $\bm{B}$ and $\bm{E}$ on both dimensions and one and two \glspl{fpc} for $\bm{C}$ on \gls{aco} and \gls{epg}, respectively.  The cut-off criterion based on the sum of total variation selects a total of eight \glspl{fpc}. The estimated \gls{mfamm} then contains five multivariate \glspl{fpc} for the smooth residual (representing $64\%$ of total variation) and three for the subject-specific random effects ($20\%$ of total variation) with $12\%$ of total variation due to measurement error. With the chosen truncation order $M_C = 0$, the crossed random component $\bm{C}$ is effectively dropped from the model as a lot of variation in the data is already explained by the included fixed effects, all of which capture characteristics of the word combinations ($8$ fixed effects for $16$ word combinations). Table \ref{APPENDIXtab:phonetic_varexp} shows the contribution of each random process to the total variation in the consonant assimilation data according to the fitted \gls{mfamm}. The first row gives the eigenvalues of the random components $\bm{B}$ and $\bm{E}$ as well as the estimated univariate error variances and the total amount of variation in the data as computed in \eqref{eq:total_variation_decomp}. The second and third row show the univariate $L^2$ norm of the multivariate eigenfunctions. The proportion of explained variance $\pi$ is given in the final three rows. The proportions displayed in the fourth and fifth row are computed according to \eqref{eq:univar_variation_decomp} and the last row is computed according to \eqref{eq:total_variation_decomp}. Note that the fitted model uses the latter multivariate cut-off criterion to determine the number of \glspl{fpc}. With different sampling mechanisms for different dimensions and thus different measurement errors, it is advisable to check whether the proportion of explained variance on each dimension \eqref{eq:univar_variation_decomp} is adequate. For the consonant assimilation data, the selected number of \glspl{fpc} explains about $97\%$ of variation on \gls{epg} but only $94\%$ on \gls{aco}, which we still deem acceptable. If a greater disparity is revealed in an application, we recommend to use the proportion of explained univariate variation as a cut-off criterion (in our case this would lead to including a sixth \gls{fpc} for component $E$).  

\begin{table} \centering 
  \caption{Variance components included in the \gls{mfamm}. First row: Estimates of eigenvalues, univariate error variances, total variation. Second (third) row: Univariate norms of estimated \glspl{fpc} on dimension \gls{aco} (\gls{epg}). Fourth (fifth) row: Proportion of univariate variation explained on dimension \gls{aco} (\gls{epg}) by eigenfunctions and error variance, total univariate variation explained. Sixth row: Proportion of multivariate variation explained by eigenfunctions and error variances, total multivariate variation explained.}
  \label{APPENDIXtab:phonetic_varexp} 
  \resizebox{\textwidth}{!}{
\begin{tabular}{l|cccccccc|cc|c} 
& $B_1$ & $B_2$ & $B_3$ & $E_1$ & $E_2$ & $E_3$ & $E_4$ & $E_5$ & $\sigma_{ACO}^2$ & $\sigma_{EPG}^2$ & Total \\\hline 
\hline 
Variation & $0.018$ & $0.009$ & $0.004$ & $0.060$ & $0.017$ & $0.012$ & $0.007$ & $0.003$ & $0.004$ & $0.014$ & $0.153$ \\ 
$\vert\vert\psi^{(ACO)}\vert\vert^2$ & $0.169$ & $0.585$ & $0.642$ & $0.153$ & $0.217$ & $0.849$ & $0.178$ & $0.713$ & $--$ & $--$ & $--$ \\ 
$\vert\vert\psi^{(EPG)}\vert\vert^2$ & $0.831$ & $0.415$ & $0.358$ & $0.847$ & $0.783$ & $0.151$ & $0.822$ & $0.287$ & $--$ & $--$ & $--$ \\ 
$\pi^{(ACO)}$ & $0.068$ & $0.126$ & $0.052$ & $0.210$ & $0.082$ & $0.226$ & $0.026$ & $0.056$ & $0.094$ & $--$ & $0.940$ \\ 
$\pi^{(EPG)}$ & $0.134$ & $0.036$ & $0.012$ & $0.467$ & $0.119$ & $0.016$ & $0.049$ & $0.009$ & $--$ & $0.127$ & $0.969$ \\ 
$\pi$ & $0.115$ & $0.062$ & $0.023$ & $0.393$ & $0.108$ & $0.077$ & $0.042$ & $0.023$ & $0.027$ & $0.091$ & $0.961$ \\ 
\hline 
\end{tabular} }
\end{table} 

\begin{figure}
\centering
\includegraphics[width=0.7\textwidth]{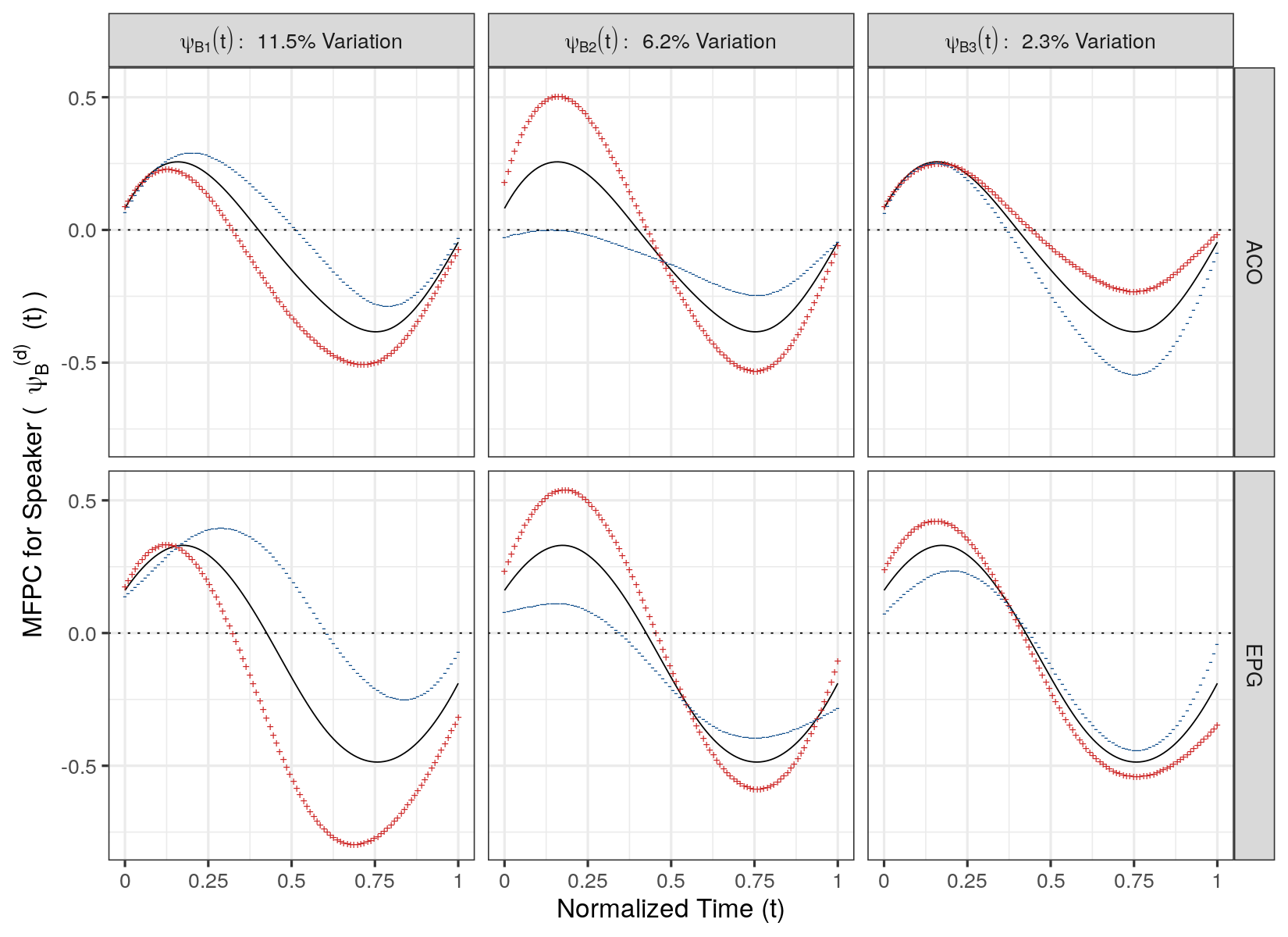}
\caption{\glspl{fpc} for the subject-specific functional random effect $\bm{B}_{i}(t)$ with the respective proportion of variance explained. The black solid line represents the mean trajectory to which a suitable multiple ($2\sqrt{\nu_{B\cdot}}$) of the \gls{fpc} is added (red $+$) and subtracted (blue $-$).}
\label{APPENDIXfig:phonetics_fpc_B}
\end{figure}

We present the estimated multivariate \glspl{fpc} for component $\bm{B}$ in Figure \ref{APPENDIXfig:phonetics_fpc_B}. The estimated leading multivariate \gls{fpc} $\bm{\psi}_{B1}(t)$ of the subject-specific random effect $\bm{B}_i(t)$ (see left panels) shows a personal tendency to pronounce the final or initial target syllable distinctly, which explains roughly $12\%$ of the total variation. It changes the relative time spent pronouncing each syllable and pulls the entire curve closer to the reference sound of the preferred consonant. The second \gls{fpc} (accounting for $6\%$ variation), however, shows the individual tendency for assimilation. It flattens or amplifies the sinus-like shape of the overall mean (black solid line) thus rendering the speaker more or less prone to distinguish between the two sounds. Note that in the univariate analysis for \gls{aco} presented in \cite{cederbaum2016}, this mode of variation is identified as the leading \gls{fpc} of the person-specific random effect. The three leading \glspl{fpc} $\bm{\psi}_{E1}(t)$, $\bm{\psi}_{B1}(t)$, and $\bm{\psi}_{E2}(t)$ all show greater univariate norms on the \gls{epg} dimension (Table \ref{APPENDIXtab:phonetic_varexp}) suggesting that this dimension is the driving source of variation in the model -- but this dimension was ignored in the analysis of \cite{cederbaum2016}. Note that only $\bm{\psi}_{B3}(t)$ impacts the dimensions differently in that more assimilation on one dimension equals less assimilation on the other. 

\begin{figure}
\centering
\includegraphics[width=0.8\textwidth]{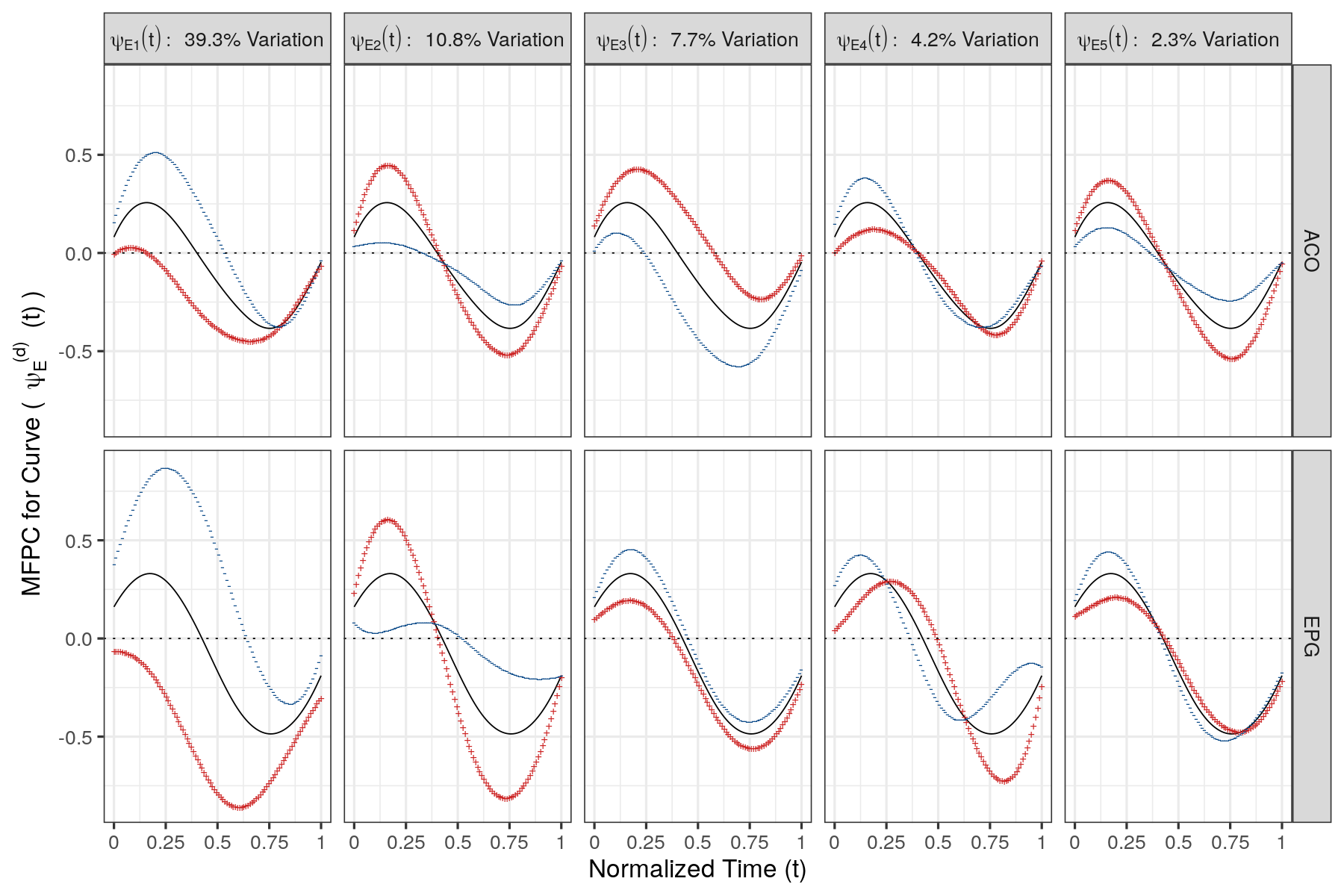}
\caption{\glspl{fpc} for the curve-specific functional random effect $\bm{E}_{ijh}(t)$ with the respective proportion of variance explained. The black solid line represents the mean trajectory to which a suitable multiple ($2\sqrt{\nu_{E\cdot}}$) of the \gls{fpc} is added (red $+$) and subtracted (blue $-$).}
\label{APPENDIXfig:phonetics_fpc_E}
\end{figure}

For the curve-specific random effect $\bm{E}_{ijh}(t)$ (Figure \ref{APPENDIXfig:phonetics_fpc_E}), the leading \gls{fpc} impacts primarily the final target consonant pulling it towards or pushing it away from its reference sound. $\bm{\psi}_{E2}(t)$ has a similar shape to $\bm{\psi}_{B2}(t)$. Note that the third leading \gls{fpc} of $E$ affects the mean function in opposite directions on to the two dimensions shifting the curve up on one dimension and down on the other.

\begin{figure}
\centering
\includegraphics[width=0.65\textwidth]{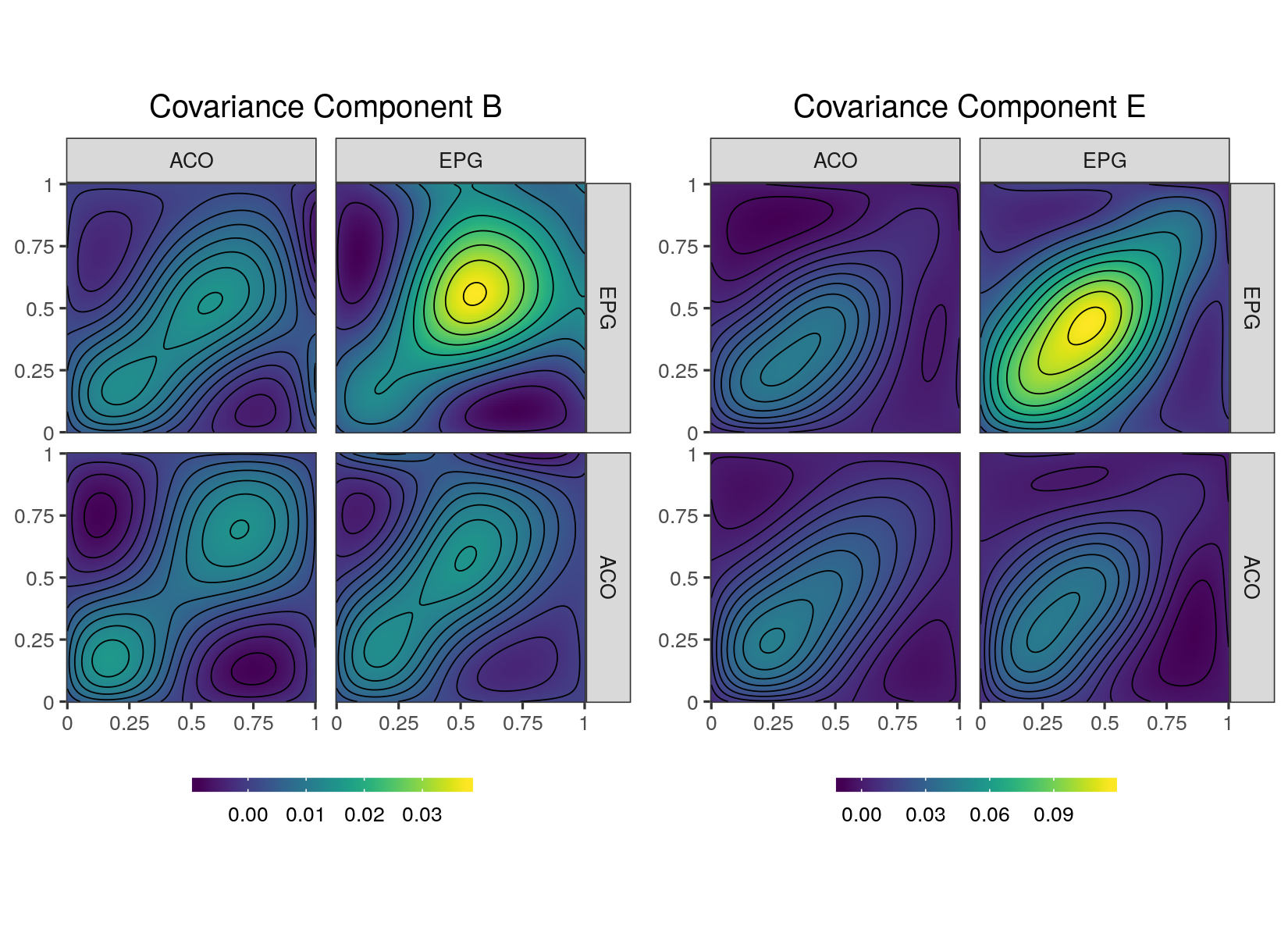}
\caption{Auto- and cross-covariance surface of the subject-specific functional random effect $\bm{B}_{i}(t)$ (left) and curve-specific random effect  $\bm{E}_{ijh}(t)$ (right). The bottom-left and top-right panels show the auto-covariance and the top-left and bottom-right panels show the cross-covariance function as a function of normalized time.}
\label{APPENDIXfig:phonetic_surv_E}
\end{figure}

Figure \ref{APPENDIXfig:phonetic_surv_E} shows the estimated auto-and cross-covariance surface of components $\bm{B}$ and $\bm{E}$ . It is evident that the cross-covariances are about as large in magnitude as the auto-covariances for the dimension \gls{aco}. In univariate analyses, however, the cross-covariance is completely ignored. Note that when no covariates are included in the model, the estimated \gls{mfamm} contains one \gls{fpc} for component $C$ (results not shown).

\subsubsection*{Analysis of the Estimated Effect Functions}

\begin{figure}
\centering
\includegraphics[width=0.9\textwidth]{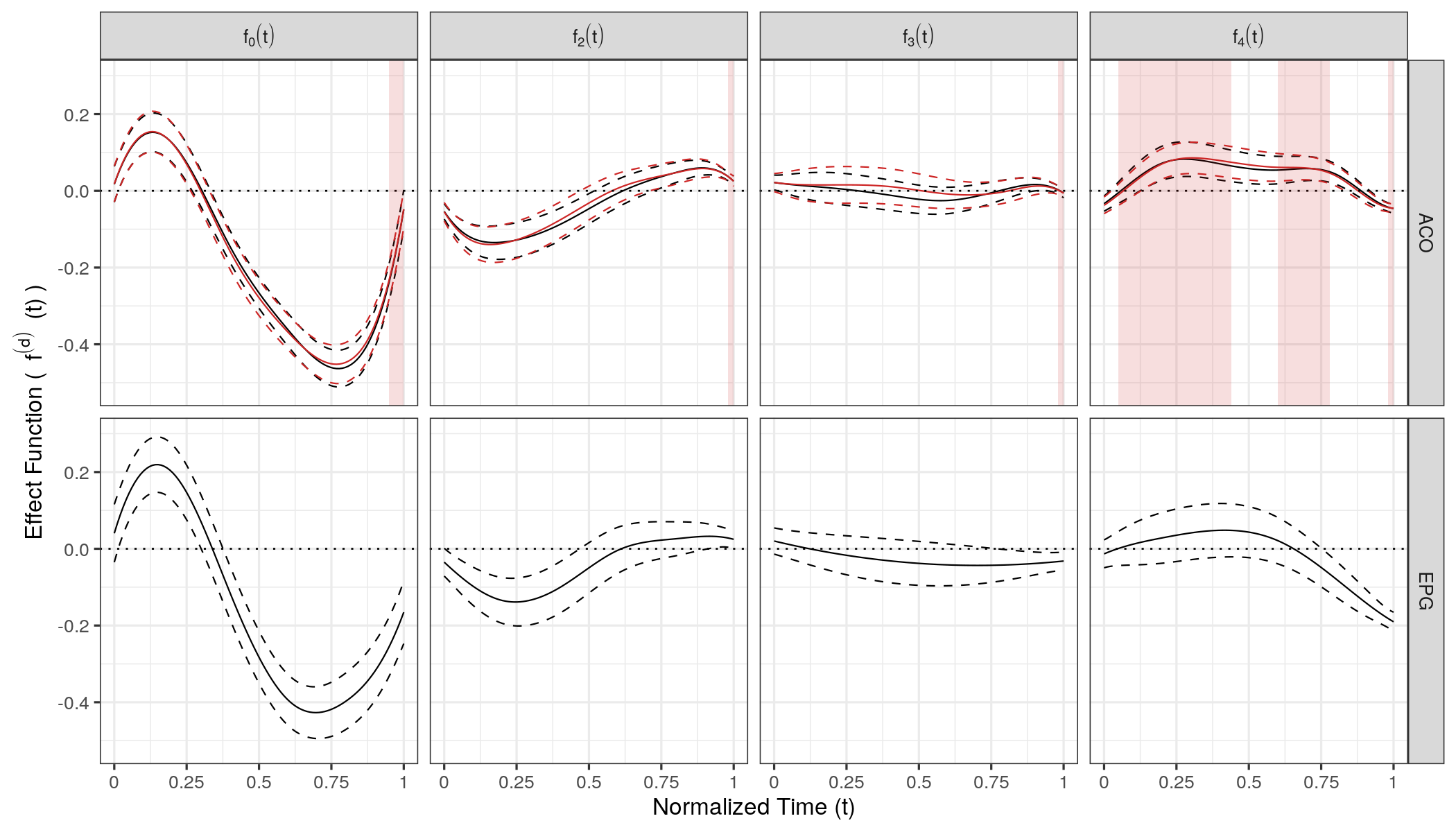}
\includegraphics[width=0.9\textwidth]{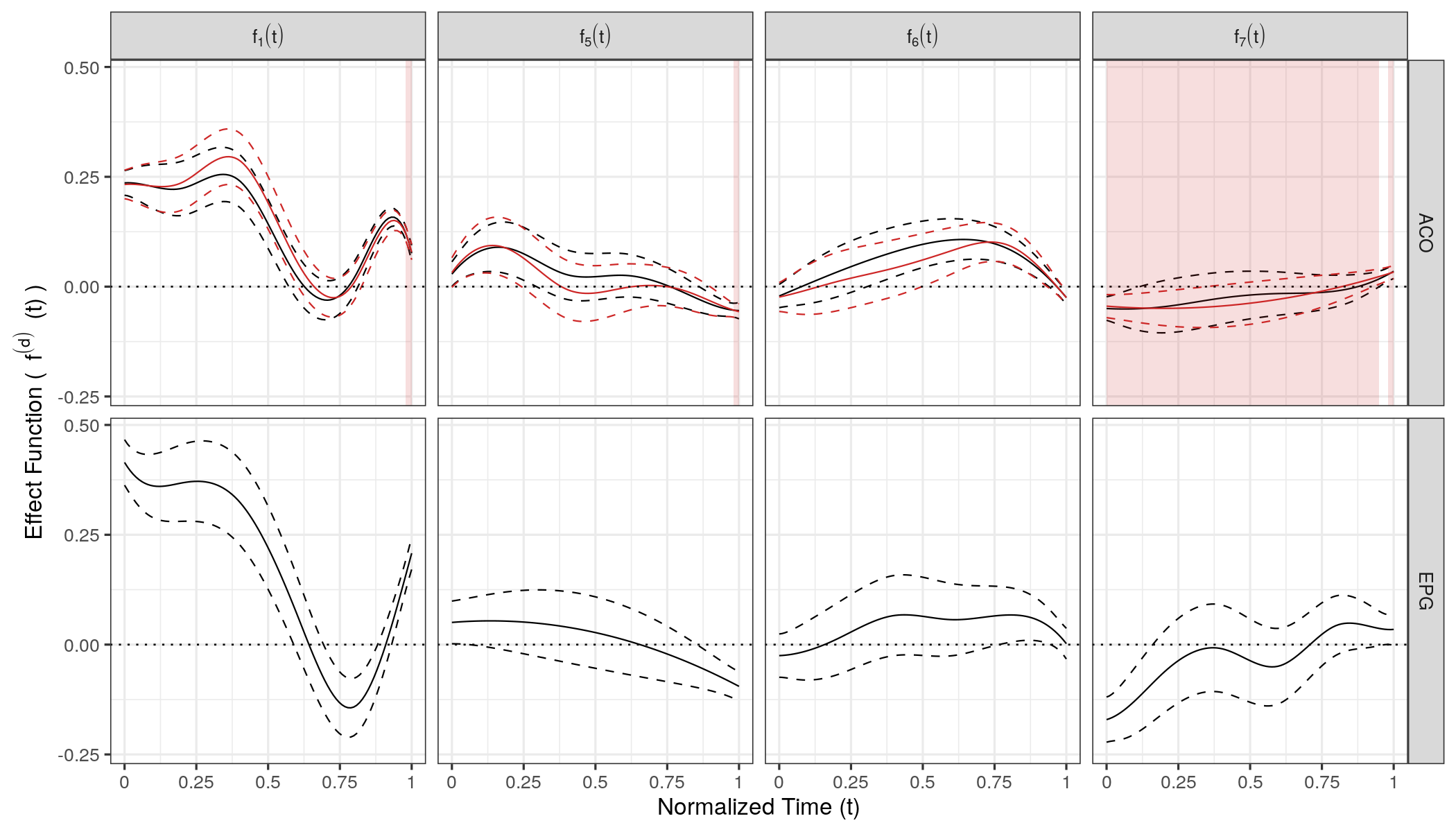}
\caption{Estimated covariate effects (black) and comparison to univariate model (red). The corresponding $95\%$ confidence intervals are given by the dashed line. Upper (from left to right): reference mean and covariate effects of \texttt{stress1}, \texttt{stress2}, \texttt{vowel} (/s\#sh/). Lower (from left to right): covariate effect of \texttt{order} and interactions of \texttt{order} with \texttt{stress1}, \texttt{stress2}, \texttt{vowel} (/sh\#s/). Areas shaded in red indicate where the standard errors of the univariate analysis are smaller.}
\label{APPENDIXfig:phonetic_covar_estim}
\end{figure}

Figure \ref{APPENDIXfig:phonetic_covar_estim} presents the estimated covariate effects (black, top plot for consonant order /s\#sh/, bottom for /sh\#s/). Overall, we find similar shapes for the estimated effects on both dimensions. In the reference group, the functional intercept $\bm{f}_0(t)$ shows signs of assimilation for consonant /s/. The effect $\bm{f}_1(t)$ of covariate \texttt{order}, however, pushes the final target syllable (in this case /sh/) towards its reference, all other things being equal. The positive effect at the end of the observed interval then pushes the initial target syllable /s/ towards the center, thus indicating a more assimilated pronunciation. Shortly before, we find a negative effect on the dimension \gls{epg} which might indicate that for a brief section, the articulatory pattern of /s/ is indeed close to its reference but this does not necessarily translate to the produced sound. Thus, similar to \cite{cederbaum2016} (red), we find that the final target syllable /sh/ is pulled towards the reference while the initial target syllable /s/ seems to be less affected. Given the similar shape on the dimension \gls{epg}, this supports their finding that assimilation is asymmetric. Since the estimated effects are similar across dimensions and similar to the univariate results, we refer to \cite{cederbaum2016} for interpretation of the other fixed effects.

\subsubsection*{Comparison to Univariate Analysis}

Compared to separate univariate \glspl{famm}, the multivariate model incorporates the dependency between the dimensions, thus reducing the number of \gls{fpc} bases in the analysis (for a similar amount of variance explained, ten univariate \glspl{fpc} would be needed). The shaded areas in Figure \ref{APPENDIXfig:phonetic_covar_estim} indicate where the standard errors of the univariate analysis presented in \cite{cederbaum2016} are smaller than the corresponding standard errors of the \gls{mfamm}. We find that overall, the multivariate analysis seems to give smaller standard errors. In order to compare the model fit between the univariate and the multivariate analysis, we can use the \gls{umse} defined in Section \ref{sec:simulation}. We then compare the fitted values with the observed values for the consonant assimilation data. The multivariate analysis yields \gls{umse} values of $0.970$ (\gls{aco}) and $0.966$ (\gls{epg}), whereas independent univariate \glspl{famm} give values of $0.978$ (\gls{aco}) and $0.961$ (\gls{epg}), respectively. Consequently, a univariate model analogously specified to the \gls{aco} model in \cite{cederbaum2016} gives a slightly better model fit on the \gls{epg} data as measured by the \gls{umse}. However, on the \gls{aco} dimension, the multivariate analysis is slightly preferred.

The added computational complexity of multivariate analyses is also negligible in our case: Fitting a univariate \gls{famm} as proposed by \cite{cederbaum2016} takes around 52 minutes on a 64-bit Linux platform and requires a considerable amount of RAM memory (32 GB is sufficient). The multivariate analysis maintains the requirements for internal memory, while the duration to fit the multivariate model (109 minutes on the aforementioned platform) only slightly increases compared to sequentially fitting two univariate models.

\subsection{Results of the Model Using a Weighted Scalar Product}
\label{app_subsec:phon_m_wei}

This section gives a short description of the considered \gls{mfamm} when a weighted scalar product is used. We only present the results of estimating the eigenfunction basis as the effect on the estimated covariate effects is negligible. Table \ref{APPENDIXtab:phonetic_varexp_weighted} shows the contribution of each random process to the total weighted variation and is structured similarly to Table \ref{APPENDIXtab:phonetic_varexp}. The number of eigenfunctions is chosen by weighted sum of total variation \eqref{eq:total_variation_decomp} which results in the same number of eigenfunctions for each random component as with the unweighted approach. However, the total share of weighted variation explained is slightly different (about $22\%$ and $65\%$  explained by the components $\bm{B}$ and $\bm{E}$, respectively). Consequently, the proportion of (weighted) variation assigned to the measurement error is smaller when applying the weighted scalar product. We also find that the univariate norms of the eigenfunctions are now more evenly distributed between the dimensions \gls{aco} and \gls{epg}, with most of the norms bigger on the \gls{aco} dimension. We thus see that weighting the scalar product shifts emphasis to the dimension \gls{aco} which has a lower measurement error. With this emphasis on \gls{aco}, the proportion of univariate variation explained on \gls{epg} falls now short of $95\%$ while \gls{aco} has a proportion of univariate variation explained of about $98\%$. For this model, too, we have specified the cut-off using the weighted sum of total variation \eqref{eq:total_variation_decomp} and the model explains $96\%$ of variation in the data.

\begin{table} \centering 
  \caption{Variance components included in the \gls{mfamm} using a weighted scalar product. First row: Estimates of eigenvalues, univariate error variances, total variation. Second (third) row: Univariate norms of estimated \glspl{fpc} on dimension \gls{aco} (\gls{epg}). Fourth (fifth) row: Proportion of univariate variation explained on dimension \gls{aco} (\gls{epg}) by eigenfunctions and error variance, total univariate variation explained. Sixth row: Proportion of multivariate variation explained by eigenfunctions and error variances, total multivariate variation explained.}
  \label{APPENDIXtab:phonetic_varexp_weighted} 
  \resizebox{\textwidth}{!}{
\begin{tabular}{l|cccccccc|cc|c} 
& $B_1$ & $B_2$ & $B_3$ & $E_1$ & $E_2$ & $E_3$ & $E_4$ & $E_5$ & $\sigma_{ACO}^2$ & $\sigma_{EPG}^2$ & Total \\\hline 
\hline 
Variation & $1.937$ & $1.625$ & $0.493$ & $6.501$ & $2.168$ & $1.823$ & $0.694$ & $0.507$ & $0.004$ & $0.014$ & $18.476$ \\ 
$\vert\vert\psi^{(ACO)}\vert\vert^2$ & $0.630$ & $0.743$ & $0.442$ & $0.574$ & $0.741$ & $0.404$ & $0.564$ & $0.455$ & $--$ & $--$ & $--$ \\ 
$\vert\vert\psi^{(EPG)}\vert\vert^2$ & $0.370$ & $0.257$ & $0.558$ & $0.426$ & $0.259$ & $0.596$ & $0.436$ & $0.545$ & $--$ & $--$ & $--$ \\ 
$\pi^{(ACO)}$ & $0.115$ & $0.114$ & $0.021$ & $0.352$ & $0.152$ & $0.070$ & $0.037$ & $0.022$ & $0.094$ & $--$ & $0.975$ \\ 
$\pi^{(EPG)}$ & $0.091$ & $0.053$ & $0.035$ & $0.352$ & $0.071$ & $0.138$ & $0.038$ & $0.035$ & $--$ & $0.127$ & $0.941$ \\ 
$\pi$ & $0.105$ & $0.088$ & $0.027$ & $0.352$ & $0.117$ & $0.099$ & $0.038$ & $0.027$ & $0.054$ & $0.054$ & $0.960$ \\ 
\hline 
\end{tabular} }
\end{table} 

\begin{figure}
\centering
\includegraphics[width=0.7\textwidth]{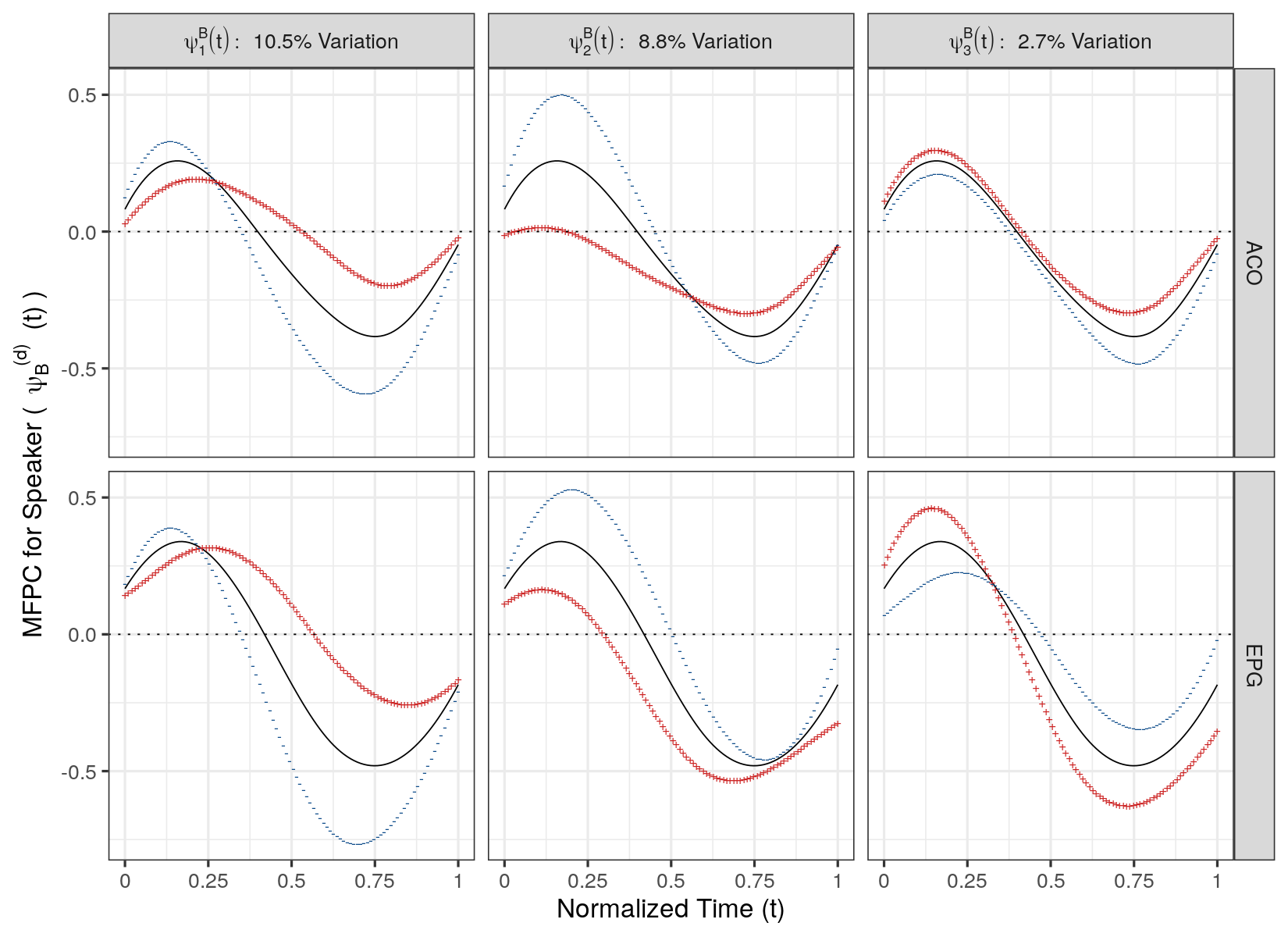}
\caption{\glspl{fpc} for the subject-specific functional random effect $\bm{B}_{i}(t)$ with the respective proportion of weighted variance explained. The black solid line represents the mean trajectory to which a suitable multiple ($2\sqrt{\nu_{B\cdot}}$) of the \gls{fpc} is added (red $+$) and subtracted (blue $-$).}
\label{APPENDIXfig:phonetics_fpc_B_wei}
\end{figure}

Figure \ref{APPENDIXfig:phonetics_fpc_B_wei} shows the estimated multivariate \glspl{fpc} of random component $\bm{B}$. We find that the leading \gls{fpc} again depicts an individual tendency to spend more time pronouncing the final or the initial target consonant. For $\bm{\psi}_{B2}(t)$, almost the same mode of variation is captured on the dimension \gls{aco} as with the unweighted scalar product. On the dimension \gls{epg}, however, we find the assimilation primarily in the final target syllable with the initial target syllable rather unaffected (more or less time spent in the pronunciation). The third \gls{fpc} for the subject-specific random effect is comparable to the scenario of the unweighted scalar product.

\begin{figure}
\centering
\includegraphics[width=0.8\textwidth]{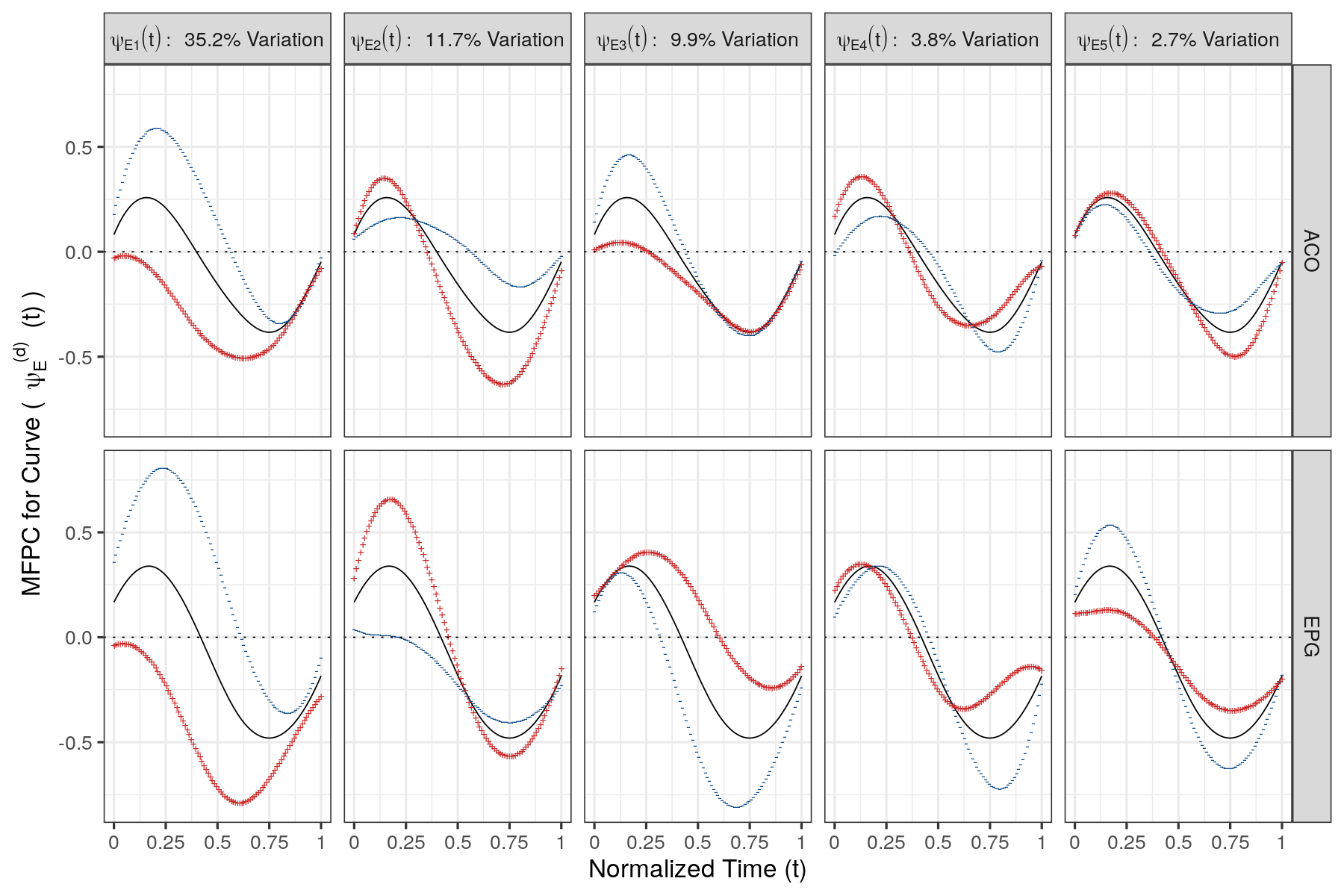}
\caption{\glspl{fpc} for the curve-specific functional random effect $\bm{E}_{ijh}(t)$ with the respective proportion of weighted variance explained. The black solid line represents the mean trajectory to which a suitable multiple ($2\sqrt{\nu_{E\cdot}}$) of the \gls{fpc} is added (red $+$) and subtracted (blue $-$).}
\label{APPENDIXfig:phonetics_fpc_E_wei}
\end{figure}

Figure \ref{APPENDIXfig:phonetics_fpc_E_wei} shows the estimated multivariate \glspl{fpc} of random component $\bm{E}$. The leading \gls{fpc} is somewhat unchanged by the weighting of the scalar product. While in the unweighted case the effect of $\bm{\psi}_{E2}$ was distributed equally across the two consonants, we find a stronger effect on the initial target consonant for dimension \gls{aco} and on the final target consonant for dimension \gls{epg} in the weighted case. This seems to be compensated by the third \gls{fpc}, where in the weighted scenario the emphasis lies on differences in the final target consonant on \gls{aco} and in the initial target consonant on \gls{epg}. The remaining \glspl{fpc} are comparable for both the weighted and unweighted scalar product.

We conclude that the model based on the weighted scalar product gives similar results. However, if some dimension were considerably more noisy than the rest, weighting the scalar product might be advisable (see results of the simulation).

\newpage
\FloatBarrier
\section{Detailed Analysis of the Simulation Study}
\label{app_sec:simulation}

We conduct an extensive simulation study in order to answer the following questions: How does the performance of the proposed \gls{mfamm} depend on different model specifications? How does the model perform in different data settings? And how does the \gls{mfamm} compare to a univariate modeling approach, where univariate regression models as proposed by \cite{cederbaum2016} are estimated on each dimension independently? Additionally, we evaluate the estimation of the covariance and the fixed effects. We simulate data based on the presented model fits \eqref{eq:snooker_model} of the snooker training data (main part) and \eqref{eq:phonetic_model} of the consonant assimilation data (Appendix \ref{APPsec:ca_data}). Six different settings of the data generating process are analysed, where for each data setting, we additionally compare up to seven different model specifications. One modeling scenario (data setting plus model specification) consists of independent model fits based on 500 generated data sets. Table \ref{tab:sim_data_model} provides an overview of the analysed data settings and model specifications, which are described in the following section. Note that we use data setting 1 and model specification A, respectively, to generate benchmark estimates useful for comparison to other settings and specifications.

\begin{table}
\centering
\caption{Summary of the different data settings and model specifications analysed in the simulation study. Each analysed combination of data setting and model specification comprises 500 model fits.}
\label{tab:sim_data_model}
%\resizebox{\textwidth}{!}{
\begin{tabular}{ccll|cl}
 \multicolumn{4}{c|}{Data Settings} &  \multicolumn{2}{c}{Model Specification} \\ \hline
 \multicolumn{2}{l}{1} & \multicolumn{2}{l|}{Consonant assimilation data} & A & True model \\
 &2 & &Strong heteroscedasticity & B & Truncation via total variation (TV) \\
 &3 & &Sparse data & C & Truncation via univariate variation (UV)\\
 &4 & &Uncentred scores & D & TV with alternate scalar product \\
 &5 & &Weighted scalar product & E & UV with alternate scalar product \\
 \multicolumn{2}{l}{6}& \multicolumn{2}{l|}{Snooker training data} & F & Scedastic misspecification \\
 && && U & Univariate models
\end{tabular}%}
\end{table}

\subsection{Simulation Description}
\label{app_subsec:sim_description}

\subsubsection*{Data Settings}

Table \ref{app_tab:sim_data_settings} provides detailed description of the data generating process for all data settings. For each data setting, 500 data sets have been simulated. For data setting $1$, we generate new observations based on the model fit \eqref{eq:phonetic_model} (Appendix \ref{APPsec:ca_data}) of the consonant assimilation data by randomly drawing the evaluation points, the random scores, and the measurement errors. Note that we center and decorrelate the random scores so that the empirical means and covariances coincide with the theoretical counterparts. As in the original data, each simulated data set contains the bivariate functional observations of $i = 1,.., 9$ individuals, each repeating the $j=1,...,16$ different word combinations five times.

Setting 2 (strong heteroscedasticity) is analogously to setting 1 but with larger difference between the measurement error variances of the two dimensions. Setting 2 mimics an application setting with (multimodal) multivariate data, where some dimensions are much noisier than others, in order to evaluate whether this imbalance interferes with the variance decomposition of the \gls{mfamm}.
Setting 3 (sparse data) focuses on the estimation quality for sparse functional data. Compared to setting 1, the number of evaluation points is reduced to three to ten measurements per dimension. In setting 4 (uncentred scores), the random scores are not decorrelated or centred (otherwise identical to setting 1). Especially for covariance components with few grouping levels (particularly $\bm{B}$), this can result in a considerable departure from the modeling assumptions, which is likely to occur in such settings. This setting explores the sensitivity of the \gls{mfamm} to violations of its assumptions. Setting 5 (weighted scalar product) is identical to setting 1 but all model components are simulated based on the estimated model using a weighted scalar product for the \gls{mfpca} (see Appendix \ref{app_subsec:phon_m_wei}). This data setting helps to understand the impact of weights in the scalar product. For setting 6 (snooker training data), we  generate new trajectory data according to the model fit \eqref{eq:snooker_model} of the snooker training data. This allows us to evaluate a higher dimensional \gls{mfamm} as well as the estimation quality of nested random effects.

\begin{table}
\caption{Description of data settings.}
\centering
\resizebox{0.95\textwidth}{!}{
\begin{tabular}{c|p{15cm}}
Data Setting & Description \\ \hline \hline
1 & Data based on consonant assimilation data (bivariate functions on $dim1, dim2$, each on $[0,1]$). $9$ individuals, each repeating $16$ crossed grouping levels $5$ times. Number of observed time points is an independent random draw for each dimension from a uniform discrete distribution of natural numbers in $[20,50]$. Observed time point is an independent random draw from uniform distribution on $[0,1]$. Mean consists of functional intercept and $7$ covariate effect functions as given in Appendix \ref{app_subsec:phon_main}. Covariates are given for each observation based on the crossed grouping level (word combination). Functional random effects $\bm{B},\bm{E}$ are based on estimated eigenfunctions of model in Appendix \ref{app_subsec:phon_main} ($3$ and $5$ \glspl{fpc}). Corresponding random scores are independent draws from $N(0,\hat{\nu}_{\cdot\cdot})$, then demeaned and decorrelated. Measurement errors are independent draws from $N(0, \hat{\sigma}^2_{\cdot})$ for each observed time point on the respective dimension. \\\hline
2 & Measurement errors are independent draws from $N(0, \hat{\sigma}^2_{1})$ for each observed time point on $dim1$ and independent draws from $N(0, 16\cdot\hat{\sigma}^2_{1})$ on $dim2$. Rest as in 1.\\\hline
3 & Number of observed time points is a random draw for each dimension from a uniform discrete distribution of natural numbers in $[3,10]$. Rest as in 1.\\\hline
4 & Random scores are independent draws from $N(0,\hat{\nu}_{\cdot\cdot})$ (no centering or decorrelation). Rest as in 1. \\\hline
5 & All estimated components (eigenfunctions, eigenvalues, covariate effect functions, measurement error variances) stem from the model based on a weighted scalar product for the \gls{mfpca} as presented in Appendix \ref{app_subsec:phon_m_wei}. Rest as in 1. \\\hline
6 & Data based on snooker training data (six-dimensional functions on $dim1,..., dim6$, each on $[0,1]$). $25$ individuals, each repeating $2$ nested grouping levels $6$ times. Number of observed points is an independent random draw from a uniform discrete distribution of natural numbers in $[10,50]$ per multivariate curve with observed time points identical over the dimensions. Observed time point is an independent random draw from uniform distribution on $[0,1]$.  Mean consists of functional intercept and $4$ covariate effect functions as given in Appendix \ref{app_sec:snooker_analysis}. Covariates \texttt{skill}, \texttt{group} are independent random draws from Bernoulli distribution with probability $0.5$ per subject and \texttt{session} is given for each observation based on the nested grouping level. Functional random effects $\bm{B},\bm{C}, \bm{E}$ are based on estimated eigenfunctions and eigenvalues of model in Appendix \ref{app_sec:snooker_analysis} ($6,5$ and $5$ \glspl{fpc}). Measurement errors are independent draws from $N(0, \hat{\sigma}^2)$ for each observed time point on each dimension.
\end{tabular}}
\label{app_tab:sim_data_settings}
\end{table}

\subsubsection*{Model Specifications}

Table \ref{app_tab:sim_model_scen} provides a detailed description of the different modeling scenarios used in the simulation study. We denote the most accurate approach of modeling as specification A (true model). This standard scenario mirrors the data generation so that there is no model misspecification. Most notably, we fix the number of  \glspl{fpc} to the number used for generating the data in order to avoid truncation effects. Though somewhat unrealistic, specification A serves to separate the impact of modeling decisions or situations with more (realistic) uncertainty for the user from the overall performance of the \gls{mfamm}.

For model specification B (\gls{tv}), the truncation orders of the \glspl{fpc} are chosen so that $95\%$ of the total variation \eqref{eq:total_variation_decomp} are explained. In scenario C (\gls{uv}), we choose the number of \glspl{fpc} so that on every dimension $95\%$ of univariate variance \eqref{eq:univar_variation_decomp} are explained. Specifications D (\gls{tv} with alternate scalar product) and E (\gls{uv} with alternate scalar product) use the cut-off criterions analogous to B and C but we alternate the scalar product on which the \gls{mfpca} is based: For data generated from a model based on an unweighted \gls{mfpca}, the scalar product used in these scenarios is weighted by $\frac{1}{\hat{\sigma}^2_d}$, and vice versa. Model specification F (scedastic misspecification) evaluates misspecifying the \gls{mfamm} with a homoscedasticity assumption. We also contrast the \gls{mfamm} with the univariate approach of fitting independent univariate models. In modeling scenarios denoted with U, we fit an independent \gls{famm} on each dimension. We use the \glspl{famm} proposed by \cite{cederbaum2016} so that we can apply the same model specifications as for the \gls{mfamm} (e.g.\ basis functions, penalties, etc.). The number of \glspl{fpc} in the model is then chosen so that $95\%$ of univariate variation is explained.

\begin{table}
\caption{Description of model specifications.}
\resizebox{\textwidth}{!}{
\begin{tabular}{c|p{15cm}}
Model Specification & Description \\ \hline \hline
A & Univariate mean, multivariate mean, univariate auto-covariance estimation based on cubic P-splines with 8, 8, 5 (marginal) basis functions  and choice of penalty as in the model used for generating the data setting (third order for 1-5, first order for 6).  No univariate truncation for the \gls{mfpca} except negative eigenvalues. \gls{mfpca} based on scalar product as in the model used for generating the data setting (weights of one except for 5). Number of multivariate \glspl{fpc} is fixed according to the data setting (1-5: ($\bm{B}$:3, $\bm{E}$:5), 6: ($\bm{B}$:6, $\bm{C}$:5, $\bm{E}$:5)). Weighted regression approach with weights obtained from univariate variance decompositions for heteroscedastic data settings (1-5).\\\hline
B & Number of \glspl{fpc} chosen so that $95\%$ of the (weighted) sum of total variation in the data is explained. Rest as in A. \\\hline
C & Number of \glspl{fpc} chosen so that on each dimension at least $95\%$ of the univariate variation in the data is explained. Rest as in A. \\\hline
D & Weighted scalar product for \gls{mfpca} with weights based on dimension specific measurement error variances if data setting is based on model using a scalar product with weights of one (all except 5) and vice versa (5). Number of \glspl{fpc} chosen so that $95\%$ of the (weighted) sum of total variation in the data is explained. Rest as in A. \\\hline
E & Weighted scalar product for \gls{mfpca} with weights based on dimension specific measurement error variances if data setting is based on model using a scalar product with weights of one (all except 5) and vice versa (5). Number of \glspl{fpc} chosen so that on each dimension at least $95\%$ of the univariate variation in the data is explained. Rest as in A. \\\hline
F & Homoscedasticity is assumed, no regression weights. Rest as in A. \\ \hline
U & Univariate modeling approach. Independent sparse \gls{flmm} models for each dimension. Number of univariate \glspl{fpc} chosen so that $95\%$ of univariate dimension is explained. Rest as in A.
\end{tabular}}
\label{app_tab:sim_model_scen}
\end{table}

\subsubsection*{Modeling Scenarios}

For each of the combinations of data settings and modeling scenarios indicated in Table \ref{app_tab:sim_combinations}, analyses were performed giving $500$ fitted models per combination. The results presented in the main part correspond to modeling scenarios 1A (Benchmark), 1B (Cut-Off Multi), 1C (Cut-Off Uni), 3A (Sparse Data), and 4A (Uncentred Scores).

\begin{table}
\caption{Overview of the analysed modeling scenarios.}
\centering
\begin{tabular}{cc||cccccc}
 & & \multicolumn{6}{c}{Data Setting}\\
\multicolumn{2}{l||}{Model Specification} & 1 & 2 & 3 & 4 & 5 & 6 \\  \hline\hline
 & A & X & X & X & X & X & X \\
 & B & X & X & & & X & \\
 & C & X & X & & & X & \\
 & D & X & X &  & & X & \\
 & E & X & X &  & & X & \\
 & F & X & X & & & & \\
 & U & X & X & & & X &
\end{tabular}
\label{app_tab:sim_combinations}
\end{table}

\subsubsection*{Evaluation Criteria}

The accuracy of the estimated model components is measured using the \gls{mse}. We analyse the accuracy of the covariance estimation, i.e.\ eigenfunctions, eigenvalues, and measurement error variances as well as the mean estimation. For evaluating the overall accuracy of the estimate $\hat {\bm{\zeta}}$ of the multivariate functional component $\bm{\zeta} = (\bm{\zeta}_1,..., \bm{\zeta}_S)^{\top}$, we define the \gls{mmse} as
\begin{align*}
    \mathrm{mrrMSE}(\boldsymbol\zeta, \hat{\boldsymbol{\zeta}}) = \sqrt{\frac{\frac{1}{S}\sum_{s=1}^{S}\vert\vert\vert\bm{\zeta}_s- \hat{\bm{\zeta}}_s\vert\vert\vert^2}{\frac{1}{S}\sum_{s=1}^{S}\vert\vert\vert\bm{\zeta}_s\vert\vert\vert^2}}
\end{align*}
with the multivariate norm based on the unweighted scalar product. Note that $S$ depends on the model component to be evaluated, e.g.\ for the fitted values $\bm{y}_{ijh}(t)$, $S$ equals $N$ but for the subject-specific random effect $\bm{B}_i(t)$, $S$ equals the number of individuals. The \gls{mmse} corresponds to the evaluation criterion used in section 5 of the main part. We also define a \gls{umse}
\begin{align*}
   \mathrm{urrMSE}\big(\boldsymbol{\zeta}^{(d)}, \hat{\boldsymbol{\zeta}}^{(d)}\big) = \sqrt{\frac{\frac{1}{S}\sum_{s=1}^{S}\vert\vert\zeta_s^{(d)}- \hat{\zeta}_s^{(d)}\vert\vert^2}{\frac{1}{S}\sum_{s=1}^{S}\vert\vert\zeta_s^{(d)}\vert\vert^2}}
\end{align*}
for $\bm{\zeta}^{(d)} = (\zeta_1^{(d)}, ..., \zeta_S^{(d)})^{\top}$, which allows to evaluate the dimension specific estimation accuracy as well as  a straightforward comparison to the univariate modeling approach. For scalar estimates such as eigenvalues and error variances, we define the \gls{mse} as
\begin{equation*}
    \mathrm{rrMSE}(\zeta, \hat{\zeta}) = \sqrt{\frac{(\zeta - \hat{\zeta})^2}{\zeta^2}}
\end{equation*}
with estimate $\hat{\zeta}$ of the scalar value $\zeta$. The \gls{mse} takes on (unbounded) positive values with smaller values indicating a better fit. As the \gls{mse} is a relative measure, small differences between estimate and true component can result in large \gls{mse} values for true component norms close to zero. Note that eigenfunctions are defined only up to a sign change. We thus flip the estimated eigenfunction (multiply it by $(-1)$) if this results in a smaller norm for the difference between the true function and its estimate. 

Additionally, we evaluate the coverage of the point-wise \glspl{cb} of the estimated fixed effects.

\FloatBarrier
\subsection{Results of the Simulation Study}
\label{app_subsec:sim_results}

\subsubsection*{Impact of Model Specifications}

\begin{figure}
\centering
\includegraphics[width=0.9\textwidth]{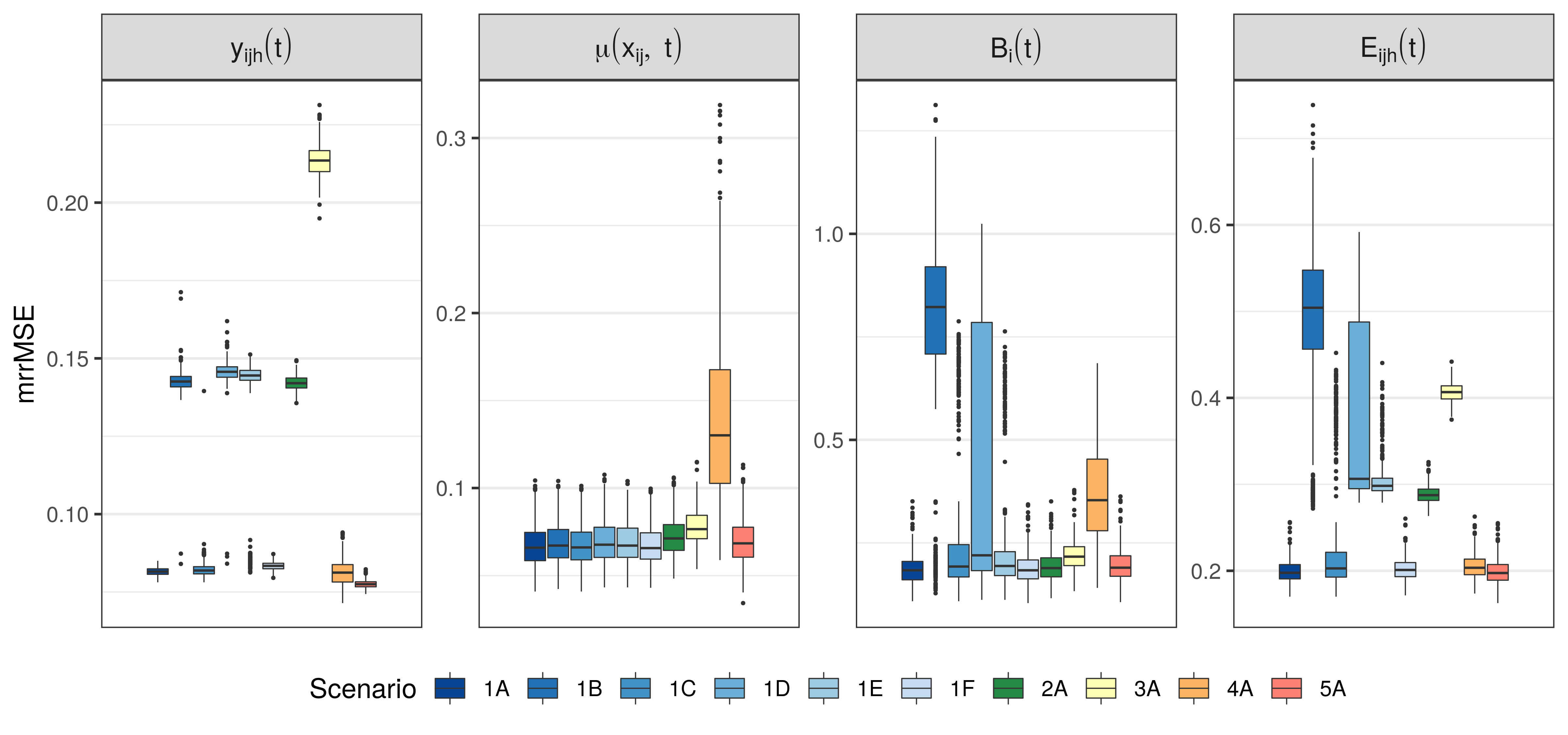}
\caption{\gls{mmse} values of the fitted curves $\bm{y}_{ijh}(t)$, the mean $\bm{\mu}(\bm{x}_j, t)$, and the random effects $\bm{B}_{i}(t)$ and $\bm{E}_{ijh}(t)$ for all modeling scenarios in data setting 1 (blue shades) and the standard modeling scenario A in data settings 2 to 5.}
\label{app_fig:sim_eval_multi}
\end{figure}

\begin{figure}
\centering
\includegraphics[width=\textwidth]{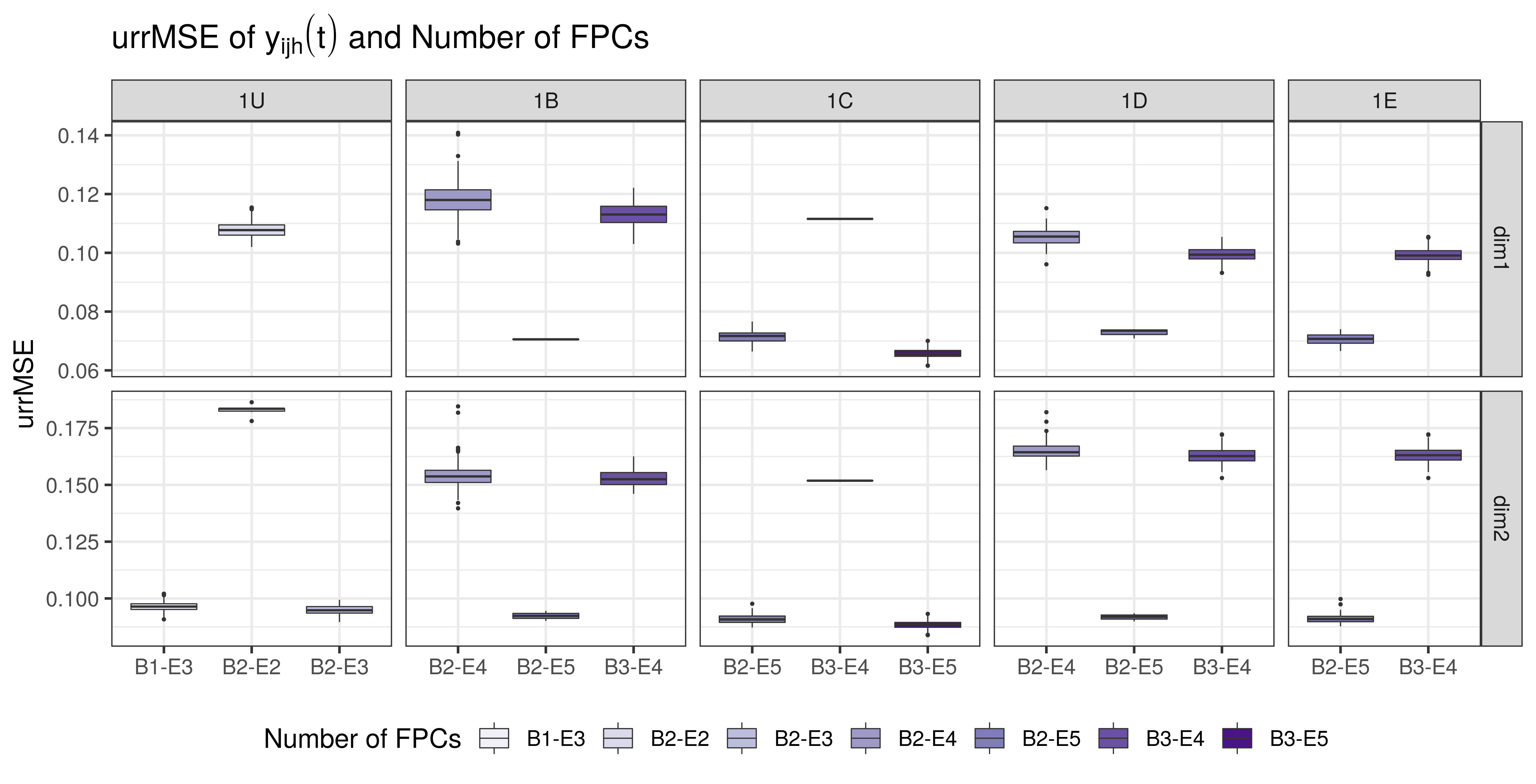}
\caption{\gls{umse} values of the fitted values $\bm{y}_{ijh}(t)$ for different scenarios. The \gls{umse} values are grouped by the number of \glspl{fpc} included in the models. B$x$-E$y$ means the model contains $x$ \glspl{fpc} for $\bm{B}_{i}(t)$ and $y$ for $\bm{E}_{ijh}(t)$.}
\label{fig:app_sim_nfpc_y}
\end{figure}

The results for setting 1 (consonant assimilation data) demonstrate the importance of the number of \glspl{fpc} in the accuracy of the estimation (see Figure \ref{app_fig:sim_eval_multi} and Table \ref{app_tab:sim_number_fpc}). With specifications A (true model) and F (scedastic misspecification), the sets of \glspl{fpc} are fixed giving overall low values for the \gls{mmse} (the misspecification in F yields only slightly worse results). Similarly, choosing the truncation order via the proportion of univariate variance explained in scenario C gives models with roughly the same number of \glspl{fpc} as is used for the data generation. The cut-off criterion based on the total amount of variance in scenario B results in more parsimonious models and thus considerably higher \gls{mmse} values. The number of selected \glspl{fpc} also explains the wider boxplots of the scenarios B-E compared to A: rather than a larger overall variance in estimation, we find separate clusters based on the included \gls{fpc} sets (see also Figure \ref{fig:app_sim_nfpc_y}). For the modeling scenarios based on a weighted scalar product (1D and 1E), the number of chosen \glspl{fpc} is  quite similar regardless of the cut-off criterion but the overall \gls{mmse} values of the fitted curves ($\bm{y}_{ijh}(t)$) are higher than for the unweighted approach. The estimate of the mean $\bm{\mu}(\bm{x}_j, t)$, however, is comparatively stable over the different model specifications.

\begin{table}
\resizebox{\textwidth}{!}{
\begin{tabular}{c||ccccccc||cccccc}
& \multirow{2}{*}{1A} & \multirow{2}{*}{1B} & \multirow{2}{*}{1C} & \multirow{2}{*}{1D} & \multirow{2}{*}{1E} & \multicolumn{2}{c||}{1U} & \multirow{2}{*}{2B} & \multirow{2}{*}{2C} & \multirow{2}{*}{2D} & \multirow{2}{*}{2E} & \multicolumn{2}{c}{2U} \\
& & & & & & dim1 & dim2 & & & & & dim1 & dim2\\ \hline\hline
\multirow{2}{*}{$\bm{B}_{i}(t)$} & \multirow{2}{*}{3.00} & \multirow{2}{*}{2.15} & \multirow{2}{*}{2.80} & \multirow{2}{*}{2.62} & \multirow{2}{*}{2.88}  & \multicolumn{2}{c||}{3.09} & \multirow{2}{*}{2.02} & \multirow{2}{*}{2.87} & \multirow{2}{*}{2.00} & \multirow{2}{*}{3.00} & \multicolumn{2}{c}{3.06} \\ 
& & & && & 2.00 & 1.09 & & & & & 2.00 & 1.06 \\ \hline\hline
\multirow{2}{*}{$\bm{E}_{ijh}(t)$} & \multirow{2}{*}{5.00} & \multirow{2}{*}{4.00} & \multirow{2}{*}{5.00} & \multirow{2}{*}{4.01} & \multirow{2}{*}{4.12} & \multicolumn{2}{c||}{4.99} & \multirow{2}{*}{3.77} & \multirow{2}{*}{4.44} & \multirow{2}{*}{3.00} & \multirow{2}{*}{4.27} & \multicolumn{2}{c}{4.72} \\ 
& & & & & & 2.00 & 2.99 & & & & & 2.00 & 2.72
\end{tabular}}
\caption{Average number of eigenfunctions selected (scenario A fixed to underlying truth) in the 500 simulation iterations per modeling scenario (column) and random effect (row). For the univariate modeling approach (scenario U) we report the total amount (top) and the number per independent model (bottom).}
\label{app_tab:sim_number_fpc}
\end{table}

Figure \ref{fig:app_sim_nfpc_y} shows the \gls{umse} values of the fitted curves $\bm{y}_{ijh}(t)$ for different modeling scenarios (1U, 1B, 1C, 1D, 1E) depending on the number of \glspl{fpc} included in the model. For example in the univariate modeling scenario 1U, all 500 models choose two \glspl{fpc} for each random effect (B2-E2) on dimension $dim1$. On $dim2$, however, different combinations of \glspl{fpc} lead to considerably different \gls{umse} values. For the multivariate modeling scenarios, we also find that the reduction in \gls{mse} values depends on which additional \gls{fpc} is included: $\bm{\psi}_{E5}$ reduces the values more than $\bm{\psi}_{B3}$.

\begin{figure}
\includegraphics[width=\textwidth]{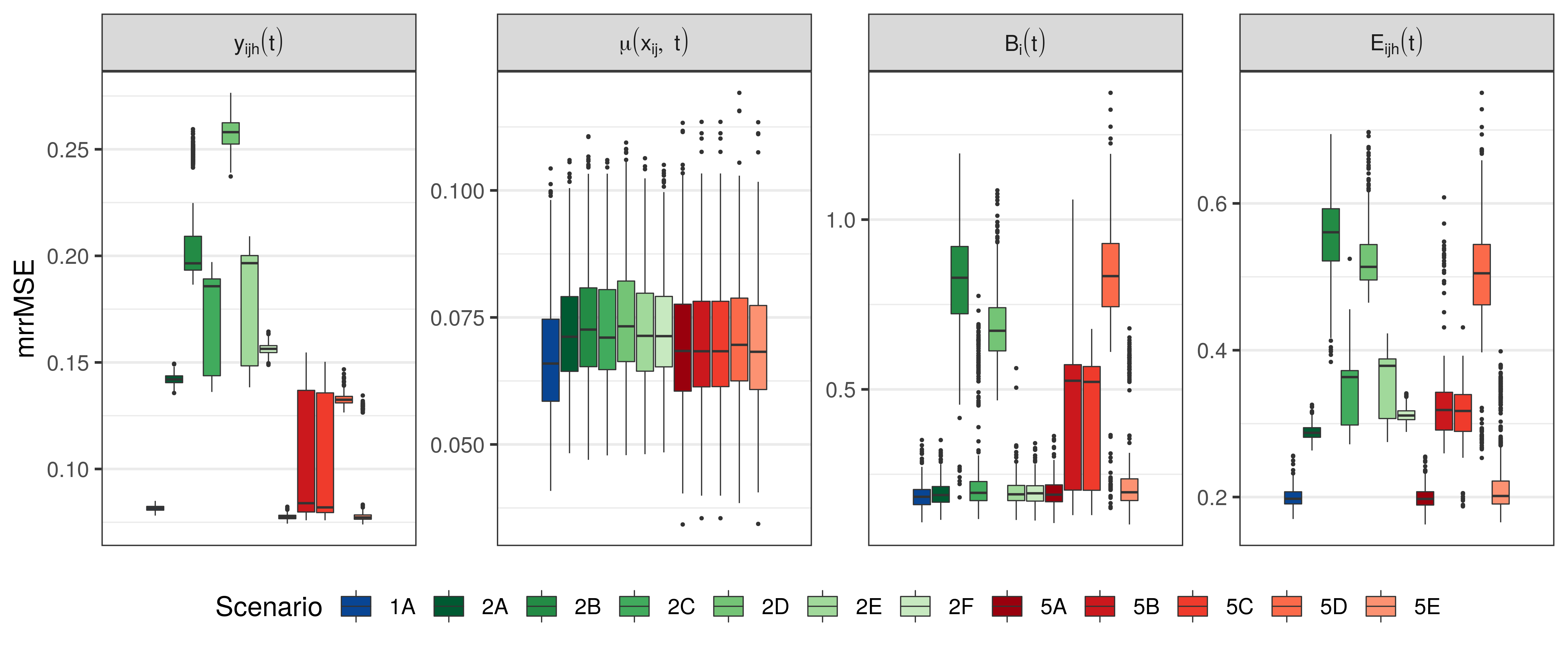}
\caption{\gls{mmse} values of the fitted curves $\bm{y}_{ijh}(t)$, the mean $\bm{\mu}(\bm{x}_i, t)$, and the random effects $\bm{B}_{i}(t)$ and $\bm{E}_{ijh}(t)$ for different scenarios.}
\label{fig:app_sim_multivariate1}
\end{figure}

Figure \ref{fig:app_sim_multivariate1} shows the \gls{mmse} values of the fitted curves $\bm{y}_{ijh}(t)$, the mean $\bm{\mu}(\bm{x}_i, t)$, and the random effects $\bm{B}_{i}(t)$ and $\bm{E}_{ijh}(t)$ for the different modeling scenarios of data settings 1, 2, and 5. Again, we find considerable differences for different numbers of \glspl{fpc} in the models when there is strong heteroscedasticity between the different dimensions (setting 2). The overall model fit seems to be better for applying an unweighted scalar product for the \gls{mfpca} (2B, 2C compared to 2D, 2E). Later, we will see that the model fit on single dimensions can be improved by weighting the scalar product. Note that for setting 2F, misspecifying the model assumption now has a larger negative impact on the fitted curves than in setting 1F. In data setting 5, the data are generated based on the model using a weighted scalar product. Then, the number of \glspl{fpc} in modeling scenarios 5B and 5C is very similar. Interestingly, basing the \gls{mfpca} on a scalar product using weights of one can lead to a similar estimation accuracy than the standard modeling scenario (5A compared to 5E).

Our simulation study thus suggests that basing the truncation orders on the proportion of explained variation on each dimension gives parsimonious and well fitting models. If interest lies mainly on the estimation of fixed effects, the alternative cut-off criterion based on the total variation in the data allows even more parsimonious models. Furthermore, an unweighted scalar product is a reasonable starting point for the \gls{mfamm}.

\subsubsection*{Model Performance on Different Data Settings}

To compare different data settings, we focus on model specification A (true model) in Figure \ref{app_fig:sim_eval_multi}. Note that the \gls{mmse} values of the other data settings cannot be directly compared as the denominator of the \gls{mmse} (slightly) changes (except for the mean of 2A, 3A, and 4A). We find that strong heteroscedasticity (2A) mainly negatively affects the fitted curves and the smooth residual. Unsurprisingly, the results for scenario 3A suggest that the estimation accuracy is lower for sparse functional data. However, the mean estimation is comparable to more densely observed data. On the other hand, the mean estimation is susceptible to violations of the assumption of uncorrelated and centred realizations of the random effects (4A). The model then has difficulties to separate intercept and random effects (see also Figure \ref{fig:app_sim_eff_umse}), which does not necessarily translate to a worse overall fit to the data. The results of scenario 5A (weighted scalar product) suggest that the accuracy of the \gls{mfamm} does not depend on the definition of the scalar product used for the \gls{mfpca}. 

\begin{figure}
\includegraphics[width=\textwidth]{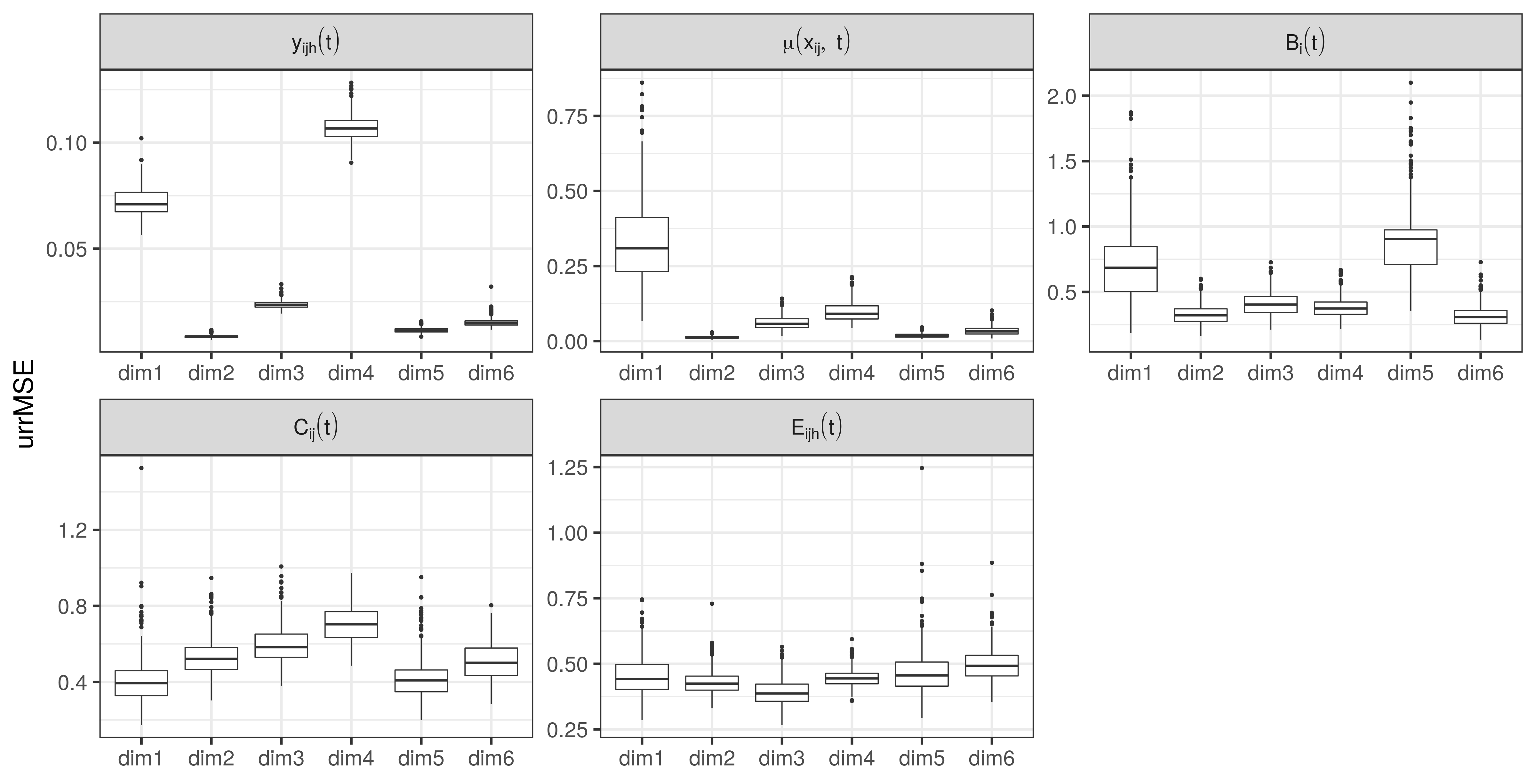}
\caption{\gls{umse} values of the fitted curves $\bm{y}_{ijh}(t)$, the mean $\bm{\mu}(\bm{x}_i, t)$, and the random effects $\bm{B}_{i}(t), \bm{C}_{ij}(t)$ and $\bm{E}_{ijh}(t)$ for modeling scenario 6A. One outlying observation with high \gls{umse} value was removed from the boxplot for $\bm{E}_{ijh}(t)$ and $dim1$.}
\label{fig:app_sim_univariate_sno}
\end{figure}

Figure \ref{fig:app_sim_univariate_sno} shows the \gls{umse} values for the aforementioned model components of modeling scenario 6A (snooker training data) where we compare the estimation accuracy across the dimensions. The fit of the functional curves $\bm{y}_{ijh}(t)$ suggests that there might be pronounced differences between the estimation accuracy of the dimensions, the dimensions $dim1$ (corresponding to $elbow.x$) and $dim4$ (corresponding to $hand.y$) giving high \gls{mse} values.

We conclude that the proposed \gls{mfamm} performs well even in more challenging data settings such as sparse functional data or data with few grouping levels for the random effects. Especially the estimation of the fixed effects seems to be stable over the different analysed settings.

\subsubsection*{Comparison to Univariate Approach}

\begin{figure}
\includegraphics[width=\textwidth]{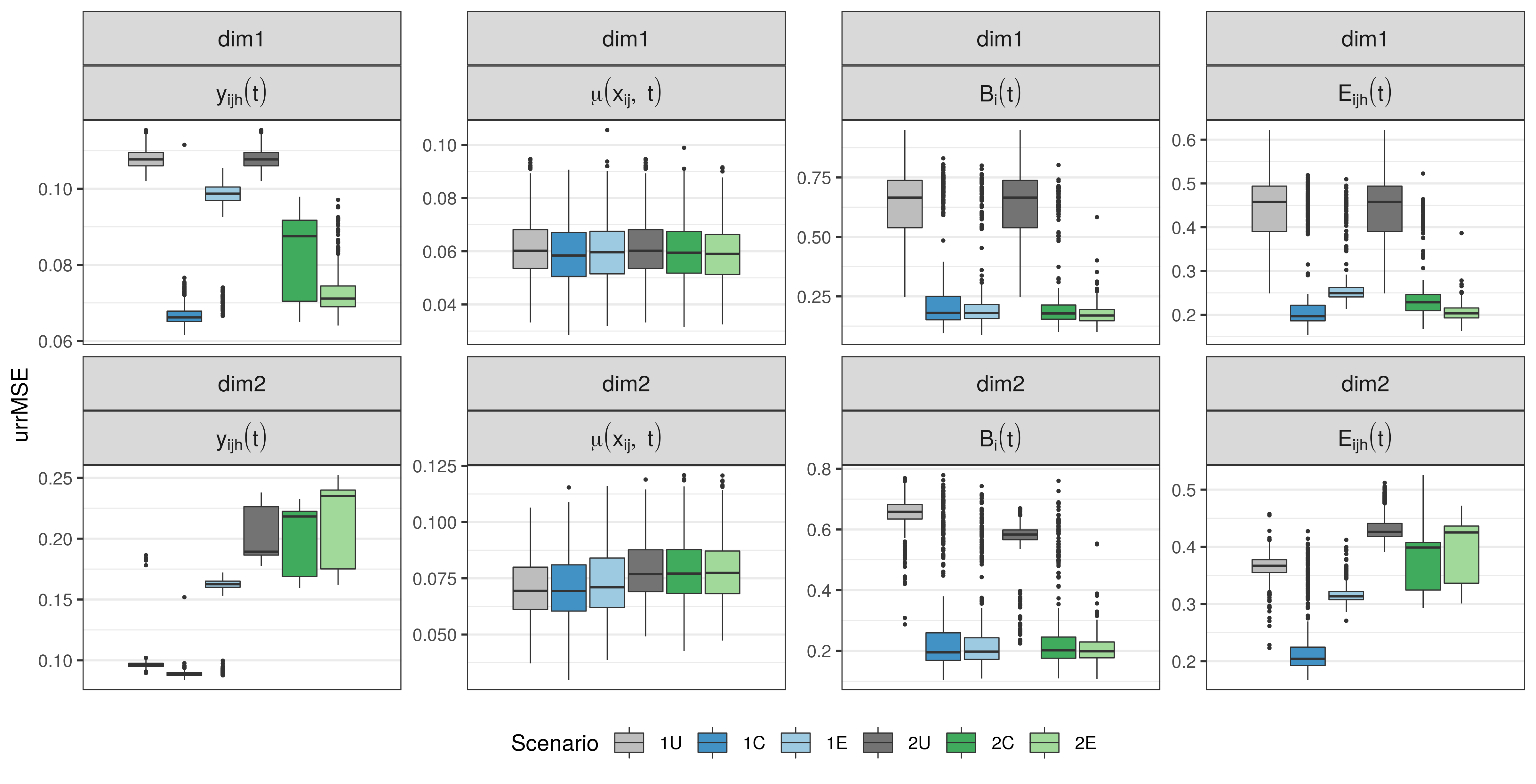}
\caption{\gls{umse} values of the fitted curves $\bm{y}_{ijh}(t)$, the mean $\bm{\mu}(\bm{x}_i, t)$, and the random effects $\bm{B}_{i}(t)$ and $\bm{E}_{ijh}(t)$ for the univariate modeling scenario U (grey) and the multivariate modeling scenarios C and E in data settings 1 (blue) and 2 (green).}
\label{fig:sim_eval_uni}
\end{figure}

We compare the univariate modeling approach U to multivariate modeling scenarios C (\gls{uv}) and E (\gls{uv} with alternate scalar product), i.e.\ we make sure that the proportion of explained univariate variation is also at least $95\%$ for each model. Table \ref{app_tab:sim_number_fpc} shows that in data settings 1 (consonant assimilation data) and 2 (strong heteroscedasticity) the number of included \glspl{fpc} tends to be higher for scenario U. Yet Figure \ref{fig:sim_eval_uni} indicates that the \gls{mfamm} yields consistently lower \gls{umse} values on $dim1$ (smaller measurement error variance). For $dim2$, the random effects seem to be estimated more accurately and the fixed effects similarly well, but especially with the weighted scalar product (scenarios E), the overall fit for $y$ can give higher \gls{umse} values. Overall, the results suggest that the unweighted scalar product is to be preferred in data situations with similar measurement error variances of the dimensions such as setting 1, where it gives reliably good results across all model components. However, downweighting the  dimension $dim2$ with larger error seems to be a reasonable modeling decision in setting 2 if interest lies primarily on $dim1$ (lowest \gls{umse} values for 2E). Again, we point out that the estimation of the fixed effects is relatively stable across approaches and slightly better than for the univariate modeling approach.

Table \ref{app_tab:st_error_comparison} compares the standard errors of the fixed effects estimation for the \gls{mfamm} in scenario 1C with the univariate modeling approach of scenario 1U. We look at the ratio of standard errors $\frac{se_m}{se_u}$, where $se_m$ denotes the standard error of the \gls{mfamm} and $se_u$ the standard error of the corresponding univariate \gls{famm}. For each dimension and partial predictor, we calculate the proportion of ratios smaller and larger than one over an equidistant grid of 100 time points and all simulation runs. Especially on $dim1$, we find that the multivariate approach gives smaller standard errors. On $dim2$, the proportions are more similar with slightly more proportions larger than one. Overall, we find more ratios smaller than one in our simulation study thus indicating that the multivariate approach can yield smaller standard errors. Table \ref{app_tab:coverage_phon_sno} further shows that the coverage of these two scenarios is very similar (see subsection on the estimation of fixed effects).

\begin{table}\centering
\caption{Proportion of standard error ratios of the \gls{mfamm} ($se_{m}$) of scenario 1C and the univariate modeling approach ($se_{u}$) of scenario 1U over all simulation runs and all evaluated time points.}
\label{app_tab:st_error_comparison}
\resizebox{\textwidth}{!}{
\begin{tabular}{cc|ccccccccc}
\hline
\hline
$d$ & Ratio & $\beta_0^{(d)}$ & $f_0^{(d)}(t)$ & $f_1^{(d)}(t)$ & $f_2^{(d)}(t)$ & $f_3^{(d)}(t)$ & $f_4^{(d)}(t)$ & $f_5^{(d)}(t)$ & $f_6^{(d)}(t)$ & $f_7^{(d)}(t)$ \\ 
\hline
\multirow{2}{*}{$dim1$} & $\frac{se_{m}}{se_{u}}<1$ & 0.67 & 0.93 & 0.65 & 0.74 & 0.76 & 0.69 & 0.85 & 0.56 & 0.42 \\ 
& $\frac{se_{m}}{se_{u}}>1$ & 0.33 & 0.07 & 0.35 & 0.26 & 0.24 & 0.31 & 0.15 & 0.44 & 0.58 \\ \hline
\multirow{2}{*}{$dim2$} & $\frac{se_{m}}{se_{u}}<1$ & 0.28 & 0.48 & 0.45 & 0.36 & 0.48 & 0.41 & 0.31 & 0.54 & 0.62 \\
& $\frac{se_{m}}{se_{u}}>1$ & 0.72 & 0.52 & 0.55 & 0.64 & 0.52 & 0.59 & 0.69 & 0.46 & 0.38 \\ \hline
\end{tabular}}
\end{table}

We conclude that the multivariate modeling approach can improve the mean estimation but is especially beneficial for the prediction of the random effects. In some cases, including weights in the multivariate scalar product might further improve the modeling.

\subsubsection*{Covariance Estimation}

\begin{figure}
\includegraphics[width=\textwidth]{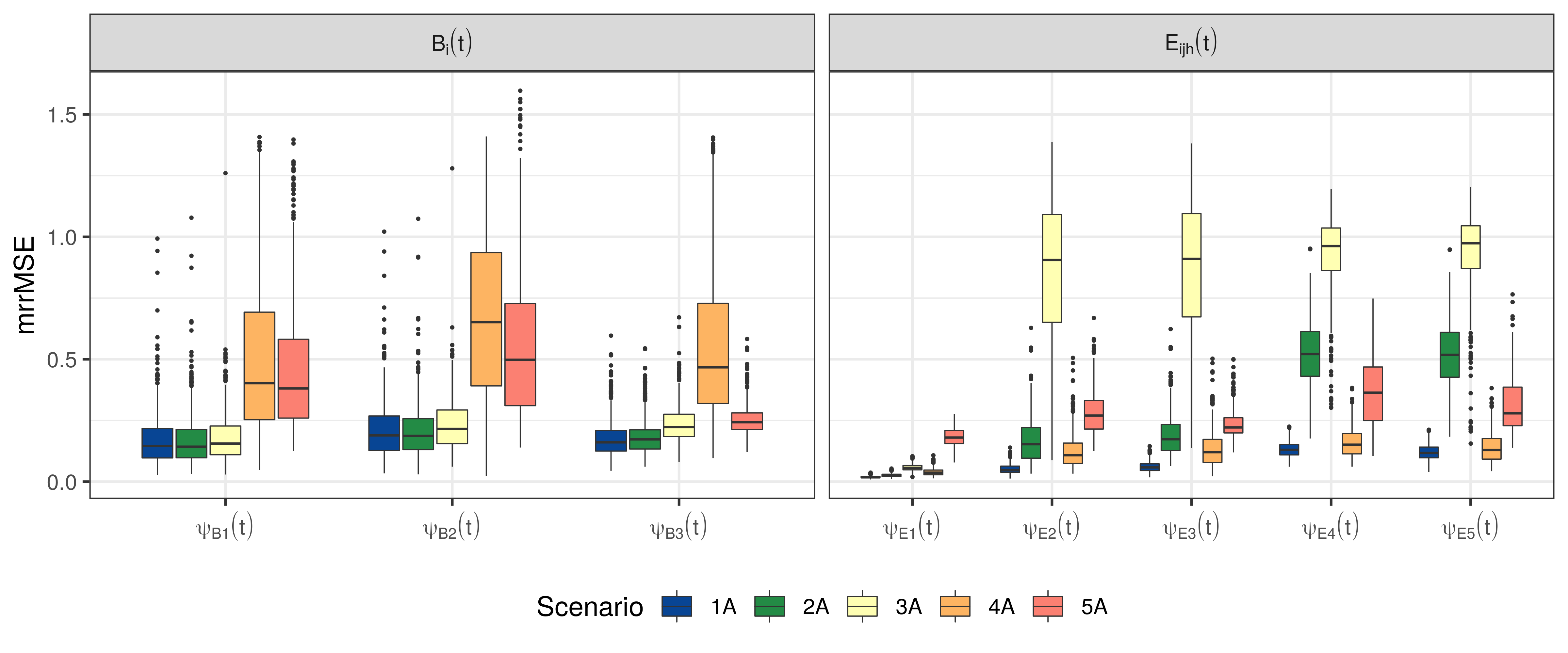}
\caption{\gls{mmse} values of the eigenfunction estimation for the random effects $\bm{B}_{i}(t)$ and $\bm{E}_{ijh}(t)$ for different scenarios.}
\label{fig:app_sim_fpc}
\end{figure}

\begin{figure}
\includegraphics[width=\textwidth]{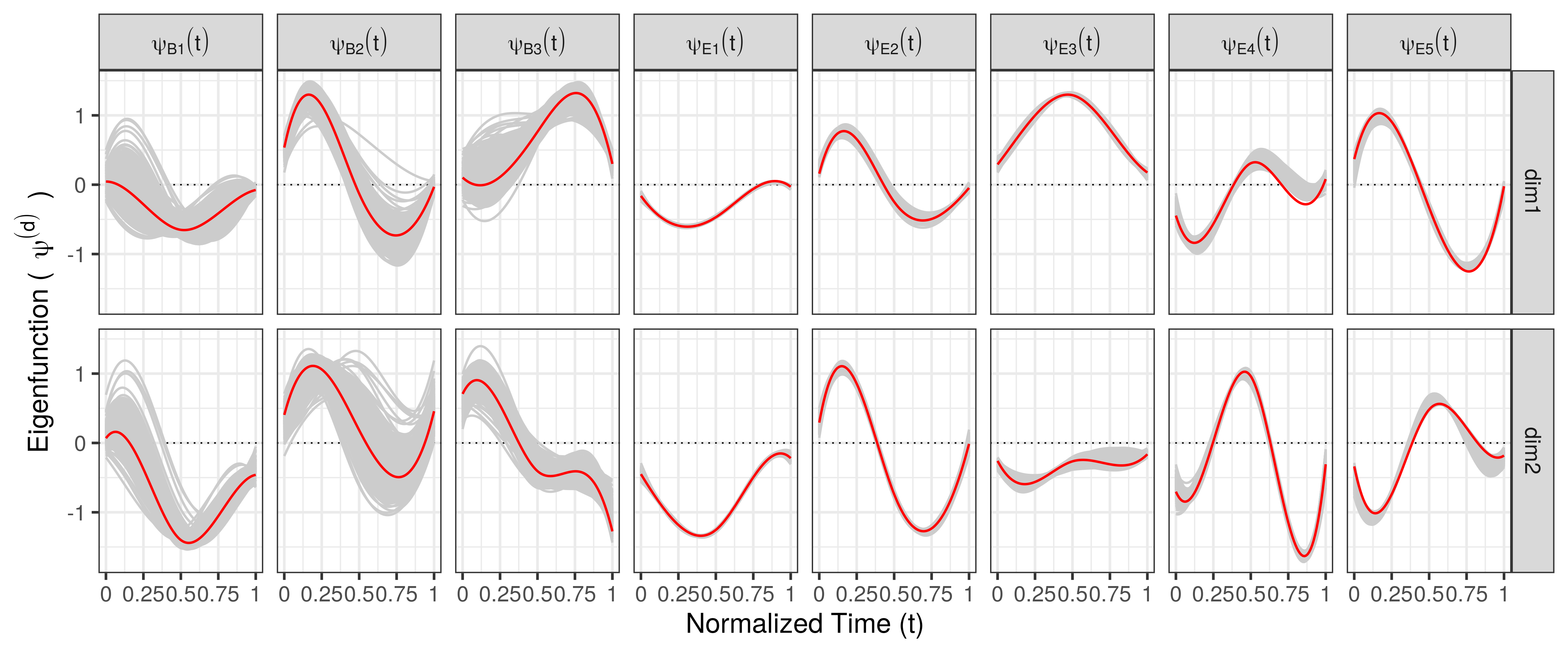}
\caption{Estimated eigenfunctions in scenario 1A for all 500 simulation iterations (grey curves). The red curves show the data generating eigenfunctions.}
\label{fig:app_sim_eigfcts}
\end{figure}

Figure \ref{fig:app_sim_fpc} shows the \gls{mmse} values for the estimated eigenfunctions of the two random effects in the standard modeling scenario A of data settings 1 to 5. In general, we find that the leading eigenfunctions tend to have lower \gls{mmse} values. Especially the leading eigenfunctions of the smooth residual show a high accuracy. There seems to be more variance in the estimation accuracy for the subject-specific random effect. This can be confirmed with Figure \ref{fig:app_sim_eigfcts}, which contains the estimated eigenfunctions of all 500 simulation iterations for modeling scenario 1A (grey curves) and compares them to the data generating function (red curve). Overall we find that the modes can be reconstructed sufficiently well, albeit considerably better for the random smooth residual. 
 
Figure \ref{fig:app_sim_fpc} also allows to compare the eigenfunction estimation across the different data settings. Keep in mind that scenario 5A is based on a different model so the \gls{mmse} values are not directly comparable. We find that the eigenfunction estimation of the smooth residual suffers from strong heteroscedasticity (2A), whereas the subject-specific random effect seems unaffected. The same effect (even more pronounced) can be observed for sparse data (3A), where fewer information about the correlation within functions is available. On the other hand, $\bm{B}_{i}(t)$ suffers from few different individuals as this can lead to a considerable departure from the modeling assumption when the independent draws of scores are not centred and decorrelated (4A). With 720 different grouping levels of the smooth residual, we typically get an empirical covariance for the random scores that is closer to its assumption and thus smaller differences in \gls{mmse} values compared to 1A. The estimation of the eigenfunctions is somewhat less accurate in scenario 5A as the weights of the scalar product have to be estimated as well. This additional uncertainty leads to a larger variance for the \gls{mmse} values and makes it harder to correctly identify the data generating modes of variation. Note that this does not affect the overall estimation of the random effects as discussed above.

\begin{figure}
\includegraphics[width=\textwidth]{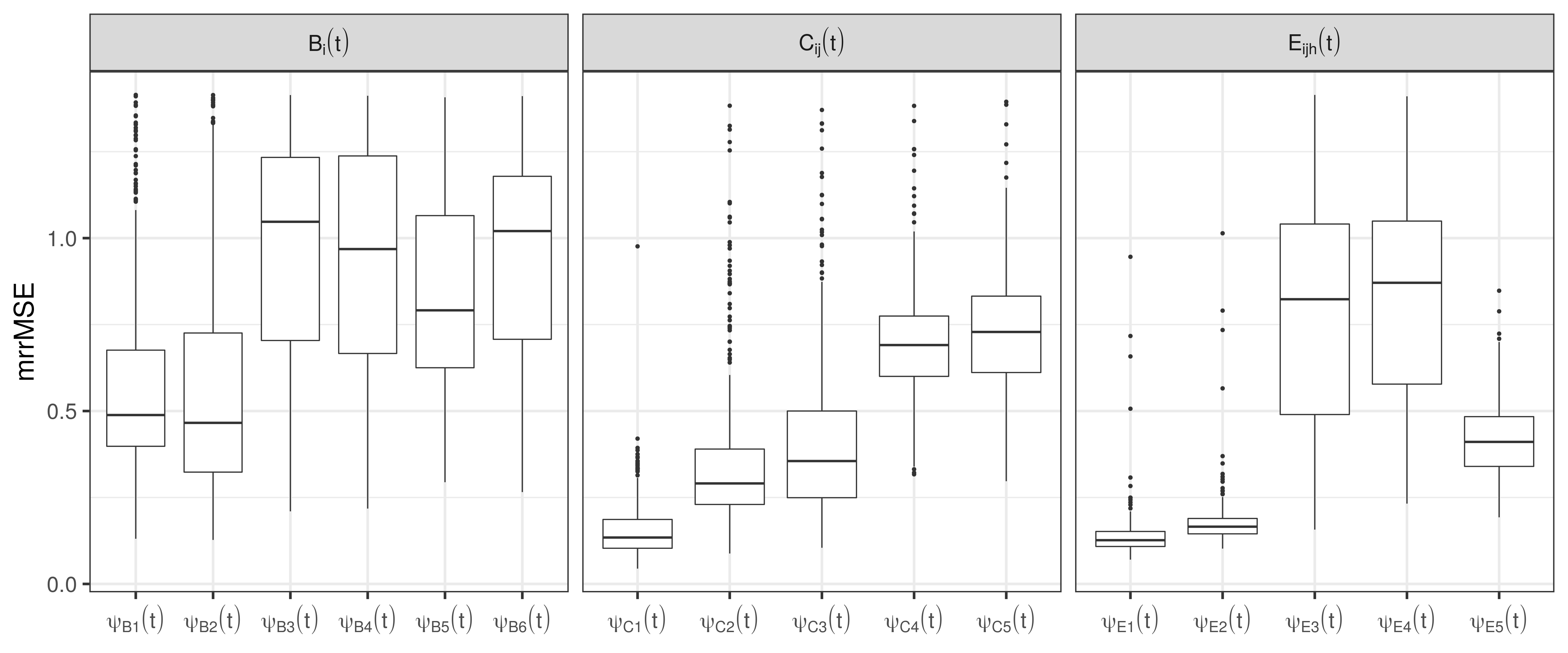}
\caption{\gls{mmse} values of the eigenfunction estimation for the random effects $\bm{B}_{i}(t), \bm{C}_{ij}(t)$, and $\bm{E}_{ijh}(t)$ for scenario 6A.}
\label{fig:app_sim_fpc_sno}
\end{figure}

We observe similar trends for the \gls{mmse} values of the estimated eigenfunctions in scenario 6A as presented in Figure \ref{fig:app_sim_fpc_sno}. Leading eigenfunctions tend to be more accurately estimated and increasing the number of grouping levels ($\bm{B}_{i}(t):25$, $\bm{C}_{ij}(t): 50$, $\bm{E}_{ijh}(t):300$) seems to have a diminishing effect on the variance of the \gls{mmse} values.

\begin{figure}
\includegraphics[width=\textwidth]{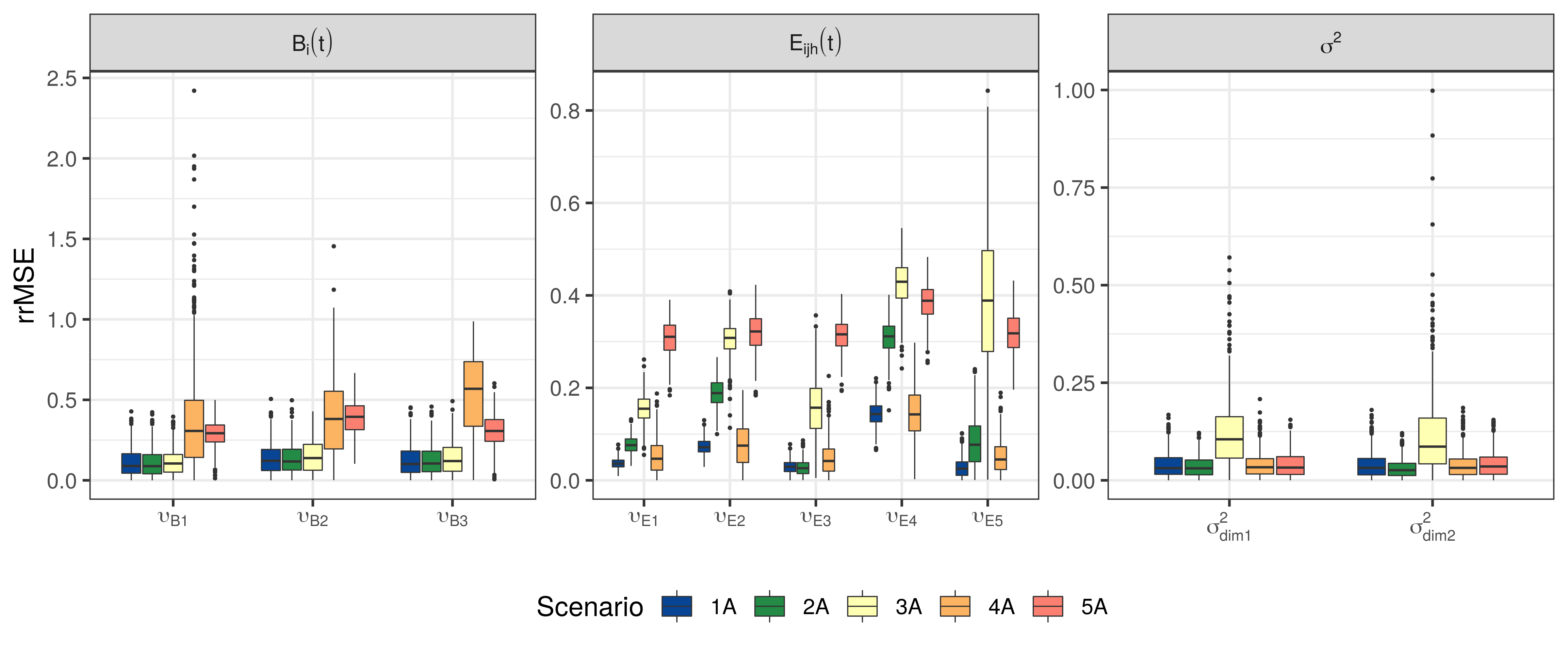}
\caption{\gls{mse} values of the eigenvalue and error variance estimation for the standard scenario A of data settings 1-5.}
\label{fig:app_sim_vars}
\end{figure}

Figure \ref{fig:app_sim_vars} shows the \gls{mse} values of the estimated multivariate eigenvalues and measurement error variances. Compared to scenario 1A, the \gls{mse} values of the smooth residual are higher for scenarios 2A (strong heteroscedasticity) and 3A (sparse data), whereas the \gls{mse} values of the subject-specific random effect are higher for scenario 4A (correlated scores). This is along the lines of the findings for the eigenfunctions and the random effects. Scenario 5A is again not directly comparable and the overall higher \gls{mse} values suggest that the uncertainty in the estimation of the weights of the scalar product influences the accuracy of the estimation of the eigenvalues. With regards to the error variance, the \gls{mse} values are comparable over the different scenarios except for the sparse data setting, where we find higher values. 

We conclude that modes of variation can be recovered well in most of the model scenarios. The nested random effect and its leading modes of variation can be well captured by the \gls{mfamm}.

\subsubsection*{Fixed Effects Estimation}

\begin{figure}
\centering
    \includegraphics[width=0.6\textwidth]{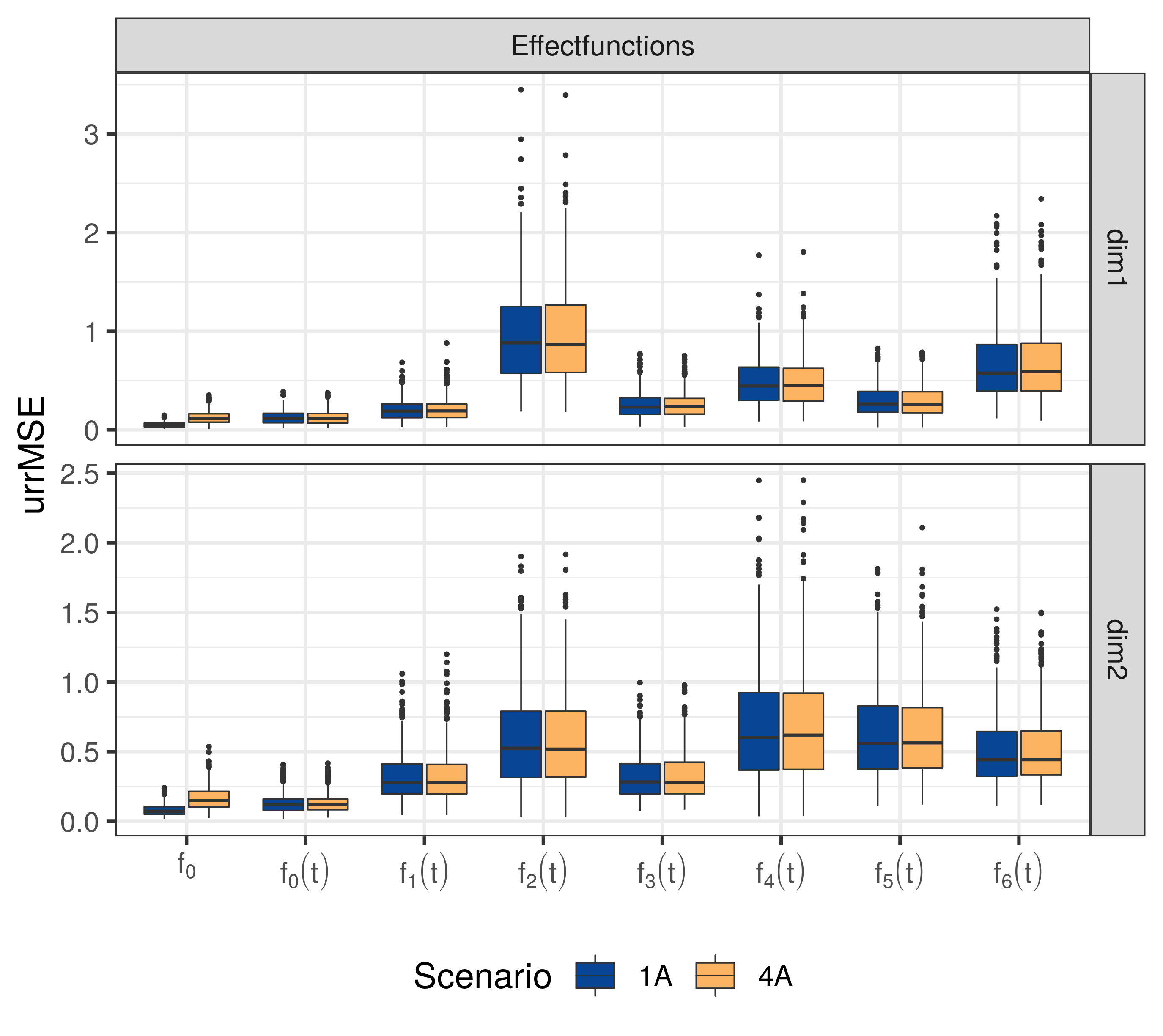}
    \caption{\gls{umse} values of the estimated effect functions in scenarios 1A and 4A.}
    \label{fig:app_sim_eff_umse}
\end{figure}

\begin{figure}
\includegraphics[width=\textwidth]{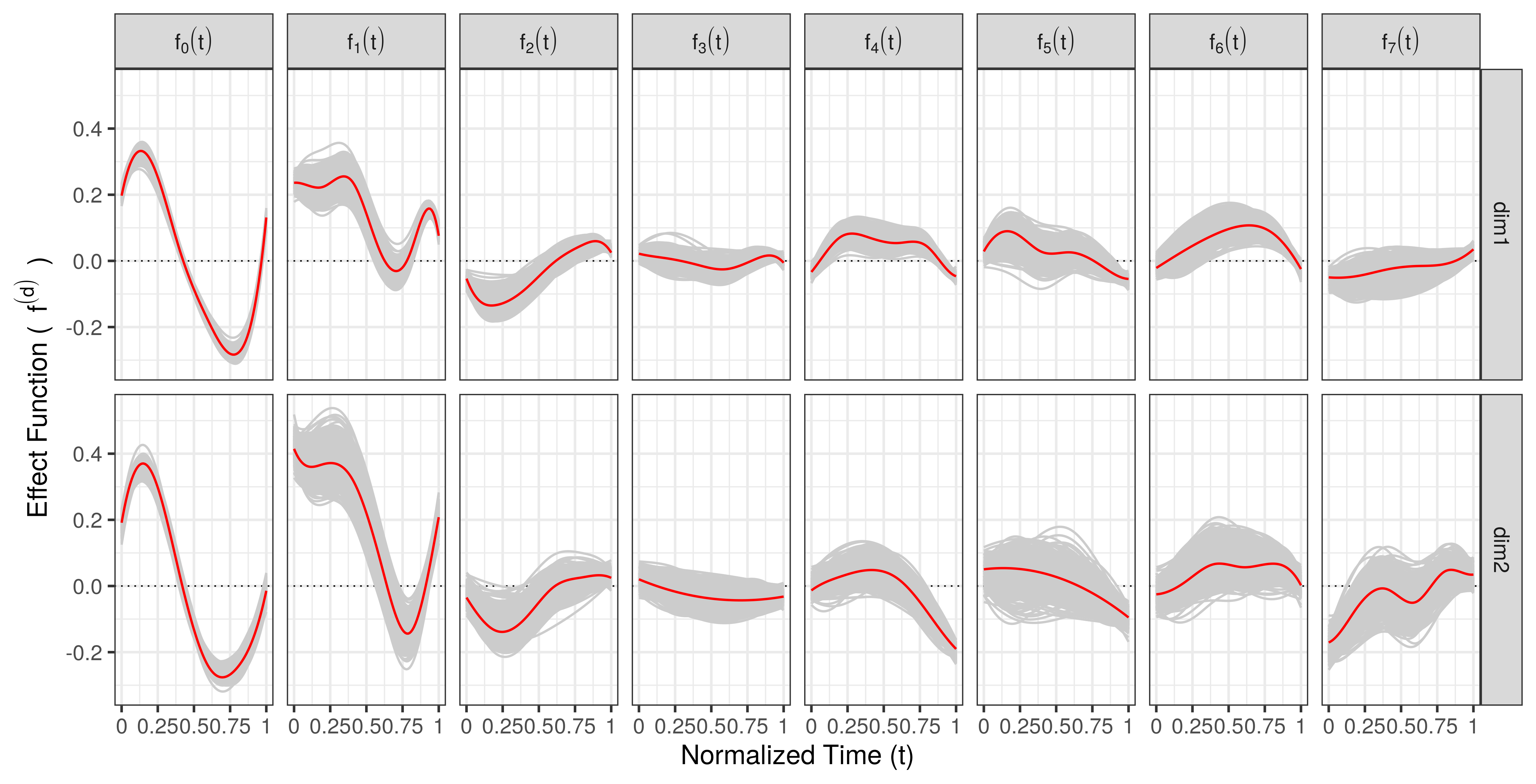}
\caption{Estimated effect functions in scenario 1A for all 500 simulation iterations (grey curves). The red curves show the data generating effect functions.}
\label{fig:app_sim_esteff}
\end{figure}

Figure \ref{fig:app_sim_eff_umse} shows the \gls{umse} values of the estimated effect functions in scenario 1A (blue). We find that the estimation of the functional intercept $\bm{f}_0(t)$ and the covariate effect of \texttt{order} $\bm{f}_1(t)$ yield low \gls{umse} values on both dimensions. The estimation of the other covariates ($\bm{f}_2(t)$ to $\bm{f}_4(t)$) and especially the interactions of \texttt{order} with the other covariates ($\bm{f}_5(t)$ to $\bm{f}_7(t)$) give larger \gls{umse} values and a higher variance of these values. The yellow boxplots show the corresponding values in scenario 4A, thus indicating that only the estimation of the intercept is affected by correlated and uncentred scores (as an empirical score mean different from zero times the corresponding eigenfunctions is captured by the intercept). Figure \ref{fig:app_sim_esteff} plots all estimated effect functions against the data generating effect functions in scenario 1A. This suggests that the \gls{mfamm} can overall capture characteristics of the true underlying effect functions.

\begin{figure}
    \includegraphics[width=\textwidth]{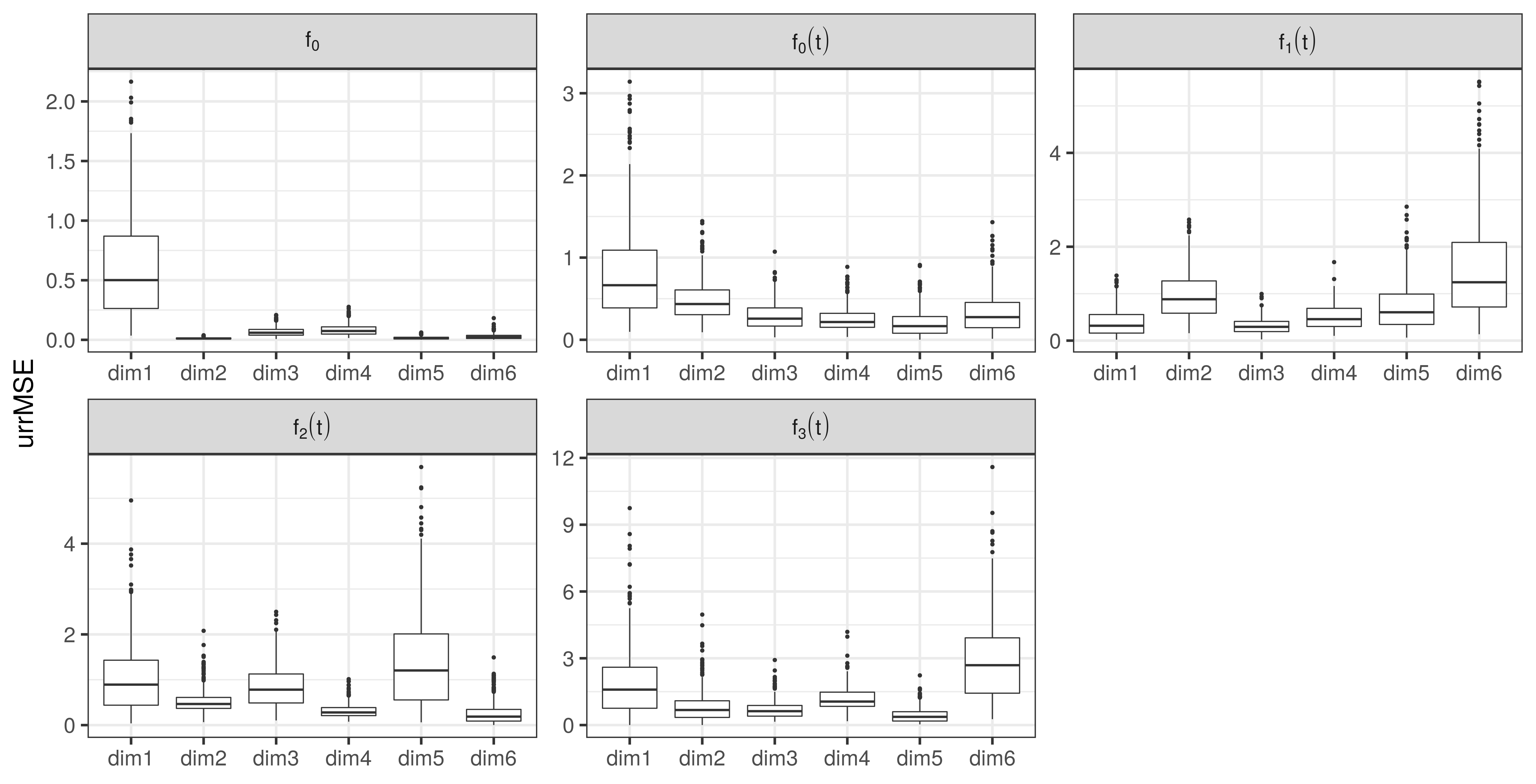}
    \caption{\gls{umse} values of the estimated effect functions in scenario 6A.}
    \label{fig:app_sim_umse_eff_sno}
\end{figure}

\begin{figure}
\includegraphics[width=\textwidth]{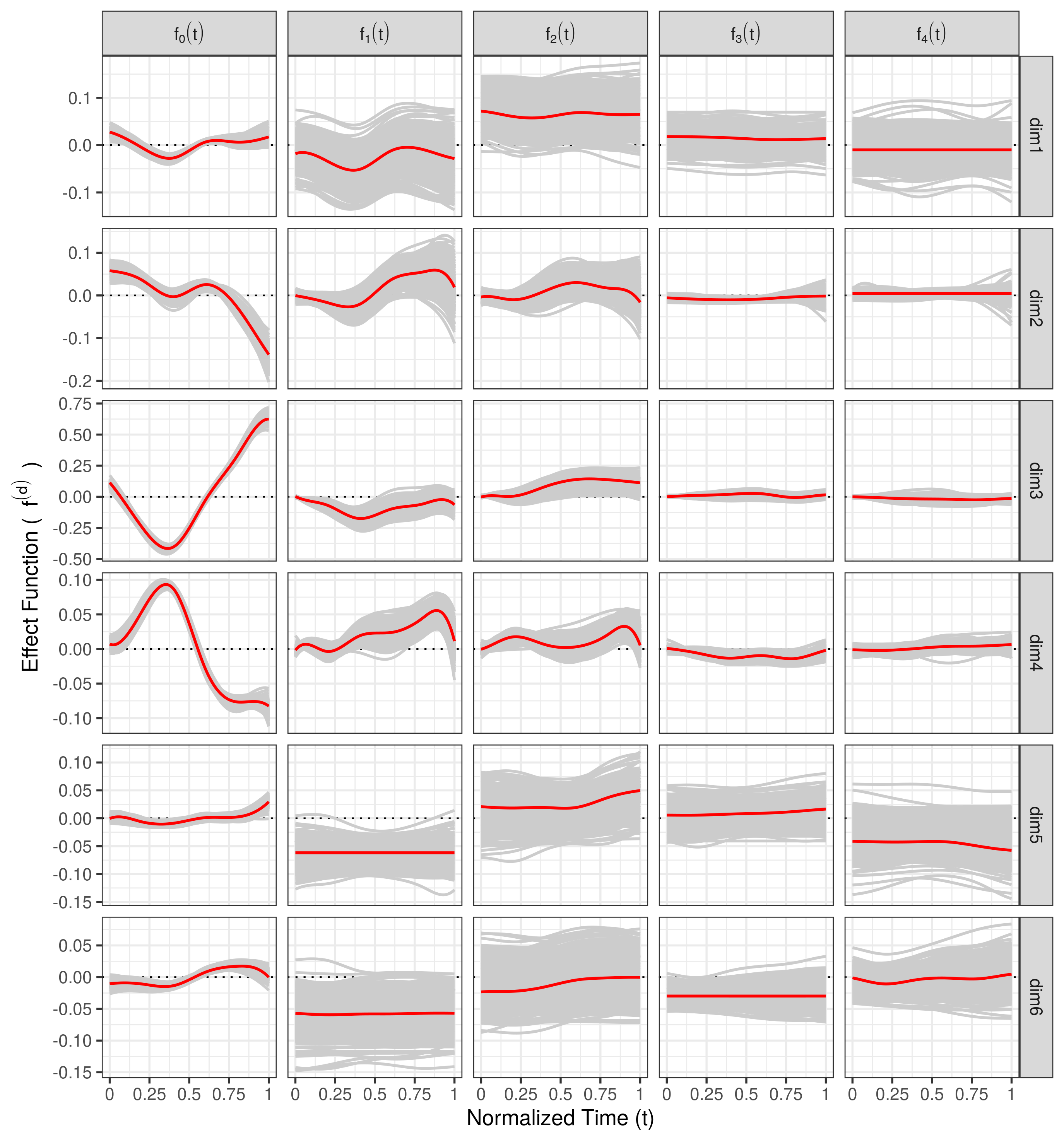}
\caption{Estimated effect functions in scenario 6A for all 500 simulation iterations (grey curves). The red curves show the data generating effect functions.}
\label{fig:app_sim_esteff_sno}
\end{figure}

Figure \ref{fig:app_sim_umse_eff_sno} shows the \gls{umse} values for each effect function in scenario 6A. Overall, the functional intercept $\bm{f}_0(t)$ and the covariate effect of \texttt{skill} $\bm{f}_1(t)$ have a higher estimation accuracy as implied by lower \gls{umse} values. The other effect functions are also somewhat less pronounced, which is why the scaling with the inverse mean norm can give large \gls{umse} values. Figure \ref{fig:app_sim_esteff_sno} shows all estimated effect functions (grey curves) and the corresponding data generating functions in red. This plot suggests that especially on the dimensions $dim1, dim5$, and $dim6$ the effect estimation shows a larger variance across the simulation runs. These dimensions correspond to those dimensions in the data set ($elbox.x, shoulder.x,$ and $shoulder.y$ in Figure \ref{APPENDIXfig:snooker_univ_obs}) where the response functions are relatively constant over $t$ compared to the other dimensions. 

For all modeling scenarios, we use the average point-wise coverage to evaluate the $95\%$ \glspl{cb} of the estimated fixed effects (see Table \ref{app_tab:coverage_phon_sno}). For scenario 1A, we find the \glspl{cb} to cover the true effect  $88-95\%$ of the time. Additional uncertainty e.g.\ about the number of \glspl{fpc} further reduces the coverage (for example 1B). Overall, the coverage of fixed effects by the corresponding \glspl{cb} is comparable over the different data settings (see Appendix Table \ref{app_tab:coverage_phon_sno}). In data settings 2 and 3 the uncertainty in the data is increased which allows for wider \glspl{cb} and thus a slightly better coverage. The averaged pointwise coverage in 4A is comparable to 1A except for the functional intercept. Here, the averaged point-wise coverage of the scalar and functional intercept lie well below $70\%$ as the data setting makes it particularly hard to identify the true underlying mean (the scores of the random effects are not centred or decorrelated and any mean difference to zero is absorbed by the intercept).  In scenario 5A, the coverage tends to be lower than for models based on an unweighted scalar product, possibly due to the added uncertainty from estimation of the weight in the scalar problem. We find averaged point-wise coverage to be considerably lower than its nominal value in scenario 6A. In this scenario, we find that in many of the simulation runs the effects are wrongly estimated as constants (cf. also Figure \ref{fig:app_sim_esteff_sno}), which can lead to undercoverage also in scalar additive mixed models \citep{Grevencomment}.

\begin{table} \centering 
  \caption{Averaged point-wise coverage of the point-wise \glspl{cb} for the estimated effect functions and the averaged coverage of the scalar intercept $\beta_0^{(d)}$ for different scenarios. The coverage is averaged over the 500 simulation iterations and over 100 evaluation points along $t$.} 
  \label{app_tab:coverage_phon_sno} 
\resizebox{\textwidth}{!}{
\begin{tabular}{cc|ccccccccc} 
\\\hline 
\hline 
$d$ & Scenario & $\beta_0^{(d)}$ & $f_0^{(d)}(t)$ & $f_1^{(d)}(t)$ & $f_2^{(d)}(t)$ & $f_3^{(d)}(t)$ & $f_4^{(d)}(t)$ & $f_5^{(d)}(t)$ & $f_6^{(d)}(t)$ & $f_7^{(d)}(t)$ \\ 
\hline
\multirow{11}{*}{$dim1$} & 1A & $93.6$ & $90.6$ & $92.6$ & $89.5$ & $93.1$ & $93.9$ & $91.3$ & $89.0$ & $90.4$ \\ 
& 1B & $93.8$ & $83.5$ & $90.6$ & $86.6$ & $89.7$ & $90.2$ & $87.9$ & $86.6$ & $88.0$ \\ 
& 1C & $93.8$ & $90.7$ & $92.5$ & $89.5$ & $93.1$ & $94.1$ & $91.3$ & $88.9$ & $90.8$ \\ 
& 1D & $93.4$ & $87.7$ & $91.3$ & $86.2$ & $89.5$ & $91.3$ & $87.9$ & $87.1$ & $88.0$ \\ 
& 1E & $92.8$ & $88.7$ & $90.7$ & $86.9$ & $90.2$ & $91.3$ & $88.0$ & $86.9$ & $87.8$ \\ 
& 1F & $94.6$ & $92.2$ & $93.7$ & $90.1$ & $94.1$ & $94.7$ & $91.5$ & $90.3$ & $93.2$ \\ 
& 1U & $92.6$ & $90.8$ & $92.5$ & $90.2$ & $93$ & $92.6$ & $91.7$ & $89.5$ & $88.2$ \\ 
& 2A & $95.4$ & $92.8$ & $94.2$ & $90.6$ & $93.9$ & $94.8$ & $92.9$ & $90.5$ & $92.0$ \\ 
& 3A & $94.4$ & $89.2$ & $93.0$ & $91.2$ & $93.0$ & $91.7$ & $90.8$ & $92.4$ & $93.5$ \\ 
& 4A & $68.4$ & $48.7$ & $92.7$ & $89.7$ & $92.7$ & $93.4$ & $91.4$ & $89.4$ & $90.2$ \\ 
& 5A & $94.0$ & $88.4$ & $91.4$ & $89.8$ & $92.8$ & $92.7$ & $89.7$ & $91.4$ & $94.2$ \\ \hline
\multirow{11}{*}{$dim2$} & 1A & $93.4$ & $87.6$ & $93.3$ & $91.8$ & $94.6$ & $92.9$ & $93.3$ & $93.5$ & $94.1$ \\ 
& 1B & $95.0$ & $86.1$ & $93.1$ & $91.9$ & $94.0$ & $93.3$ & $93.6$ & $92.2$ & $92.5$ \\ 
& 1C & $93.8$ & $87.5$ & $93.2$ & $91.9$ & $94.9$ & $93.0$ & $93.4$ & $93.5$ & $94.0$ \\ 
& 1D & $91.2$ & $86.5$ & $90.7$ & $89.4$ & $89.4$ & $89.5$ & $90.0$ & $88.2$ & $90.0$ \\ 
& 1E & $91.0$ & $86.6$ & $90.8$ & $90.0$ & $90.0$ & $89.3$ & $90.2$ & $88.5$ & $90.0$ \\ 
& 1F & $93.4$ & $87.8$ & $92.7$ & $92.3$ & $93.7$ & $92.5$ & $93.2$ & $92.9$ & $94.0$ \\ 
& 1U & $93.2$ & $87.1$ & $92.3$ & $93.6$ & $94.7$ & $92.0$ & $94.0$ & $93.9$ & $93.6$ \\ 
& 2A & $94.8$ & $89.4$ & $92.5$ & $93.2$ & $94.5$ & $93.4$ & $93.8$ & $93.4$ & $92.7$ \\ 
& 3A & $94.6$ & $90.1$ & $92.7$ & $92.8$ & $95.8$ & $94.1$ & $94.0$ & $95.2$ & $91.9$ \\ 
& 4A & $68.6$ & $54.7$ & $93.2$ & $92.3$ & $94.3$ & $92.7$ & $93.3$ & $93.7$ & $93.8$ \\ 
& 5A & $90.4$ & $83.5$ & $88.1$ & $88.6$ & $93.9$ & $87.5$ & $86.7$ & $91.9$ & $91.4$ \\  
\hline \\[-1.8ex] 
\end{tabular} }
\resizebox{0.7\textwidth}{!}{
\begin{tabular}{cc|cccccc} 
\\\hline 
\hline 
$d$ & Scenario & $\beta_0^{(d)}$ & $f_0^{(d)}(t)$ & $f_1^{(d)}(t)$ & $f_2^{(d)}(t)$ & $f_3^{(d)}(t)$ & $f_4^{(d)}(t)$\\ 
\hline
$dim1$ & \multirow{6}{*}{6A} & $87.4$ & $84.9$ & $90.8$ & $80.7$ & $74.4$ & $76.0$ \\
$dim2$ &  & $87.6$ & $88.3$ & $87.9$ & $86.9$ & $75.2$ & $81.4$ \\
$dim3$ &  & $82.4$ & $85.3$ & $87.1$ & $84.2$ & $81.6$ & $71.7$ \\ 
$dim4$ &  & $82.4$ & $82.2$ & $83.6$ & $78.5$ & $78.3$ & $74.1$ \\ 
$dim5$ &  & $85.4$ & $82.1$ & $84.3$ & $84.3$ & $78.0$ & $76.3$ \\ 
$dim6$ &  & $86.2$ & $74.1$ & $83.0$ & $84.2$ & $79.3$ & $65.7$ \\ 
\hline \\ 
\end{tabular} }
\end{table} 

\begin{figure}
\includegraphics[width=\textwidth]{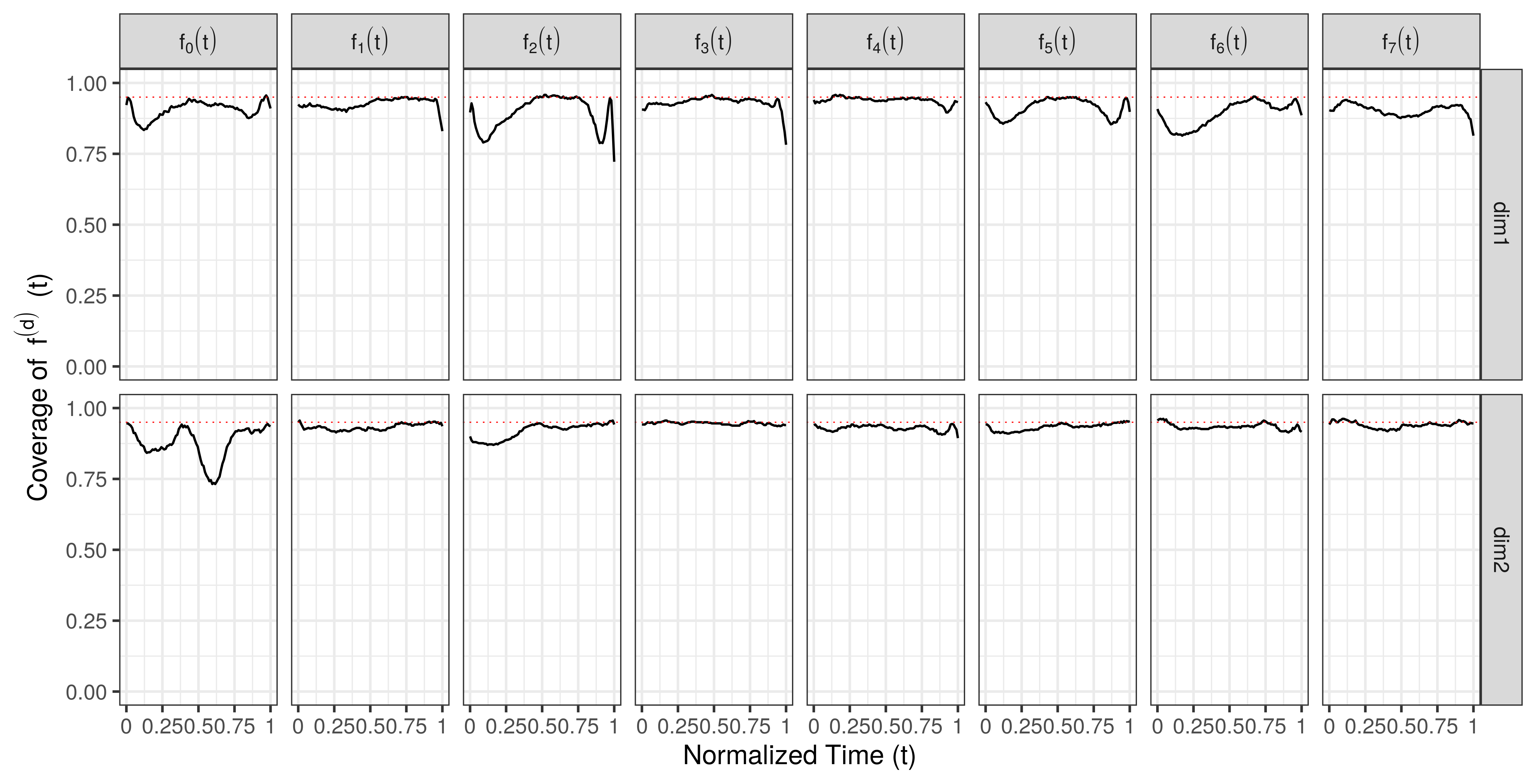}
\caption{Point-wise averaged coverage over the functional index for the estimated effect functions in scenario 1A. The nominal value is $95\%$ (red dotted line).}
\label{fig:app_sim_coverage}
\end{figure}

Figure \ref{fig:app_sim_coverage} shows the point-wise coverage average over the 500 simulation iterations in scenario 1A. It suggests that coverage tends to lie below the nominal value in areas close to the border of the functions.

\end{document}